\documentclass[submission,copyright,noncommercial]{eptcs}
\providecommand{\event}{arXiv preprint}

\usepackage[top=3cm,bottom=3cm,left=2cm,right=2cm]{geometry}

\usepackage{iftex}

\ifpdf
  \usepackage{underscore}         
  \usepackage[T1]{fontenc}        
\else
  \usepackage{breakurl}           
\fi

\title{Tracking relevance syntactically: A nested approach to weak relevant logic}

\author{
Fabio De Martin Polo
\institute{Scuola Universitaria Superiore IUSS\\ Class of Human and Life Sciences}
\email{\quad fabio.demartinpolo@iusspavia.it}
\and
Han Gao
\institute{Institute of Computer Science\\
Czech Academy of Sciences}
\email{\quad gao@cs.cas.cz}
}

\newcommand{\titlerunning}{A Nested Approach to Weak Relevant Logic}
\newcommand{\authorrunning}{F. De Martin Polo \& H. Gao}

\hypersetup{
  bookmarksnumbered,
  pdftitle    = {\titlerunning},
  pdfauthor   = {\authorrunning},
  pdfsubject  = {EPTCS},               
  pdfkeywords = {keyword1, keyword2} 
}

\usepackage{graphicx}
\usepackage{amsmath, amssymb}
\usepackage{amsfonts}
\usepackage{mathtools}
\usepackage{amsthm}
\newtheorem{theorem}{Theorem}
\newtheorem{lemma}{Lemma}
\newtheorem{definition}{Definition}
\newtheorem{example}{Example}
\newtheorem{proposition}{Proposition}
\newtheorem{corollary}[theorem]{Proposition}
\newtheorem{fact}{Fact}
\usepackage{ebproof}
\usepackage{float}
\usepackage{capt-of} 
\usepackage{rotating}
\usepackage{pifont}
\usepackage{mdframed}

\usepackage{enumitem}

\makeatletter
\newsavebox{\@bra}
\newsavebox{\@brb}
\DeclarePairedDelimiterX\myinnerp[1]{.}{.}{%
	\delimsize\langle%
	\hspace*{0.3mm}\hspace*{0.55mm}\savebox{\@bra}{\(\displaystyle\left\langle\vphantom{#1}\right.\)}\hspace*{-1.035\wd\@bra}%
	\delimsize\langle%
	#1%
}
\makeatother

\usepackage{tikz}
\usetikzlibrary{arrows.meta, positioning}
\usepackage{ragged2e}
\usepackage{xhfill}
\usepackage{listings}
\usepackage{dsfont}
\usepackage[linesnumbered,ruled,vlined]{algorithm2e}
\usepackage{ebproof}
\usepackage{adjustbox}

\usepackage{virginialake}
\usepackage{framed}
\usepackage{thm-restate}

\usepackage{makecell}
\usepackage{stmaryrd}

\usepackage{enumitem}
\usepackage{tabularx}

\usepackage{xcolor}
\definecolor{lightgray}{gray}{0.9} 

\usepackage{cellspace} 
\newlist{enumeratee}{enumerate}{5}
\setlist[enumeratee]{label*=\arabic*.,leftmargin=*,font=\it}

\def\+#1{\mathcal{#1}}

\newcommand{\nec}{\Box}

\newcommand{\logicb}{\mathsf{B}\xspace}

\newcommand{\bplus}{\mathsf{B}^+}
\newcommand{\bplusk}{\mathsf{B}^+{\mathsf{K}^\nec}\xspace}

\newcommand{\cbplus}{\mathbf{C}_{\bplus}\xspace}

\newcommand{\hbplus}{\+H_{\bplus}\xspace}

\usepackage{pdflscape}

\newcommand{\sand}{\mathbin{/\mkern-5mu/}}

\newcommand\iblock[1]{\langle {#1}\rangle}

\newcommand{\axfont}[1]{\mathsf{#1}}
\newcommand{\ax}[1]{\axfont{ax{#1}}}
\newcommand{\modp}{\axfont{mp}}

\newcommand{\cut}{(\mathsf{cut})}

\usepackage{color}

\newcommand{\idone}{({\sf id})}

\def\lef#1{#1^\bullet}
\def\rig#1{#1^\circ}
\def\dow#1{{#1}^\ast}
\def\remove#1{\widetilde{{#1}}}

\newcommand{\ruletor}{\rig \to}
\newcommand{\ruletol}{\lef \to}
\newcommand{\ruleandr}{\rig \land}
\newcommand{\ruleandl}{\lef \land}
\newcommand{\ruleorl}{\lef \lor}
\newcommand{\ruleorr}{\rig \lor}

\newcommand{\ruleliftib}{(\mathsf{lift})}

\newcommand{\powerset}{\raisebox{.15\baselineskip}{\Large\ensuremath{\wp}}}

\newcommand{\rulecl}{\mathsf{c}}
\newcommand{\rulemerge}{\mathsf{m}}
\newcommand{\rulewl}{\mathsf{w}}

\newcommand{\rulec}{\mathsf{c}_{\ell}}

\newcommand{\rulecutl}{\mathsf{cut}_{\ell}}

\newcommand\csseq[1]{G\{#1\}}
\newcommand\dowcsseq[1]{\dow{G}\{#1\}}

\newcommand\var[1]{\mathsf{Var}(#1)}

\newcommand{\idl}{\mathsf{id}_\ell}
\newcommand{\ruletorl}{\rig{\to}_\ell}
\newcommand{\ruletoll}{\lef{\to}_\ell}

\newcommand{\cip}{$\mathrm{CIP}$}

\newcommand{\dep}[2]{\mathsf{dp}_{#1}(#2)}
\newcommand{\indep}[3]{\mathsf{dp}_{#1}^{#2}(#3)}

\newcommand\focus[1]{\triangleright{(#1)}}
\newcommand\dsseq[1]{H\{#1\}}

\newcommand\act[1]{\mathsf{act}{(#1)}}
\newcommand\actl[2]{\mathsf{act}^L_0{(#1)}{\{ #2\}}}
\newcommand\actr[2]{\mathsf{act}^R_0{(#1)}{\{ #2\}}}

\newcommand\vactl[1]{\mathsf{Var}^L_0{(#1)}}
\newcommand\vactr[1]{\mathsf{Var}^R_0{(#1)}}

\theoremstyle{remark}
\newtheorem{rem}{Remark}

\begin{document}

\maketitle

\begin{abstract}

The variable-sharing property (VSP) is regarded as a key feature of relevant logics, while the Craig interpolation property (CIP), formulated with respect to relevant implication, implies the VSP in constant-free propositional languages. However, to the best of our knowledge, no constant-free distributive relevant logic has previously been shown to enjoy the CIP. This absence of positive results has contributed to a widespread impression among relevant logicians that relevance and Craig interpolation are incompatible. 
In this work, we address this tension by proving that the negation-free fragment of base relevant logic $\logicb$, denoted $\bplus$, enjoys both properties. 
Our proof is purely syntactic and is based on a novel nested sequent calculus featuring binary relevant blocks as primitive structural constructs. Unlike conventional nested sequent calculi, where unary nesting reflects the binary accessibility relation of Kripke semantics, our calculus employs a binary nesting operator that directly reflects the ternary relational structure of Routley-Meyer semantics for relevant logics. 
After establishing soundness and cut-free completeness of the calculus, we prove several meta-logical properties for the logic. 
Most notably, we provide the first syntactic proof of the VSP (along with a variety of its stronger variants) and the first proof of the CIP for $\bplus$, both derived directly from the cut-free system using a structural notion internalized within 
nested sequents. 

\end{abstract}

\section{Introduction}\label{sec:intro}

Relevant logics form a family of non-classical logics introduced to address the paradoxes of material and strict implication \cite{belnap.anderson}. They admit a modular presentation, obtained by starting from a minimal set of axioms and inference rules to construct systematic extensions. Within this family, it is customary to distinguish between \textit{weak} and \textit{strong} systems, a distinction that typically hinges on whether the axiom of contraction $(A \to (A \to B)) \to (A \to B)$ is present. Common to these logics is the \textit{variable-sharing property}, according to which an implication $A \to B$ is 
a theorem of the given logic 
only if $A$ and $B$ share at least one propositional variable. 
This property is regarded as an intrinsic feature of relevant logics~\cite{Standefer2025vsp} and has been extensively investigated by relevant logicians~\cite{Brady:1984-depthrl}, together with a variety of stronger forms, such as the \textit{strong relevance property} and the \textit{depth relevance property}~\cite{Brady:1984-depthrl,logan2021strong}. 
At the same time, interest in Craig interpolation in non-classical logic, has led to the question of whether relevant logics enjoying the variable sharing property, are compatible with it. This question, however, has so far been answered negatively for some well known strong relevant logics like $\mathsf{T}, \mathsf{E}$ and $\mathsf{R}$ \cite{urquhart1993failure}, but positively 
for the `quasi relevant' logic $\mathsf{RM}$ extended with Ackermann's constant $\mathsf{t}$ and several extensions of it (see \cite[\S 29.3.3; \S 29.4]{belnap.anderson}; \cite{MarchioniMetcalfe2012}). 

It is noted in~\cite{standefer2026relevant} that it is not currently known whether any 
relevant logics between $\mathsf{B}$ and $\mathsf{R}$ 
enjoy Craig interpolation formulated in terms of the relevant implication; more specifically, to the best of our knowledge, no positive result concerning interpolation in this sense has been established for a constant-free distributive relevant logic with the standard variable-sharing property. 
Indeed, as noted in~\cite{FussnerTedder2026}, this has even given rise to an informal consensus that \textit{(weak) relevant logics don't have (Craig) interpolation}: the relationship between the two properties is regarded as incompatible, in the sense that logics enjoying the variable-sharing property are taken to lack interpolation. 
However, despite the lack of positive results in both strong and weak relevant logics, the specific situations differ. 
As remarked above, Urquhart~\cite{urquhart1993failure} showed that interpolation fails for the strong relevant logics $\mathsf{T}$, $\mathsf{E}$ and $\mathsf{R}$, 
whereas for the weak systems, the issue about interpolation is unclear. In particular, it was recently mentioned that whether the base logic $\logicb$ and its positive fragment $\bplus$ have the CIP remains open~\cite{Knudstorp2026}. 

In this paper, we address this tension and 
show that for the base positive relevant logic $\bplus$ the situation is indeed contrary to the strong relevant logics: 
going beyond Urquhart's negative results, we use a novel nested sequent calculus for $\bplus$ to provide, first, an alternative proof of the variable sharing property and, second, a proof of Craig interpolation property for it. 
Our results answers the question of whether $\bplus$ enjoys (Craig) interpolation, which also provide a partial answer to the question concerning the relationship between interpolation and the variable sharing property \cite[\S 3.5]{fussner2025interpolation}.

The interest 
in interpolation  
for $\bplus$ is also driven by the fact that, in most circumstances, $\bplus$ is taken to be the foundational system of the relevant family. It admits conservative extensions to richer languages, such as the full relevant logic $\sf B$ via De Morgan negation $\sim$ \cite{rl.rivals}, and $\sf CB$ via Boolean negation $\neg$ \cite{Meyer2004}, generally serving as the baseline for systematic investigations into positive relevant logics, their diverse negation extensions (see \cite{RoblesMendez2018RM} for a survey), and their stronger companions. However, besides this foundational role, it is worth mentioning that $\bplus$ has also been studied as an object of interest in its own right, in particular, in connection with combinators \cite{DunnMeyer1997Combinators,PalMeyer2005}, computational complexity and decidability \cite{Vigano2000}, and algorithmic reasoning \cite{Gochet2005}. 

\smallskip

\noindent \textbf{Methodology.} \ 
The main results of the paper are based on a novel nested calculus for $\bplus$. 
Regarding technical developments, relational and algebraic studies of weak relevant logics have been thoroughly conducted, leading to a mature understanding of their underlying semantics. However, despite these semantic advances, their proof theory remains comparatively underdeveloped (for a general overview, see e.g. \cite[Sec.~7]{FabioDMP2025}; see also \cite{bimbobook}). 
While existing proof-theoretic treatments of relevant logic rely primarily on labelled systems (see \cite{urquhart.proof.semil.,vigano,kurokawa.negri,FabioDMP2023}) and display calculi~(cf. e.g., \cite{gore1998substructural,belnap1982display}), we move beyond these approaches by introducing a novel syntactic proof system based on a generalized form of nested sequents. 

Nested sequent calculi are a generalization of ordinary Gentzen-style sequent systems in which sequents may occur \textit{inside} other sequents.
This additional structural expressiveness allows nested calculi to represent relational information such as accessibility relations between worlds in Kripke semantics directly at the proof-theoretic level, without resorting to labels or external annotations. 
As a result, nested sequents provide a uniform and flexible framework for the proof theory of a wide range of logics, including modal logic \cite{Brunner:2009}, intuitionistic logic \cite{Fitting:2014}, and their combinations \cite{Strassburger:2013-ik-cut,kuznets2019maehara,Galmiche:Salhi:2010}. This design also appears in systems closely related to relevant logics, most notably conditional logics \cite{AlendaOlivettiPozzato2016}. More recently, \cite{CSL-2024,IJCAR-2024,gao2026bi-ik} extend this tradition by introducing \emph{bi-nested} calculi for certain intuitionistic modal logics, where two kinds of nesting blocks coexist in the syntax, reflecting respectively the preorder underlying intuitionistic implication and the modal accessibility relation in the semantics.

However, despite these variations and other differences in their syntax, all the systems mentioned above employ \emph{unary} blocks. Accordingly, in those works, a nested sequent occurring inside a block is interpreted as a direct successor of the outer sequent in the corresponding semantics. 
In this paper, we drop this restriction and develop the nested framework in a more general setting. We introduce \emph{binary} blocks in the syntax of nested sequents: rather than containing a single sequent, semantically corresponding to a direct successor, a block contains a pair of successors. This syntactic feature captures the ternary relation underlying the semantics of relevant logics.
In this context, it is worth noting that an important virtue of the calculus for $\bplus$ introduced here is exactly its tight connection to the ternary relational semantics characteristic of relevant logics \cite{Meyer2004}. Most notably, unlike labelled calculi, which typically require semantic frame conditions to be encoded as primitive structural rules, our system employs only a single structural rule corresponding to monotonicity. All other rules reflecting frame conditions are shown to be admissible. As a result, in comparison with existing calculi for $\bplus$, the proof system proposed in this paper achieves completeness with a minimal increase in syntactic complexity and a parsimonious set of inference rules. 

\smallskip

\noindent \textbf{Contribution of this paper.} \ 
The contributions of this paper include a novel notion of nested sequents with applications in $\bplus$, ranging from a new calculus, syntactic cut-admissibility, first syntactic proofs of relevance properties and first proof of the CIP. 
To be precise, first, we propose a nested sequent calculus $\cbplus$ for $\bplus$ built on binary nested blocks, which is sound and complete and admits cut. Then by means of the cut-free calculus, we provide the first syntactic proofs of the variable-sharing property and of the depth relevance property for $\bplus$, via a notion of reachability between nodes of a nested sequent. The strong variable-sharing property follows with almost no extra work, the calculus already recording the polarities. Moreover, we prove Craig interpolation for $\bplus$. To the best of our knowledge this is the first positive result on Craig interpolation of a weak relevant logic with the variable-sharing property. It is worth mentioning that 
our proof is constructive but departs from the traditional Maehara's method: 
it extracts an interpolant from a given derivation and considers only the innermost structure with the unique partition given by the single-output feature, which is shown to be adequate for constructing an interpolant for an implicational formula. 

\smallskip

\noindent \textbf{Structure of the paper.} \ 
The rest of the paper is organized as follows. Section~\ref{sec:logic} fixes the language and recalls
$\bplus$ with its Routley-Meyer semantics. Section~\ref{sec:CBcut} introduces nested
sequents with binary `relevant' blocks, presents a novel label-free calculus, and proves its soundness and completeness. 
Section~\ref{sec:relevance-property} develops the notion of reachability and proves the variable-sharing and depth relevance properties by means of the calculus. 
Section~\ref{sec:cip} establishes the Craig interpolation property for $\bplus$. Finally Section~\ref{sec:con} concludes with a discussion of directions for future research.
\section{The negation-free relevant logic $\bplus$}\label{sec:logic}

In this section, we introduce a Hilbert-style proof system together with a relational semantics for the negation-free fragment of the relevant logic $\mathsf{B}$ (cf. \cite{rl.rivals}), denoted $\mathsf{B}^+$. 

Now, let us begin by defining the propositional language considered throughout this paper. 

\begin{definition}
    Let $\mathsf{At}$ be a countable set of propositional atoms, ranged over by $p,q,\dots$. The propositional language is generated by the set of connectives $\mathsf{Con} = \{\land, \lor, \to\}$. 
We denote by $\mathsf{Frm}$ 
the set of propositional formulas, which are inductively generated by the following grammar:
$$
\begin{aligned}
    A \coloneqq  p \in \mathsf{At}
\mid (A \land A)
\mid (A \lor A)
\mid (A \to A)
\end{aligned}
$$
The \textit{complexity} of a formula $A$, denoted by $\lvert A\rvert$, is defined as follows: $\lvert p\rvert:=0$ (for every $p\in\mathsf{At}$) and $\lvert A\heartsuit B\rvert:=\lvert A\rvert+\lvert B\rvert+1$ (for $\heartsuit\in\{\land,\lor,\to\}$). 
Moreover, we denote the set of propositional atoms occurring in $A$ by $\var{A}$. For a set $\Gamma$ of formulas, let $\var{\Gamma}:=\bigcup_{A\in\Gamma}\var{A}$. 
\end{definition}


Next, we examine $\bplus$, presenting its Hilbert system denoted $\hbplus$, together with the 
Routley-Meyer semantics. 
$\hbplus$ is defined by the axioms and closed under the inference rules 
of Figure~\ref{fig:axioms-bplus}. 
As usual, a formula $A$ is a \textit{theorem} of $\hbplus$, written $\vdash_{\hbplus} A$, iff there exists a finite sequence of formulas $A_1, \dots, A_n$ such that $A_n = A$ and each $A_i$ is either an instance of an axiom of $\hbplus$ or is obtained from preceding formulas in the sequence by an inference rule.

\begin{figure}[t]
\begin{center}
	\begin{minipage}{.6\textwidth}
        \begin{adjustbox}{max width=\textwidth}
		  \begin{tabular}{ccl}
                $\ax{1}$ & & $A \to A$\\
                $\ax{2}$ & & $A_1  \land A_2 \to A_i$ \quad where $i\in\{1,2\}$\\
                $\ax{3}$ & & $(A \to B)  \land (A \to C)  \to (A \to (B \land C))$ \\
                $\ax{4}$ & & $A_i  \to (A_1  \lor A_2)$ \quad where $i\in\{1,2\}$\\
                $\ax{5}$ & & $(A \to C)  \land (B \to C)  \to ((A \lor B)  \to C)$ \\
                $\ax{6}$ & & $A \land (B \lor C)  \to (A \land B)  \lor (A \land C)$\\
			    & \\
			\multicolumn{3}{c}{
            \begin{prooftree}
                \hypo{A}
              \hypo{A \to B}
              \infer2[$\modp$]{B}
            \end{prooftree} 
            \qquad
            \begin{prooftree}
                \hypo{A}
              \hypo{B}
              \infer2[$\sf adj$]{A\wedge B}
            \end{prooftree}
              \qquad
              \begin{prooftree}
                \hypo{A\to B}
              \hypo{C\to D}
              \infer2[$\sf aff$]{(B \to C)  \to (A \to D)}
            \end{prooftree}
			}
		\end{tabular}
        \end{adjustbox}
	\end{minipage}%
    \caption{The Hilbert system $\hbplus$ for $\bplus$. }
	\label{fig:axioms-bplus}
\end{center}
\end{figure}

The Routley-Meyer semantics for $\bplus$ 
features a ternary relation among states (see \cite{routleymeyer73,rl.rivals,Meyer2004}). 
Before turning to the formal details, it is worth noting that relational semantics for propositional relevant logics admits several variants. One may, for instance, distinguish a subset of \emph{regular states} within the set of all states, or alternatively collapse this structure to a single distinguished regular state, typically denoted by $0$ (cf. e.g., \cite{RoutleyMeyer1972,giambrone.red,slaney.red}). 
The former is used in this paper. 

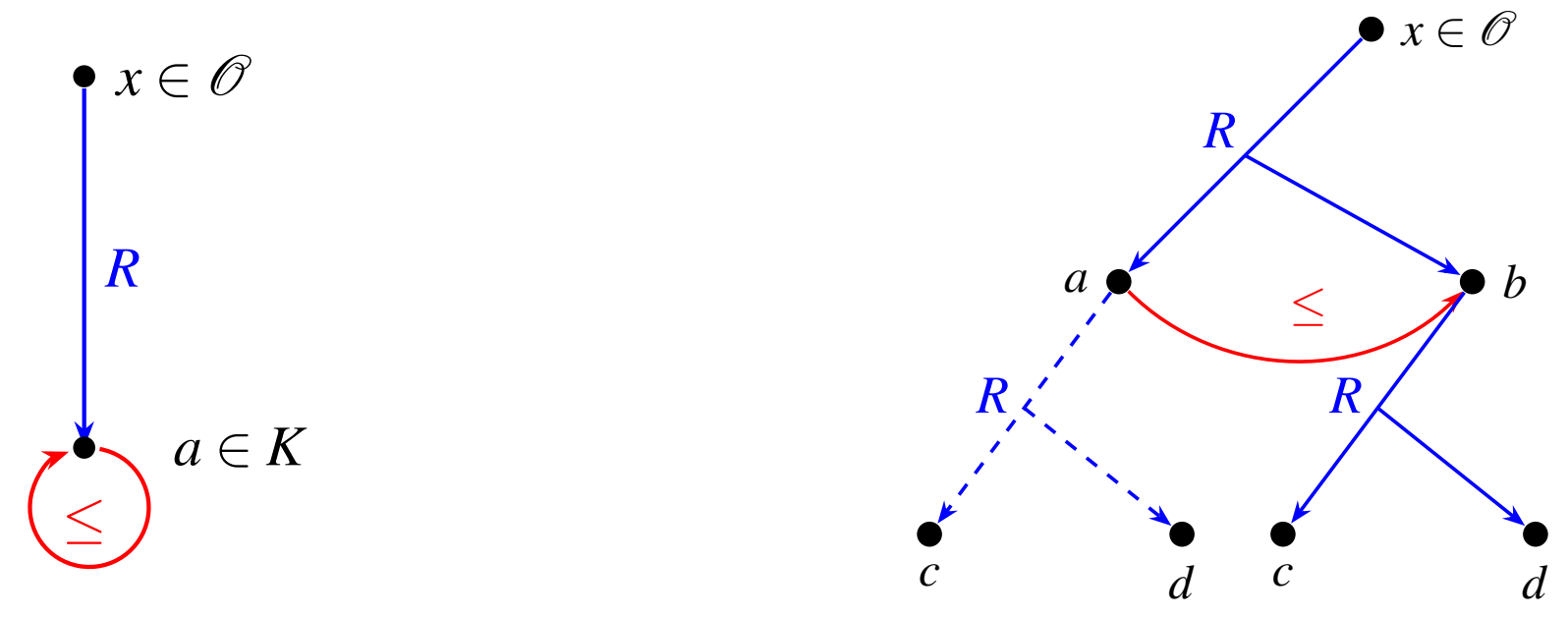
\begin{figure}[t]
\begin{center}
\begin{minipage}{.4\textwidth}
\centering
\scalebox{1}{%
\begin{tikzpicture}[
  dot/.style={circle, fill=black, inner sep=1.5pt},
  every label/.style={font=\normalsize}
]
\node[dot, label={right:$x \in \+O$}] (x) at (0, 0) {};

\draw[-{Stealth[length=5pt]}, thick,blue]
  (x) -- node[right, font=\normalsize] {$R$} (0, -2.5);

\node[dot, label={right:$\quad a \in K$}] (a) at (0, -2.5) {};

\draw[-{Stealth[length=5pt]}, thick, red]
  (0, -3.1) ++(80:0.6) arc[start angle=80, end angle=-250, radius=0.4];

\node[font=\normalsize,red] at (0, -3) {$\leq$};
\end{tikzpicture}%
}
\end{minipage}
\begin{minipage}{.4\textwidth}
\centering
\scalebox{0.85}{%
\begin{tikzpicture}[
  dot/.style={circle, fill=black, inner sep=2pt},
  lbl/.style={font=\large}
]
 
\node[dot, label={right:$x\in\+O$}] (x)  at (0, -1) {};
\node[dot, label={left:$a$}]  (a)  at (-2.0, -3) {};
\node[dot, label={right:$b$}] (b)  at (.8, -3) {};
 
\node[dot, label={below:$c$}]  (c1) at (-3.5, -5) {};
\node[dot, label={below:$d$}] (d1) at (-1.5, -5) {};
 
\node[dot, label={below:$c$}]  (c2) at (-0.7, -5) {};
\node[dot, label={below:$d$}] (d2) at (1.3, -5) {};
 
\coordinate (f1) at (-1, -2);
\draw[-{Stealth[length=5pt]}, thick, blue] (x) -- (a);
\draw[-{Stealth[length=5pt]}, thick, blue] (f1) -- (b);
\node[lbl,blue] at (-1.2, -1.8) {$R$};
\node[font=\normalsize,red] at (-.5, -3.2) {$\leq$};
\draw[-{Stealth[length=5pt]},thick, red] (a) to[bend right=45] (b);
 
\coordinate (f2) at (-2.75, -4);
\draw[-{Stealth[length=5pt]}, thick, dashed, blue] (a) -- (c1);
\draw[-{Stealth[length=5pt]}, thick, dashed, blue] (f2) -- (d1);
\node[lbl,blue] at (-3, -3.9) {$R$};
 
\coordinate (f3) at (.05, -4);
\draw[-{Stealth[length=5pt]}, thick, blue] (b) -- (c2);
\draw[-{Stealth[length=5pt]}, thick, blue] (f3) -- (d2);
\node[lbl,blue] at (-.2, -3.9) {$R$};
 
\end{tikzpicture}%
}
\end{minipage}
\captionof{figure}{Frame conditions F1 (left) and F2 (right). The Y-like fork structure in blue represents the ternary relation, where the successor is a pair of nodes. The arrows in red represent the $\leq$-order between the two worlds in the same pair and the pair is a successor of a regular world. In the left graph, the two nodes in the pair coincide, make the relation $\leq$ reflexive. In the right graph, the dashed arrows mean that the same worlds $c,d$ also form a pair of successors of $a$. }
\label{fig:frame-condition}
\end{center}
\end{figure}

\begin{definition}[$\bplus$-frames and $\bplus$-models]\label{def:frameB+}
A \emph{$\bplus$-frame} is a triple $\mathcal{F} = \langle K, \+O, R\rangle$, where $K$ is a non-empty set of worlds, $\+O \subseteq K$ is a non-empty set of \emph{regular worlds}, and $R$ is a ternary relation on $K$. For any $a,b \in K$, we define a binary relation $\leq$ as follows:
$a \leq b  \text{~~iff~~} \exists x(x\in\+O~\&~Rxab)$. The following conditions are imposed on $\bplus$-frames (cf. Figure \ref{fig:frame-condition}). For all $a,b,c,d \in K$: 
\begin{quote}
    (F1) \ $a \leq a$ \\ (F2) \ if $a \leq b$ and $R b c d$, then $R a c d$.
\end{quote}
A \emph{$\bplus$-model} is a pair $\mathcal{M} = \langle \mathcal{F}, V \rangle$, where $\mathcal{F}$ is a $\bplus$-frame and $V : \mathsf{At} \longrightarrow \powerset(K)$ is a valuation function satisfying the following condition: for all $p \in \mathsf{At}$ and all $a,b\in K$, if $a \leq b$ and $a \in V(p)$, then $b \in V(p)$. 
\end{definition}

\begin{definition}\label{def:modelB+}
The satisfaction relation in $\bplus$ for propositional formulas is defined inductively in the following way. 
For all $a \in K$ and $A,B\in \mathsf{Frm}$:
\begin{align*}
	&\mathcal{M},a\Vdash  p	& \text{iff}&\qquad a\in V(p), \text{ for }p\in\mathsf{At}\tag{i}\label{1}\\
&\mathcal{M},a\Vdash  A\land B 	&\text{iff}&\qquad \mathcal{M},a\Vdash A\ \text{and}\ \mathcal{M},a\Vdash B\tag{ii}\label{2}\\
		&\mathcal{M},a\Vdash  A\lor B	&\text{iff}&\qquad \mathcal{M},a\Vdash A\ \text{or}\ \mathcal{M},a\Vdash B\tag{iii}\label{3}\\
&\mathcal{M},a\Vdash  A\to B	&\text{iff}&\qquad \forall b,c\in K,\ \text{if}\ Rabc\ \text{and}\ \mathcal{M},b\Vdash A,\ \text{then}\ \mathcal{M},c\Vdash B\tag{iv}\label{4}
\end{align*}

A formula $A$ is \emph{satisfied} in $\mathcal{M}$ iff $\mathcal{M},x \Vdash A$ for all $x \in \+O$.  
A formula $A$ is \emph{valid} in a frame $\mathcal{F}$ iff it is satisfied in every $\bplus$-model based on $\mathcal{F}$.
A formula $A$ is $\bplus$-\emph{valid} iff it is valid in every $\bplus$-frame $\mathcal{F}$.
\end{definition}

Now, we consider some fundamental results on $\bplus$, the proofs of which are included in \cite{routleymeyer73,rl.rivals}.

\begin{lemma}\label{lem:monotonicity}
For all $a,b\in K$ and $A\in \mathsf{Frm}$, if $a\leq b$ and $a\Vdash A$, then $b\Vdash A$.
\end{lemma}

The result below is usually known as \textit{Verification Lemma} (or \textit{Entailment Lemma}), and it is important for the following reason: the argument showing that $A \to B$ holds in a model, which appeals
to it, differs in an important respect from the argument showing that the conditional holds at every
point of the model, which appeals to the satisfaction clause for $\to$. In the former case, one assumes
that $A$ holds at an arbitrary point and shows that $B$ holds at that point as well; this does not
establish that the conditional holds at every point of the model. To establish the latter, one assumes
$Rabc$ and $b\Vdash A$, for arbitrary $a,b,c\in K$, and shows that $c\Vdash B$. The distinction matters
because in relevant logics one rarely needs to show that a conditional holds at every point of a model \cite{standefer2022connective}.

\begin{lemma}\label{lem:verification} 
A formula $A\to B$ is satisfied in a model $\mathcal{M}$ iff for all $a \in K$, if $a \Vdash A$, then $a \Vdash B$.
\end{lemma}

Finally, we state soundness and completeness \cite{routleymeyer73,rl.rivals}:

\begin{theorem}
$A$ is 
a theorem of $\hbplus$ if and only if $A$ is $\bplus$-valid.
\end{theorem}

Given that $\hbplus$ is sound and complete, we will simply refer to a theorem of $\hbplus$ as a theorem of $\bplus$ in this paper. 
\section{Nested sequents for relevant implication}\label{sec:CBcut}

In this section, we develop the proof theory of $\bplus$ by extending the standard nested framework 
to capture the ternary relation of Routley-Meyer frames. 
We define the first nested sequent calculus for $\bplus$ based on our new notion of nesting that relies on the use of \emph{binary blocks}, i.e., blocks with two components. We then present the semantic interpretation of the proposed framework and prove soundness of the calculus. Finally, we establish completeness by means of a syntactic proof of $\cut$-admissibility.

First, we define the syntax of nested sequents based on $\mathsf{Frm}$ where formulas are decorated with polarities $\bullet$ and $\circ$. 
To generate the nesting structure, we make use of a structural operator $\iblock{\cdot\sand\cdot}$, referred to as \textit{relevant block}, with two arguments (also called \textit{components}).

\begin{definition}\label{def:nested-sequents}
Nested sequents are generated by the following grammar:
$$
\begin{aligned}
    \Gamma::\Lambda,\Pi & \quad & \Lambda::\lef{A_1},\ldots,\lef{A_n}, \iblock{\Lambda_{11}\sand \Lambda_{12}},\ldots,\iblock{\Lambda_{k1}\sand\Lambda_{k2}} & \quad & \Pi::\rig A~|~\iblock{\Lambda\sand \Gamma}~|~\iblock{\Gamma\sand \Lambda}
\end{aligned}
$$
\end{definition}

Expressions in the above definition containing an output formula $\rig A$ are called \emph{full} nested sequent (sequent for short) while those with only $\bullet$-formulas occurring nestedly are called \emph{left-sided} sequents. According to the definition, a sequent consists of two parts, the left part which only contains $\bullet$-decorated formulas, and the right part which is a single formula $\rig A$ or a relevant block where exactly one of the two arguments is a full sequent. We also need the notion of \emph{generic sequents} when it is not necessary to specify whether it is full or left-sided. 
To make it clear, we shall use the following naming for different types of sequents in this paper: 
$$
\begin{aligned}
    \text{full~sequents:~} \Gamma\quad & \quad \text{left-sided~sequents:~} \Lambda \quad & \text{generic~sequents:~}\Sigma,\Phi,\Psi
\end{aligned}
$$

\begin{rem}\label{rem:conditional}
    To the best of our knowledge, binary nested sequents representing trees of pairs of sequents have not been considered in the literature. 
    In \cite{AlendaOlivettiPozzato2016}, a binary structure $[A:S]$ is considered, where $A$ and $S$ go in pairs 
    and it looks like our binary nested block. However, in their definition, $A$ is a decorated formula which is neither involved in the nested structure, nor decomposed by rule applications, whereas $S$ is a nested sequent. 
    Therefore, this kind of modified nested sequents still lies in the category of so-called nested sequents with \emph{unary} blocks. 
\end{rem}

As usual in nested calculi, we have the notion of context in order to specify where the rule is applied within the nested structure. We follow the terminology and notation used in \cite{CSL-2024}. 
 By $\csseq{ \ }$ we denote a nested sequent holding an empty component with the unique palceholder $\{ \ \}$ whereas $\csseq{\Sigma}$ refers to a nested sequent which is fulfilled with $\Sigma$ to $G\{ \ \}$. Here $\Sigma$ can be either full or left-sided according to whether $G\{ \ \}$ contains an output formula. This is slightly different from the initial use of contexts in e.g. \cite{Strassburger:2013-ik-cut,Brunner:2009}, but will be useful for further studies of meta-logical properties in the following sections. A discussion between the general distinctions of the two uses of contexts is provided in \cite{gao2026bi-ik}. 

We write $\iblock{\Phi \sand -}$ and $\iblock{- \sand \Phi}$ if either component in the relevant pair is empty. Since a full nested sequent only contains one output formula and our calculus will only operate on full sequents, to simplify notations, we will generically use $\Sigma,\Phi,\Psi,\ldots$ to denote both sequents and left-sided sequents. 

\begin{definition}\label{def:nested-operators}
    We define two operators for a (left-sided) sequent $\Sigma$: (1) $\remove{\Sigma}$ to remove all the possible blocks in $\Sigma$; (2) $\dow\Sigma$ to remove the unique output formula. 
\end{definition}

Notice that if $\Sigma$ is a left-sided sequent, then $\dow\Sigma=\Sigma$. The $*$-operator is adopted to context as well, by $\dowcsseq{ \ }$ we mean removing the unique output formula in $G$. 

\begin{figure}[t]
\begin{center}
\begin{tikzpicture}[
  every node/.style = {draw, shape=rectangle, rounded corners, fill=gray!20, minimum width=1.5cm},
  node distance = 2cm and 1cm
]

  \node (root) {$\lef r, \lef {p \vee q}$};
  \node (p_blue) at (-3,-2) {$\lef p$};
  \draw (root) -- (p_blue) coordinate[pos=0.5] (blue_fork);
  \node (pvq_blue) at (-0.8,-2) {$\lef{r\to s}$};
  \draw (blue_fork) -- (pvq_blue);

  \node (ptq_red) at (3,-2) {$\lef {p \to q}$};
  \draw (root) -- (ptq_red) coordinate[pos=0.5] (red_fork);
  \node (q_red) at (0.8,-2) {$\rig q$};
  \draw (red_fork) -- (q_red);
\end{tikzpicture}
\captionof{figure}{Graphic representation of $\lef r, \lef{p\vee q}, \iblock{\lef p\sand \lef{r\to s}},\iblock{\rig q\sand\lef {p\to q}}$. The root corresponds to the top-level component, which contains two left-formulas $\lef r, \lef{p\vee q}$; the two blocks then represent two independent pairs of children of the root in this relational structure. }
\label{graph:sequent}
\end{center}
\end{figure}
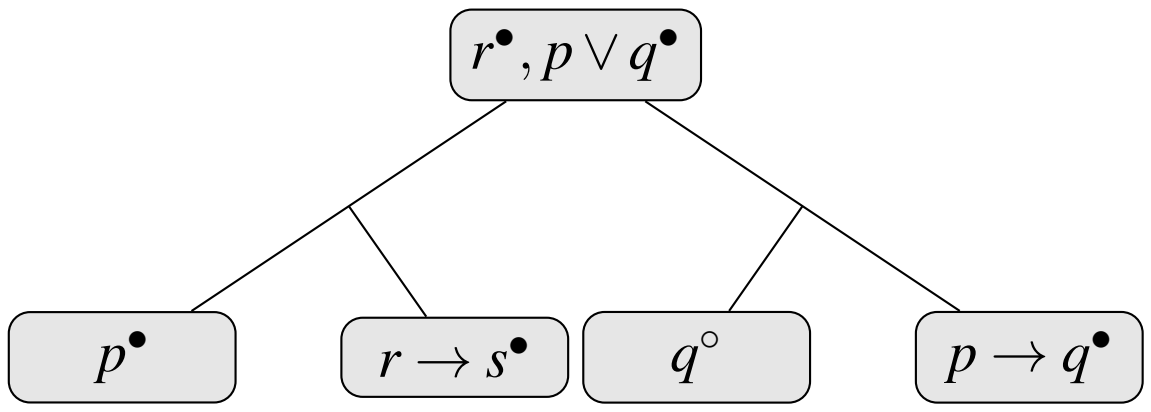

\begin{example}
    By definition, $\Sigma=\lef{p\vee q}, \iblock{\lef p\sand \lef{p\vee q} },\iblock{\rig q\sand\lef {p\to q}}$ is a full nested sequent, $\dow\Sigma=\lef{p\vee q}, \iblock{\lef p\sand \lef{p\vee q}},\linebreak  \iblock{ - \sand\lef {p\to q}}$ and $\remove{\Sigma
    }=\lef{p\vee q}$. A graphic representation of $\Sigma$ is found in Figure \ref{graph:sequent}. 
\end{example}

\begin{figure}[t]
\begin{center}
\begin{framed}
    \begin{adjustbox}{max width = \textwidth}
    \begin{tabular}{cc}
        \multicolumn{2}{l}{Initial sequents:}\\
        \multicolumn{2}{c}{
            \begin{prooftree}
            \hypo{}
            \infer1[$\idone$]{\csseq{\Sigma,\lef p, \rig p}}
            \end{prooftree}
        }\\[.6cm]
        \multicolumn{2}{l}{Logical rules:}\\
        \begin{prooftree}
            \hypo{\csseq{\Sigma,\lef A,\lef B}}
            \infer1[$\ruleandl$]{\csseq{\Sigma,\lef{A\wedge B}}}
        \end{prooftree}
        &
        \begin{prooftree}
            \hypo{\csseq{\Sigma,\rig A}}
            \hypo{\csseq{\Sigma,\rig B}}
            \infer2[$\ruleandr$]{\csseq{\Sigma,\rig{A\wedge B}}}
        \end{prooftree}
        \\[.6cm]
        \begin{prooftree}
            \hypo{\csseq{\Sigma,\lef A}}
            \hypo{\csseq{\Sigma,\lef B}}
            \infer2[$\ruleorl$]{\csseq{\Sigma,\lef{A\vee B}}}
        \end{prooftree}
        &
        \begin{prooftree}
            \hypo{\csseq{\Sigma,\rig{A}}}
            \infer1[$\ruleorr$]{\csseq{\Sigma,\rig{A\vee B}}}
        \end{prooftree}
        \quad
        \begin{prooftree}
            \hypo{\csseq{\Sigma,\rig{B}}}
            \infer1[$\ruleorr$]{\csseq{\Sigma,\rig{A\vee B}}}
        \end{prooftree}
        \\[.6cm]
        \begin{prooftree}
            \hypo{\dowcsseq{\dow\Sigma,\lef{A\to B},\iblock{\dow\Phi_1,\rig A\sand\dow\Phi_2}}}
            \hypo{\csseq{\Sigma,\lef{A\to B},\iblock{\Phi_1\sand\Phi_2,\lef B}}}
            \infer2[$\ruletol$]{\csseq{\Sigma,\lef{A\to B},\iblock{\Phi_1\sand\Phi_2}}}
        \end{prooftree}
        &
        \begin{prooftree}
            \hypo{\csseq{\Sigma,\iblock{\lef A\sand \rig B}}}
            \infer1[$\ruletor$]{\csseq{\Sigma,\rig{A\to B}}}
        \end{prooftree}
        \\[.6cm]
        \multicolumn{2}{l}{Structural rules:}\\
        \multicolumn{2}{c}{
        \begin{prooftree}
            \hypo{\Sigma,\iblock{\lef A,\Phi_1\sand\lef A,\Phi_2}}
            \infer1[$\ruleliftib$]{\Sigma,\iblock{\lef A,\Phi_1\sand\Phi_2}}
        \end{prooftree}
        }
    \end{tabular}
    \end{adjustbox}

    \captionof{figure}{$\cbplus$, a nested sequent calculus for $\bplus$}
    \label{fig:calculus-bplus}
\end{framed}
\end{center}
\end{figure}

A nested sequent calculus for the logic $\bplus$ is given in Figure \ref{fig:calculus-bplus}. We provide some explanations for these rules as below. 

\begin{rem}\label{rem:interpretationRules}
In $\cbplus$ 
all logical rules 
can be applied at any component within any nested block 
while $\ruleliftib$ is only applicable to blocks belonging to the top-level component. 
This distinction is 
useful 
for separating the semantic roles of the relations $a \leq b$ and $Rabc$ within the calculus. 
As will be shown later in the semantic interpretation, 
the top-level component of a nested sequent is 
associated with the regular worlds of the semantics, while the relevant block immediately succeeding it corresponds to two worlds, say $a,b \in K$, to be read as $a \leq b$. 
By contrast, the components occurring inside nested sequents are associated with \textit{irregular} worlds in $K$. Thus $Rxab$ (where $x\in\+O$) and $Rabc$ 
are represented graphically in a nested sequent as 
$$
\begin{aligned}
    \overbrace{\Sigma}^{x\in \+O},
\iblock{\overbrace{\Phi_1}^{a\in K} \sand \overbrace{\Phi_2}^{b\in K}}
&
\quad\text{and}
&
G\{\overbrace{\Sigma}^{a},
\iblock{\overbrace{\Phi_1}^{b} \sand \overbrace{\Phi_2}^{c}}\}
\end{aligned}
$$
The rules $\ruleliftib,\ruletol,\ruletor$ involving propagation or production of relevant blocks make the two arguments contained in the same block not interchangeable. 
The rule $\ruleliftib$ -- some variants thereof being familiar from intuitionistic logic (cf. \cite{Fitting:2014}) -- expresses the monotonicity property semantically captured by Lemma \ref{lem:monotonicity}. 
The rules for $\to$ are inspired by the semantic clause for implication. 
From a backward view, 
$\ruletor$ states that an implication is false at a world $a$ when there exist worlds $b,c$ with $Rabc$, where $b$ satisfies the antecedent and $c$ falsifies the consequent. Syntactically, this is captured by the relevant block $\csseq{\Sigma,\iblock{\lef A \sand \rig B}}$ in the premise, whose two components represent $b$ and $c$. 
Dually, $\ruletol$ reflects the semantic use of already opened worlds: any relevant block associated with a sequent containing $\lef{A\to B}$ may be used. The two premises then represent the worlds $b$ and $c$, with $\rig A$ in the first component and $\lef B$ in the second. To keep the single output structure, we have the similar `pruning' for the left premise as \cite{Strassburger:2013-ik-cut} to remove the existing output formula and put $\rig A$ as the unique output formula. 

\end{rem}



As usual, a \textit{derivation} is a rooted tree defined in the standard way, and its \textit{height} is the length of its longest branch. 
We say that a formula $A$ is $\cbplus$-\textit{derivable} whenever the sequent $\rig A$ is $\cbplus$-derivable.

\begin{definition}[Formula occurrences] 
    We say an occurrence of a formula occurring in the conclusion of an application of a logical rule $(r)$ is \emph{side} if it is 
    not involved in 
    the rule application and kept in at least one of the premise(es); \emph{weak} if it is not involved in the rule application and does not occur in any of the premise(es). Occurrences of formulas which are neither side nor weak are called \emph{principal}. 
    For example, in $(\rig {\vee})$, the explicit $\rig {A\vee B}$ is principal and formulas in $\Sigma$ and $\csseq{ \ }$ are all side. 
    For the only structural rule $\ruleliftib$, we call all the formula occurrences by side formulas. 
\end{definition}

The following proposition is shown by induction on the complexity of $A$. 

\begin{proposition}\label{proposition:identity}
Generalized initial sequents of the form $\csseq{\Sigma,\lef A,\rig A}$ are derivable in $\cbplus$.  
\end{proposition}

\begin{proof}
    We only demonstrate the non-trivial case when $A=B\to C$, 
    \begin{center}
        \begin{prooftree}
            \hypo{}
            \infer1[IH]{\dowcsseq{\dow\Sigma,\lef{B\to C}, \iblock{\lef B,\rig B\sand - }}}
            \hypo{}
            \infer1[IH]{\csseq{\Sigma,\lef{B\to C}, \iblock{\lef B\sand\lef C,\rig C}}}
            \infer2[$\ruletol$]{\csseq{\Sigma,\lef{B\to C}, \iblock{\lef B\sand\rig C}}}
            \infer1[$\ruletor$]{\csseq{\Sigma,\lef{B\to C}, \rig{B\to C}}}
        \end{prooftree}
    \end{center}
This completes the proof. 
\end{proof}

\subsection{Semantic interpretation and soundness}

In what follows, we establish the soundness of $\cbplus$. To this end, the satisfaction relation introduced in the previous section is extended from formulas to nested sequents, in accordance with the interpretation outlined in Remark~\ref{rem:interpretationRules}.



\begin{definition}[Semantic interpretation]\label{def:seq-semantics}
    Let $\Sigma$ be a sequent and $\+M=(\+O,K,R,V)$ be a $\bplus$-model and $x,y,z\in K$. 
    
    For a left-sided sequent $\Sigma=\lef {A_1},\ldots,\lef{A_m}, \iblock{\Phi_{11}\sand\Phi_{12}},\ldots, \iblock{\Phi_{j1}\sand\Phi_{j2}}$, we define 
    $x\Vdash \Sigma$ iff 
    \begin{itemize}
        \item for any $i\in\{1,\ldots,m\}$, $x\Vdash \lef {A_i}$; and 
        \item for any $j\in\{1,\ldots,n\}$, there exist $y,z$ with $Rxyz$, $y\Vdash \Phi_{j1}$ and $z\Vdash \Phi_{j2}$
    \end{itemize}

    For a full sequent, we define the interpretation according to the location of the output formula:  
    \begin{itemize}
        \item $x\Vdash \Sigma,\rig A$ iff $x\Vdash \Sigma$ implies $x\Vdash A$
        \item $x\Vdash\Sigma,\iblock{\Phi_1\sand\Phi_2}$ (where $\Phi_1$ is full) iff if $x\Vdash \Sigma$ then for any $y,z$ with $Rxyz$, $z\Vdash \Phi_2$ implies $y\Vdash \Phi_1$
        \item $x\Vdash\Sigma,\iblock{\Phi_1\sand\Phi_2}$ (where $\Phi_2$ is full) iff if $x\Vdash \Sigma$ then for any $y,z$ with $Rxyz$, $y\Vdash \Phi_1$ implies $z\Vdash \Phi_2$
    \end{itemize}
    
    We say a sequent $\Sigma$ is $\bplus$-satisfiable if there is a $\bplus$-model $\+M$ and a regular world $w$ in it such that $\+M,w\Vdash \Sigma$; 
    and $\Sigma$ is 
    $\bplus$-\emph{valid} if $\+M,w\Vdash \Sigma$ for any $\bplus$-model $\+M$ and any regular world $w$ in it.  
\end{definition}

\begin{rem}
    Notice that a block is interpreted differently depending on whether it contains the output formula and where that formula is precisely located.  
    In particular, if a block does not contain the unique output formula, it is interpreted existentially; otherwise, it is interpreted universally.
    Similar practice for interpreting nested sequents with single output is also used in the context of intuitionistic modal logic \cite{Strassburger:2013-ik-cut}, where an unary block is interpreted as $\Diamond$ (existential) if it does not contain the unique output formula; and as $\square$ (universal) otherwise. 
\end{rem}

Intuitively, the satisfaction of a full sequent says if the left-sided part is satisfied, so is the unique output formula. 
According to the definition, 
for a model $\+M=(\+O,K,R, V)$ and $w\in \+O$, if $\+M,w\nVdash \csseq{\Sigma,\rig A}$, then there exists some $x\in K$ such that $x\nVdash \Sigma,\rig A$. 

To show the soundness of the calculus, we first verify each rule of $\cbplus$ is valid. For a rule $(r)$ in $\cbplus$, we say $(r)$ is \emph{valid} if whenever all of its premise are valid, the conclusion is also valid. 

\begin{lemma}\label{lem:valid-cbolus}
    All the rules in $\cbplus$ are $\bplus$-valid. 
\end{lemma}

\begin{proof}
    We prove the basic case when $\csseq{ \ }$ is empty, the general case is then proved by induction on the structure of the context. We only show the cases of $\lef \to,\rig \to$ and $\ruleliftib$ as other cases are straightforward.  
    In all the cases, we assume the premise is valid but the conclusion is not and prove the validity of the rule by 
    contradiction. 
    \begin{enumerate}
 \item[$\ruletol$:]  Since $\nVdash \Sigma',\lef{A\to B},\iblock{\Phi_1\sand\Phi_2}$, there exists $\+M=(\+O,K,R,V)$ and $x\in\+O$ such that $\+M,x\nVdash \Sigma,\lef{A\to B},\iblock{\Phi_1\sand\Phi_2}$. This further implies $\+M,x\Vdash\dow{\Sigma},\iblock{\dow\Phi_1\sand\dow\Phi_2}$. 
 According to the location of the output formula of the conclusion, we gave the following sub-cases: 
 \begin{itemize}[leftmargin=*]
     \item The output formula is in $\Phi_1$. Since $\Phi_1$ is full, $\dow{\Sigma}=\Sigma$, $\dow\Phi_2=\Phi_2$, 
     and there exist $y,z\in K$ such that $Rxyz$, $z\Vdash\Phi_2$ and $y\nVdash\Phi_1$. Since $x\Vdash A\to B$, by definition we have that for all $u',v'$ with $Rxu'v'$, if $u'\Vdash A$ then $v'\Vdash B$, which equivalently means $u'\nVdash A$ or $v'\Vdash B$. Thus, either $y\nVdash A$ or $z\Vdash B$. 
     If $y\nVdash A$, then $\+M,x\nVdash \Sigma,\iblock{\dow\Phi_1,\rig A\sand\Phi_2}$ as $y\nVdash \Phi_1,\rig A$ and $z\Vdash\Phi_2$, contradicting the validity of the left premise; if $v'\Vdash B$, then $\+M,x\nVdash \Sigma,\iblock{\Phi_1\sand\Phi_2,\lef B}$, contradicting the validity of the right premise. 
     \item The output formula is in $\Phi_2$. 
     Since $\Phi_2$ is full, 
     there exist $y,z\in K$ such that $Rxyz$, $y\Vdash\Phi_1$ and $z\nVdash\Phi_2$. Since $x\Vdash A\to B$, by definition we have that for all $u',v'$ with $Rxu'v'$, if $u'\Vdash A$ then $v'\Vdash B$, which equivalently means $u'\nVdash A$ or $v'\Vdash B$. Thus, either $y\nVdash A$ or $z\Vdash B$. 
     If $y\nVdash A$, then $\+M,x\nVdash \Sigma,\iblock{\dow\Phi_1,\rig A\sand\Phi_2}$, contradicting the validity of the left premise; if $v'\Vdash B$, then $\+M,x\nVdash \Sigma,\iblock{\Phi_1\sand\Phi_2,\lef B}$ as $y\Vdash \Phi_1$ and $z\nVdash\Phi_2,\lef B$, contradicting the validity of the right premise. 
     \item The output formula is in $\Sigma$. 
     Since $\Sigma$ is full, we have $x\nVdash \Sigma$ and $x\Vdash \iblock{\Phi_1\sand\Phi_2}$. 
     By definition, there exists $y,z$ such that $y\Vdash \Phi_1$ and $z\Vdash \Phi_2$. Similarly as the above two cases, we have either $y\nVdash A$ or $z\Vdash B$. If $y\nVdash A$, then $y\nVdash \Phi_1, \rig A$, which will further falsify the left premise; if $z\Vdash B$, then $x\Vdash \iblock{\Phi_1\sand\Phi_2,\lef B}$, which makes $x\nVdash \Sigma,\lef{A\to B},\iblock{\Phi_1\sand\Phi_2,\lef B}$, contradicting the validity of the right premise. 
 \end{itemize}
  
    \item[$\ruletor$:] 
        Since $\nVdash \Sigma,\rig{A\to B}$, there exists a model $\+M=(\+O,K,R,V)$ and $x\in \+O$ such that $\+M,x\nVdash \Sigma,\rig{A\to B}$, which further implies $\+M,x\Vdash\Sigma$ and $\+M,x\nVdash A\to B$. 
        Hence there exist $y,z\in K$ such that $Rxyz$, $y\Vdash A$, and $z\nVdash B$. Meanwhile, since $\Vdash \Sigma,\iblock{\lef A\sand\rig B}$, we have 
        $\+M, x\Vdash \iblock{\lef A\sand \rig B}$. 
        Since the forcing of $\iblock{\lef A\sand\rig B}$ at $x$ requires that for all $y',z'$ with $Rx y' z'$, if $y'\Vdash A$ then $z'\Vdash B$, 
        so by $y\Vdash A$ we have $z\Vdash B$, a contradiction. 

    \item[$\ruleliftib$:] 
    Since $\nVdash \Sigma,\iblock{\lef A,\Phi_1\sand\Phi_2}$, there exists $\+M=(\+O,K,R,V)$ and $x\in\+O$ such that $\+M,x\nVdash \Sigma,\iblock{\lef A, \Phi_1\sand\Phi_2}$. 
    According to the location of the output formula of the conclusion, we gave the following sub-cases: 
    \begin{itemize}[leftmargin=*]
    \item The output formula is in $\Phi_1$. Since $\Phi_1$ is full, there exist $y, z$ with $Rxyz$ such that $z\Vdash \Phi_2$ and $y\nVdash \Phi_1,\lef A$. By definition, this further implies $y\Vdash \dow\Phi_1,\lef A$ and hence $y\Vdash A$. 
    Since $x\in\+O$, we have $y\leq z$, 
    by Lemma~\ref{lem:monotonicity}, $z\Vdash A$ as well. This makes $z\Vdash \Phi_2,\lef A$ and hence $x\nVdash \Sigma,\iblock{\lef A, \Phi_1\sand\lef A, \Phi_2}$, a contradiction. 
    \item The output formula is in $\Phi_2$. Since $\Phi_2$ is full, there exist $y, z$ with $Rxyz$ such that $y\Vdash \lef A, \Phi_1$ and $z\nVdash \Phi_2$. By definition, this further implies $y\Vdash \lef A$. 
    Similarly as the previous case, we have $z\Vdash A$ as well. This further makes $z\nVdash \Phi_2,\lef A$ and hence $x\nVdash \Sigma,\iblock{\lef A, \Phi_1\sand\lef A, \Phi_2}$, a contradiction. 
    \item The output formula is in $\Sigma$. Since $\Sigma$ is full, both $\Phi_1,\Phi_2$ are left-sided, we have 
    $x\Vdash \iblock{\lef A, \Phi_1\sand\Phi_2}$, which means there exist $y, z$ with $Rxyz$ such that $y\Vdash \lef A, \Phi_1$ and $z\Vdash \Phi_2$. By definition, this further implies $y\Vdash \lef A$. 
    Similarly as the previous case, it follows $z\Vdash A$. This further makes $z\Vdash \Phi_2,\lef A$ and hence $x\nVdash \Sigma,\iblock{\lef A, \Phi_1\sand\lef A, \Phi_2}$, a contradiction. 
    \end{itemize}
    \end{enumerate}
    This completes the proof. 
\end{proof}

The soundness of $\cbplus$ then follows as usual by induction on the structure of a derivation. 

\begin{theorem}[Soundness of $\cbplus$]\label{thm:cbplus-soundness}
Let $\Sigma$ be a sequent. If $\Sigma$ is derivable in $\cbplus$, then it is $\bplus$-valid.
\end{theorem}


\subsection{Completeness and $\cut$-admissibility}\label{subsec:Cut-CB}
In this subsection, our aim is twofold. First, we establish the completeness of the calculus with cut rule by deriving the axioms of the Hilbert system $\hbplus$ within $\cbplus + \cut$. On the basis of these results, and by proving some preliminary structural properties of $\cbplus$ together with the invertibility of certain rules, we then show that $\cut$ is an admissible rule. As a consequence, the calculus $\cbplus$ itself is complete. 

\begin{definition}[Cut rule]
    The $\cut$-rule is defined as 
    $$
        \begin{aligned}
            \begin{prooftree}
                \hypo{\dowcsseq{\dow\Sigma,\rig A}}
                \hypo{\csseq{\Sigma,\lef A}}
                \infer2[$\cut$]{\csseq{\Sigma}}
            \end{prooftree}
        \end{aligned}
    $$
\end{definition}


\begin{restatable}[Completeness of $\cbplus+\cut$]{theorem}{completenesscbplus}\label{prop:completeness}
    All the axioms and rules in $\hbplus$ are derivable in $\cbplus+\cut$. Consequently, if a formula $A\in\mathsf{Frm}$ is valid, then it is provable in $\cbplus+\cut$. 
\end{restatable}

\begin{proof}
    The derivations of the axioms and rules are found in the appendix. 
\end{proof}

The rest of this section is devoted to a syntactic proof of cut-admissibility. To this end, we begin by introducing the following preliminary notions.

\begin{definition}[(hp-)admissibility and (hp-)invertibility]\label{def:hp-adm.inv}
A rule $\+R$ is called \emph{admissible} in a calculus if, whenever all of its premises are derivable at height $n$, its conclusion is also derivable with height at most $m$. If $m\leq n$, then $\+R$ is called \emph{height-preserving admissible} (\emph{hp-admissible} for short). 
A rule $\+R$ is called \emph{height-preserving invertible} (\emph{hp-invertible} for short) in a calculus if, whenever its conclusion is derivable with height at most $m$, all of its premises are derivable with height at most $m$ as well. 
\end{definition}

As usual, we show some basic facts about sequent calculi. We first prove that the weakening rule is hp-admissible and then use this to show that some of the logical rules of $\cbplus$ are hp-invertible. Finally, we prove that the contraction and merge rules are hp-admissible as well. 

\begin{proposition}\label{prop:hp-admi-weakening}
    The following weakening rule is hp-admissible in $\cbplus$:
    $$
    \begin{aligned}
        \begin{prooftree}
        \hypo{\csseq{\Sigma}}
        \infer1[$\rulewl$]{\csseq{\Lambda,\Sigma}}
        \end{prooftree}
    \end{aligned}
    $$
\end{proposition}

\begin{proof}
    By induction on the height $n$ of the derivation.
\end{proof}

\begin{proposition}\label{prop:invertibility}
    The rules $\ruleandl,\ruleorl,\rig\wedge$, $\ruletor$ and $\ruleliftib$ are hp-invertible in $\cbplus$. 
\end{proposition}

\begin{proof}
The case of $\ruleliftib$ as the inverse of the rule is just weakening. For other rules, we proceed with induction on the height $n$ of the derivation.  \\
\noindent\textit{Base case}: $n=0$. Let $\Lambda$ be an initial sequent. 
Notice that the presence of $\rig p$ entails that $\Lambda$ is full, with $\rig p$ being the unique output formula, which makes the conclusion of $\rig \to$ and $\rig \wedge$ not applicable. 
Thus we only need to consider the case of $\lef \vee, \lef \wedge$. 
The sequent $\Lambda'$ obtained by replacing an input conjunction (or disjunction) with its subformulas is also directly derivable with height $0$. In particular, if $\Lambda=\csseq{\Sigma, \lef{A\land B}}$ is an instance of $\idone$ derived with height $0$, then $\Lambda'=\csseq{\Sigma, \lef A, \lef B}$ is also an instance of $\idone$ derivable with height $0$. Similarly for $\csseq{\Sigma, \lef{A\lor B}}$. \\
\noindent\textit{Inductive step: $n>0$.} 
We only display the case where $r=\ruletor$ is the rule to be inverted as the proof for $r \in \{\ruleandl, \ruleorl, \rig\wedge, \ruletor\}$ follows the same pattern.
\\
First, suppose the formula $\rig{C\to D}$ is side and distinguish cases based on which rule $r'$ was principally applied in the derivation. 
Let $r'=\ruletol$ with $\lef{A\to B}$ principal. We distinguish where the side output formula $\rig{C\to D}$ appears. We display some salient examples as all other cases can be dealt with analogously.
 \begin{itemize}[leftmargin=*]
     \item Let $\Sigma=\Sigma', \rig{C\to D}$.
 \end{itemize}
\begin{center}
   \begin{adjustbox}{max width=.8\textwidth}
\begin{tabular}{c}
 \begin{prooftree}
            \hypo{}
        \ellipsis{}{\dowcsseq{\dow{\Sigma'},\lef{A\to B},\iblock{\dow\Phi_1,\rig A\sand\dow\Phi_2}}}
            \hypo{}
        \ellipsis{}{\csseq{\Sigma', \rig{C\to D},\lef{A\to B},\iblock{\Phi_1\sand\Phi_2,\lef B}}}
            \infer2[$r'$]{\csseq{\Sigma', \rig{C\to D},\lef{A\to B},\iblock{\Phi_1\sand\Phi_2}}}
        \end{prooftree}
 \hspace{0.3cm}$ \leadsto$\vspace{0.3cm}\\
 
         \begin{prooftree}
            \hypo{}
        \ellipsis{}{\dowcsseq{\dow{\Sigma'},\lef{A\to B},\iblock{\dow\Phi_1,\rig A\sand\dow\Phi_2}}}
        \infer1[$\rulewl$]{\dowcsseq{\dow{\Sigma'},\iblock{\lef{C}\sand -},\lef{A\to B},\iblock{\dow\Phi_1,\rig A\sand\dow\Phi_2}}}
            \hypo{}
        \ellipsis{}{\csseq{\Sigma', \rig{C\to D},\lef{A\to B},\iblock{\Phi_1\sand\Phi_2,\lef B}}}
        \infer1[IH]{\csseq{\Sigma', \iblock{\lef{C}\sand \rig{D}},\lef{A\to B},\iblock{\Phi_1\sand\Phi_2,\lef B}}}
            \infer2[$r'$]{\csseq{\Sigma', \iblock{\lef{C}\sand \rig D},\lef{A\to B},\iblock{\Phi_1\sand\Phi_2}}}
        \end{prooftree}
         \end{tabular}
\end{adjustbox}
\end{center}

 \begin{itemize}[leftmargin=*]
     \item Let $\Phi_1=\Phi_1', \rig{C\to D}$.
 \end{itemize}
\begin{center}
\begin{adjustbox}{max width=.8\textwidth}
\begin{tabular}{c}
 \begin{prooftree}
            \hypo{}
        \ellipsis{}{\dowcsseq{\dow{\Sigma},\lef{A\to B},\iblock{\dow{\Phi_1'},\rig A\sand\dow\Phi_2}}}
            \hypo{}
        \ellipsis{}{\csseq{\Sigma', \lef{A\to B},\iblock{\Phi_1',\rig{C\to D},\sand\Phi_2,\lef B}}}
            \infer2[$r'$]{\csseq{\Sigma', \lef{A\to B},\iblock{\Phi_1',\rig{C\to D},\sand\Phi_2}}}
        \end{prooftree}
 \hspace{0.3cm}$ \leadsto$\vspace{0.3cm}\\
 
         \begin{prooftree}
            \hypo{}
        \ellipsis{}{\dowcsseq{\dow{\Sigma},\lef{A\to B},\iblock{\dow{\Phi_1'},\rig A\sand\dow\Phi_2}}}
        \infer1[$\rulewl$]{\dowcsseq{\dow{\Sigma},\lef{A\to B},\iblock{\dow{\Phi_1'}, \iblock{\lef{C}\sand -},\rig A\sand\dow\Phi_2}}}
            \hypo{}
        \ellipsis{}{\csseq{\Sigma,\lef{A\to B},\iblock{\Phi_1',\rig{C\to D},\sand\Phi_2,\lef B}}}
        \infer1[IH]{\csseq{\Sigma,\lef{A\to B},\iblock{\Phi_1',\iblock{\lef{C}\sand \rig{D}}\sand\Phi_2,\lef B}}}
            \infer2[$r'$]{\csseq{\Sigma,\lef{A\to B},\iblock{\Phi_1',\iblock{\lef{C}\sand \rig D},\sand\Phi_2}}}
        \end{prooftree}
         \end{tabular}
\end{adjustbox}
\end{center}

\noindent Notice that all applications of $\rulewl$ are hp-admissible by Proposition \ref{prop:hp-admi-weakening}. \\
 Finally, we consider the case in which the rule $\ruletor$ is applied principally to the formula $\rig{C\to D}$. Assume we obtained a derivation of a sequent of the form $\Lambda=\csseq{\Sigma, \rig{C\to D}}$ of height $n+1$. Analyzing the the sequent immediately preceding $\Lambda$, namely $\Lambda'=\csseq{\Sigma, \iblock{\lef C\sand \rig D}}$, we notice that this premise possesses a derivation height reduced to $n$, thus obtaining the desired derivation directly. 
\end{proof}

\begin{proposition}\label{prop:hp-admi-structural}
    The following structural rules are hp-admissible in $\cbplus$:
    $$
    \begin{aligned}
        \begin{prooftree}
        \hypo{\csseq{\lef A,\lef A,\Sigma}}
        \infer1[$\rulecl$]{\csseq{\lef A,\Sigma}}
        \end{prooftree}
        &
        \quad
        &
        \begin{prooftree}
    \hypo{\csseq{\Sigma,\iblock{\Phi_1\sand\Phi_2},\iblock{\Psi_1\sand\Psi_2}}}
        \infer1[$\rulemerge$]{\csseq{\Sigma,\iblock{\Phi_1,\Psi_1\sand\Psi_2,\Phi_2}}}
        \end{prooftree}
    \end{aligned}
    $$
\end{proposition}

\begin{proof}
The proof proceeds by a routine induction on the height $n$ of the derivation. We only present the case for $\rulemerge$, as the case for contraction is similar. 
If $n=0$, the conclusion follows straightforwardly.
Otherwise, $n>0$. Let us consider the last rule $r$ applied in the derivation of the premise. We distinguish several cases.

\noindent \textbf{Case 1.} $r$ is applied in $\Sigma$ or $\csseq{ \ }$. We only demonstrate the nontrivial sub-cases when $r\in\{\lef\to,\rig \to,\ruleliftib\}$ and is applied in $\Sigma$. Other sub-cases are analogous. 

\begin{itemize}[leftmargin=*]
\item Let $r=\ruletol$ and $\Sigma=\Sigma',\lef{A\to B}$. Assume $r$ propagates the subformulas of $\lef{A\to B}$ in the block $\iblock{\Phi_1\sand\Phi_2}$:
\begin{center}
\begin{adjustbox}{max width=.8\textwidth}
\begin{tabular}{c}
\begin{prooftree}
\hypo{}
\ellipsis{}{\csseq{\dow{\Sigma'},\lef{A\to B},\iblock{\dow\Phi_1, \rig A\sand\dow\Phi_2},\iblock{\dow\Psi_1\sand\dow\Psi_2}}}
\hypo{}
\ellipsis{}{\csseq{\Sigma',\lef{A\to B},\iblock{\Phi_1\sand\Phi_2, \lef B},\iblock{\Psi_1\sand\Psi_2}}}
\infer2[$r$]{\csseq{\Sigma',\lef{A\to B},\iblock{\Phi_1\sand\Phi_2},\iblock{\Psi_1\sand\Psi_2}}}
        \infer1[$\rulemerge$]{\csseq{\Sigma',\lef{A\to B},\iblock{\Phi_1,\Psi_1\sand\Psi_2,\Phi_2}}}
        \end{prooftree}
        \hspace{0.3cm}$\leadsto$\vspace{0.3cm}\\
         \begin{prooftree}
\hypo{}
\ellipsis{}{\csseq{\dow{\Sigma'},\lef{A\to B},\iblock{\dow\Phi_1, \rig A\sand\dow\Phi_2},\iblock{\dow\Psi_1\sand\dow\Psi_2}}}
\infer1[IH]{\csseq{\dow{\Sigma'},\lef{A\to B},\iblock{\dow\Phi_1, \rig A,\dow\Psi_1\sand\dow\Phi_2, \dow\Psi_2}}}
\hypo{}
\ellipsis{}{\csseq{\Sigma',\lef{A\to B},\iblock{\Phi_1\sand\Phi_2, \lef B},\iblock{\Psi_1\sand\Psi_2}}}
\infer1[IH]{\csseq{\Sigma',\lef{A\to B},\iblock{\Phi_1, \Psi_1\sand\Phi_2, \lef B,\Psi_2}}}
\infer2[$r$]{\csseq{\Sigma',\lef{A\to B},\iblock{\Phi_1,\Psi_1\sand\Psi_2,\Phi_2}}}
        \end{prooftree}
           \end{tabular}     
        \end{adjustbox}
\end{center}

\item Let $r=\ruletol$ and $\Sigma=\Sigma',\lef{A\to B}$. Assume $r$ propagates the subformulas of $\lef{A\to B}$ in the block $\iblock{\Psi_1\sand\Psi_2}$:
\begin{center}
\begin{adjustbox}{max width=.8\textwidth}
\begin{tabular}{c}
\begin{prooftree}
\hypo{}
\ellipsis{}{\csseq{\dow{\Sigma'},\lef{A\to B},\iblock{\dow\Phi_1\sand\dow\Phi_2},\iblock{\dow\Psi_1,\rig A\sand\dow\Psi_2}}}
\hypo{}
\ellipsis{}{\csseq{\Sigma',\lef{A\to B},\iblock{\Phi_1\sand\Phi_2},\iblock{\Psi_1\sand\Psi_2, \lef B}}}
\infer2[$r$]{\csseq{\Sigma',\lef{A\to B},\iblock{\Phi_1\sand\Phi_2},\iblock{\Psi_1\sand\Psi_2}}}
        \infer1[$\rulemerge$]{\csseq{\Sigma',\lef{A\to B},\iblock{\Phi_1,\Psi_1\sand\Psi_2,\Phi_2}}}
        \end{prooftree}
        \hspace{0.3cm}$\leadsto$\vspace{0.3cm}\\
         \begin{prooftree}
\hypo{}
\ellipsis{}{\csseq{\dow{\Sigma'},\lef{A\to B},\iblock{\dow\Phi_1\sand\dow\Phi_2},\iblock{\dow\Psi_1,\rig A\sand\dow\Psi_2}}}
\infer1[IH]{\csseq{\dow{\Sigma'},\lef{A\to B},\iblock{\dow\Phi_1,\dow\Psi_1, \rig A\sand\dow\Phi_2, \dow\Psi_2}}}
\hypo{}
\ellipsis{}{\csseq{\Sigma',\lef{A\to B},\iblock{\Phi_1\sand\Phi_2},\iblock{\Psi_1\sand\Psi_2, \lef B}}}
\infer1[IH]{\csseq{\Sigma',\lef{A\to B},\iblock{\Phi_1, \Psi_1\sand\Phi_2,\Psi_2, \lef B}}}
\infer2[$r$]{\csseq{\Sigma',\lef{A\to B},\iblock{\Phi_1,\Psi_1\sand\Psi_2,\Phi_2}}}
        \end{prooftree}
           \end{tabular}     
        \end{adjustbox}
\end{center}

\item Let $r=\ruletol$ and $\Sigma=\Sigma',\lef{A\to B},\iblock{\Omega_1\sand \Omega_2}$. Assume $r$ propagates the subformulas of $\lef{A\to B}$ in the block $\iblock{\Omega_1\sand \Omega_2}$:
\begin{center}
\begin{adjustbox}{max width=.9\textwidth}
\begin{tabular}{c}
      \begin{prooftree}
\hypo{}
\ellipsis{}{\csseq{\dow{\Sigma'},\lef{A\to B},\iblock{\dow\Omega_1,\rig A\sand\dow\Omega_2},\iblock{\dow\Phi_1\sand\dow\Phi_2},\iblock{\dow\Psi_1\sand\dow\Psi_2}}}
\hypo{}
\ellipsis{}{\csseq{\Sigma',\lef{A\to B},\iblock{\Omega_1\sand \Omega_2,\lef B},\iblock{\Phi_1\sand\Phi_2},\iblock{\Psi_1\sand\Psi_2}}}
\infer2[$r$]{\csseq{\Sigma',\lef{A\to B},\iblock{\Omega_1\sand \Omega_2},\iblock{\Phi_1\sand\Phi_2},\iblock{\Psi_1\sand\Psi_2}}}
        \infer1[$\rulemerge$]{\csseq{\Sigma',\lef{A\to B},\iblock{\Omega_1\sand \Omega_2},\iblock{\Phi_1,\Psi_1\sand\Psi_2,\Phi_2}}}
        \end{prooftree}
        \hspace{0.3cm}$\leadsto$\vspace{0.3cm}\\
         \begin{prooftree}
\hypo{}
\ellipsis{}{\csseq{\dow{\Sigma'},\lef{A\to B},\iblock{\dow\Omega_1,\rig A\sand\dow\Omega_2},\iblock{\dow\Phi_1\sand\dow\Phi_2},\iblock{\dow\Psi_1\sand\dow\Psi_2}}}
\infer1[IH]{\csseq{\dow{\Sigma'},\lef{A\to B},\iblock{\dow\Omega_1,\rig A\sand\dow\Omega_2},\iblock{\dow\Phi_1,\dow\Psi_1\sand\dow\Phi_2, \dow\Psi_2}}}
\hypo{}
\ellipsis{}{\csseq{\Sigma',\lef{A\to B},\iblock{\Omega_1\sand \Omega_2,\lef B},\iblock{\Phi_1\sand\Phi_2},\iblock{\Psi_1\sand\Psi_2}}}
\infer1[IH]{\csseq{\Sigma',\lef{A\to B},\iblock{\Omega_1\sand \Omega_2,\lef B},\iblock{\Phi_1, \Psi_1\sand\Phi_2,\Psi_2}}}
\infer2[$r$]{\csseq{\Sigma',\lef{A\to B},\iblock{\Omega_1\sand \Omega_2},\iblock{\Phi_1,\Psi_1\sand\Psi_2,\Phi_2}}}
        \end{prooftree}
   \end{tabular}     
        \end{adjustbox}
\end{center}

\item Let $r=\ruletor$ and $\Sigma=\Sigma',\rig{A\to B}$.
\begin{center}
\begin{adjustbox}{max width=.9\textwidth}
\begin{tabular}{c}
      \begin{prooftree}
\hypo{}
\ellipsis{}{\csseq{\Sigma',\iblock{\rig A\sand \lef B},\iblock{\Phi_1\sand\Phi_2},\iblock{\Psi_1\sand\Psi_2}}}
\infer1[$r$]{\csseq{\Sigma',\rig{A\to B},\iblock{\Phi_1\sand\Phi_2},\iblock{\Psi_1\sand\Psi_2}}}
        \infer1[$\rulemerge$]{\csseq{\Sigma',\rig{A\to B},\iblock{\Phi_1,\Psi_1\sand\Psi_2,\Phi_2}}}
        \end{prooftree}
        \hspace{0.3cm}$\leadsto$\hspace{0.3cm}
 \begin{prooftree}
\hypo{}
\ellipsis{}{\csseq{\Sigma',\iblock{\rig A\sand \lef B},\iblock{\Phi_1\sand\Phi_2},\iblock{\Psi_1\sand\Psi_2}}}
\infer1[IH]{\csseq{\Sigma',\iblock{\rig A\sand \lef B},\iblock{\Phi_1,\Psi_1\sand\Psi_2,\Phi_2}}}
        \infer1[$r$]{\csseq{\Sigma',\rig{A\to B},\iblock{\Phi_1,\Psi_1\sand\Psi_2,\Phi_2}}}
        \end{prooftree}
        \end{tabular}     
        \end{adjustbox}
\end{center}

\item Let $r=\ruleliftib$, $\csseq{ \ }$ be empty, and $\Sigma=\Sigma',\iblock{\Omega_1, \lef A\sand\Omega_2}$.
\begin{center}
\begin{adjustbox}{max width=.9\textwidth}
\begin{tabular}{c}
      \begin{prooftree}
\hypo{}
\ellipsis{}{{\Sigma',\iblock{\Omega_1,\lef A\sand\Omega_2,\lef A},\iblock{\Phi_1\sand\Phi_2},\iblock{\Psi_1\sand\Psi_2}}}
\infer1[$r$]{{\Sigma',\iblock{\Omega_1,\lef A\sand\Omega_2},\iblock{\Phi_1\sand\Phi_2},\iblock{\Psi_1\sand\Psi_2}}}
        \infer1[$\rulemerge$]{{\Sigma',\iblock{\Omega_1,\lef A\sand\Omega_2},\iblock{\Phi_1,\Psi_1\sand\Psi_2,\Phi_2}}}
        \end{prooftree}
        \hspace{0.3cm}$\leadsto$\hspace{0.3cm}
 \begin{prooftree}
\hypo{}
\ellipsis{}{{\Sigma',\iblock{\Omega_1,\lef A\sand\Omega_2,\lef A},\iblock{\Phi_1\sand\Phi_2},\iblock{\Psi_1\sand\Psi_2}}}
\infer1[IH]{{\Sigma',\iblock{\Omega_1,\lef A\sand\Omega_2,\lef A},\iblock{\Phi_1,\Psi_1\sand\Psi_2,\Phi_2}}}
        \infer1[$r$]{{\Sigma',\iblock{\Omega_1,\lef A\sand\Omega_2},\iblock{\Phi_1,\Psi_1\sand\Psi_2,\Phi_2}}}
        \end{prooftree}
        \end{tabular}     
        \end{adjustbox}
\end{center}

All other cases for rules $r$ can be dealt with similarly.
\end{itemize}

\noindent \textbf{Case 2.} $r$ is applied in the block $\iblock{\Phi_1\sand\Phi_2}$. We only display some salient examples.
\begin{itemize}[leftmargin=*]
\item Let $r=\ruleliftib$, $\csseq{ \ }$ be empty, and $\Phi_1=\Phi_1',\lef A$.

\begin{center}
\begin{tabular}{c}
\begin{prooftree}
\hypo{}
\ellipsis{}{{\Sigma,\iblock{\Phi_1', \lef A\sand\Phi_2, \lef A},\iblock{\Psi_1\sand\Psi_2}}}
\infer1[$r$]{{\Sigma,\iblock{\Phi_1', \lef A\sand\Phi_2},\iblock{\Psi_1\sand\Psi_2}}}
        \infer1[$\rulemerge$]{{\Sigma,\iblock{\Phi_1',\lef A,\Psi_1\sand\Psi_2,\Phi_2}}}
        \end{prooftree}
        \hspace{0.5cm}$\leadsto$\hspace{0.5cm}
 \begin{prooftree}
\hypo{}
\ellipsis{}{{\Sigma,\iblock{\Phi_1',\lef A\sand\Phi_2,\lef A},\iblock{\Psi_1\sand\Psi_2}}}
\infer1[IH]{{\Sigma,\iblock{\Phi_1',\lef A,\Psi_1\sand\Psi_2,\lef A,\Phi_2}}}
        \infer1[$r$]{{\Sigma,\iblock{\Phi_1,\lef A,\Psi_1\sand\Psi_2,\Phi_2}}}
        \end{prooftree}
        \end{tabular}     
\end{center}

\item If $r$ is one of the other rules, the proof proceeds straightforwardly. As an example for the propositional rules, let $r=\ruletol$ and $\Phi_2 = \Phi_2', \lef{A\to B}, \iblock{\Omega_1\sand\Omega_2}$.

\begin{center}
\begin{adjustbox}{max width=.9\textwidth}
\begin{tabular}{c}
     \begin{prooftree}
\hypo{}
\ellipsis{}{\dowcsseq{\dow\Sigma,\iblock{\dow\Phi_1\sand\dow{\Phi_2'},\lef{A\to B},\iblock{\dow\Omega_1,\rig A\sand\dow\Omega_2}},\iblock{\dow\Psi_1\sand\dow\Psi_2}}}
\hypo{}
\ellipsis{}{\csseq{\Sigma,\iblock{\Phi_1\sand\Phi_2',\lef{A\to B},\iblock{\Omega_1\sand\Omega_2,\lef B}},\iblock{\Psi_1\sand\Psi_2}}}
\infer2[$r$]{\csseq{\Sigma,\iblock{\Phi_1\sand\Phi_2',\lef{A\to B},\iblock{\Omega_1\sand\Omega_2}},\iblock{\Psi_1\sand\Psi_2}}}
        \infer1[$\rulemerge$]{\csseq{\Sigma,\iblock{\Phi_1,\Psi_1\sand\Psi_2,\Phi_2',\lef{A\to B},\iblock{\Omega_1\sand\Omega_2}}}}
        \end{prooftree}
        \hspace{0.2cm}$\leadsto$\vspace{0.3cm}\\
         \begin{prooftree}
\hypo{}
\ellipsis{}{\dowcsseq{\dow\Sigma,\iblock{\dow\Phi_1\sand\dow{\Phi_2'},\lef{A\to B},\iblock{\dow\Omega_1,\rig A\sand\dow\Omega_2}},\iblock{\dow\Psi_1\sand\dow\Psi_2}}}
\infer1[IH]{\dowcsseq{\dow\Sigma,\iblock{\dow\Phi_1,\dow\Psi_1\sand\dow{\Phi_2'},\dow\Psi_2,\lef{A\to B},\iblock{\dow\Omega_1,\rig A\sand\dow\Omega_2}}}}
\hypo{}
\ellipsis{}{\csseq{\Sigma,\iblock{\Phi_1\sand\Phi_2',\lef{A\to B},\iblock{\Omega_1\sand\Omega_2,\lef B}},\iblock{\Psi_1\sand\Psi_2}}}
\infer1[IH]{\csseq{\Sigma,\iblock{\Phi_1,\Psi_1\sand\Phi_2',\Psi_2,\lef{A\to B},\iblock{\Omega_1\sand\Omega_2,\lef B}}}}
\infer2[$r$]{\csseq{\Sigma,\iblock{\Phi_1,\Psi_1\sand\Psi_2,\Phi_2',\lef{A\to B},\iblock{\Omega_1\sand\Omega_2}}}}
        \end{prooftree}
\end{tabular}     
\end{adjustbox}
\end{center}
\end{itemize}

\noindent \textbf{Case 3.} $r$ is applied in the block $\iblock{\Psi_1\sand\Psi_2}$. The proof proceeds as in Case 2. 
\end{proof}

   Further structural rules 
   are considered for establishing the completeness of 
   $\cbplus$, but they need not be taken as primitive. 
    We consider the following two rules:
    $$
    \begin{aligned}
    \begin{prooftree}
    \hypo{\iblock{\Phi_1\sand\Phi_2}}
    \infer1[{\sf R1}]{\csseq{\Phi_1,\Phi_2}}
    \end{prooftree}
    &
    \qquad
    &
    \begin{prooftree}          
    \hypo{\Sigma,\iblock{\Phi_1,\iblock{\Psi_1\sand\Psi_2}\sand\Phi_2}}
    \infer1[$\sf R2$]{\Sigma,\iblock{\Phi_1\sand\Phi_2,\iblock{\Psi_1\sand\Psi_2}}}
    \end{prooftree}
    \end{aligned}
    $$
   Rule $\sf R1$ reflects frame condition (F1), expressing the reflexivity of $\leq$ by allowing any nested component to be treated as a whole relevant block. Rule $\sf R2$ corresponds to frame condition (F2) and allows a relevant block occurring in the second component of a top-level block to be moved to the first component. 
   Moreover, the semantics given in Definition~\ref{def:seq-semantics} aligns well with these rules, ensuring their validity in full generality and making any restriction on the position of the output formula unnecessary. 

We show both $\sf R1$ and $\sf R2$ are `correct' structural rules in $\cbplus$ in the sense that they are semantically valid and syntactically hp-admissible in the calculus. 

\begin{proposition}
    The rules $\sf R1$ and $\sf R2$ are $\bplus$-valid.
\end{proposition}

\begin{proof}
First, we prove the validity of $\sf R1$. Since $\iblock{\Phi_1\sand\Phi_2}$ is full, the output formula occurs either in $\Phi_1$ or $\Phi_2$. We only show the former case as the latter is similar. 
Assume the premise is valid but the conclusion is not. Then there exist a model $\+M=(\+O,K,R,V)$ and $a\in K$ such that $a\Vdash \Phi_2$ and $a\not\Vdash \Phi_1$. Since $a\le a$, by (F1) there exists $x\in\+O$ such that $Rxaa$. By the semantic clause for $\iblock{\Phi_1\sand\Phi_2}$, $x\Vdash\iblock{\Phi_1\sand\Phi_2}$ requires that for all $y,z\in K$, if $Rxyz$ and $z\Vdash\Phi_2$ then $y\Vdash\Phi_1$. Taking $y=a$ and $z=a$, since $Rxaa$, $a\Vdash \Phi_2$ and $a\not\Vdash \Phi_1$, we obtain $x\not\Vdash\iblock{\Phi_1\sand\Phi_2}$, contradicting the validity of the premise. 

Next, we move towards the validity of $\sf R2$. We only show two cases when the output formula occurs in $\Sigma$ and $\Phi_1$ as the other cases are similar. 
    \begin{itemize}
         \item The output formula occurs in $\Sigma$. Assume the conclusion is not valid, this means there exists a model $\+M=(\+O,K,R,V)$ and $w\in\+O$ such that 
         $w\nVdash \Sigma$ and 
         $w\Vdash \dow\Sigma,\iblock{\Phi_1\sand\Phi_2,\iblock{\Psi_1\sand\Psi_2}}$. By definition, we have $w\Vdash \iblock{\Phi_1\sand\Phi_2,\iblock{\Psi_1\sand\Psi_2}}$, which is a left-sided sequent. 
        This further implies there exist $a,b,c,d\in K$ such that $Rwab, a\Vdash \Phi_1, b\Vdash\Phi_2$ and $Rbcd, c\Vdash \Psi_1$ and $d\Vdash \Psi_2$. 
        Since the premise is valid, we have $w\Vdash \Sigma,\iblock{\Phi_1,\iblock{\Psi_1\sand\Psi_2}\sand\Phi_2}$. 
        Since $w\nVdash \Sigma$, we have $w\nVdash \iblock{\Phi_1,\iblock{\Psi_1\sand\Psi_2}\sand\Phi_2}$. 
        This implies that for any $a',b',c',d'$ with $Rwa'b'$ and $Ra'c'd'$, either $a'\nVdash \Phi_1$ or $b'\nVdash \Phi_2$ or $c'\nVdash \Psi_1$ or $d'\nVdash\Psi_2$.  
        Taking $a,b,c,d$ given by the invalidity of the conclusion, we have $a\Vdash \Phi_1$ and $b\Vdash \Phi_2$; moreover, from $Rbcd$ and $Rwab$, by (F2) we have $Racd$, hence $c\nVdash \Psi_1$ or $d\nVdash\Psi_2$. 
        Since $c\Vdash \Psi_1$ and $d\Vdash \Psi_2$, a contradiction. 
        
        \item The output formula occurs in $\Phi_1$. Assume the conclusion is not valid, this means there exists a model $\+M=(\+O,K,R,V)$ and $w\in\+O$ such that $w\Vdash \Sigma$ and $w\nVdash \iblock{\Phi_1\sand\Phi_2,\iblock{\Psi_1\sand\Psi_2}}$. 
        By Definition~\ref{def:seq-semantics} there exist 
        $a,b\in K$ such that $Rwab, a\nVdash\Phi_1$ and $b\Vdash \Phi_2,\iblock{\Psi_1\sand\Psi_2}$. Since $\iblock{\Psi_1\sand\Psi_2}$ is left-sided, this further implies for any $c,d\in K$ with $Rbcd$ it holds $c\Vdash \Psi_1$ and $d\Vdash \Psi_2$. 
        Since the premise is valid we have $w\Vdash \Sigma,\iblock{\Phi_1,\iblock{\Psi_1\sand\Psi_2}\sand\Phi_2}$. 
        Since $b\Vdash \Phi_2$, we have $a\Vdash \Phi_1,\iblock{\Psi_1\sand\Psi_2}$. 
        Recall that $a\nVdash \Phi_1$ and $\iblock{\Psi_1\sand\Psi_2}$ is left-sided, hence $a\nVdash \iblock{\Psi_1\sand\Psi_2}$. 
        This implies that for any $c',d'$ with $Rac'd'$, either 
        $c'\nVdash \Psi_1$ or $d'\nVdash\Psi_2$. 
        Taking $c,d$ obtained above, from $Rbcd$ and $Rwab$ we obtain $Racd$ by (F2), 
        since $c\Vdash\Psi_1$ and $d\Vdash\Psi_2$, a contradiction. 
    \end{itemize}
    This completes the proof. 
\end{proof}

\begin{proposition}\label{prop:hp-admi-r1-r2}
    The rules $\sf R1$ and $\sf R2$ are hp-admissible in $\cbplus$ and $\cbplus+\cut$. 
\end{proposition}

\begin{proof}
We prove both rules are hp-admissible in $\cbplus+\cut$ and the proof is a modular extension for the same result in $\cbplus$. 
First, we provide a detailed proof of the admissibility of $\sf R2$.  
The proof proceeds by induction on the height $n$ of the derivation. If $n=0$, the conclusion follows straightforwardly. 
If $n>0$, let us consider the last rule $r$ applied in the derivation of the premise. Notice that each rule can happen deep in the displayed component, to simplify the proof, we only show the basic case when it happens on the surface. 
Deep cases are treated similarly. 

\noindent\textbf{Case 1.} $r$ is applied in $\Sigma$. If $r$ is one of the propositional rules, the case is straightforward: by considering a derivation of the premise ending with an application of $r$, the induction hypothesis yields a derivation of the desired conclusion, to which $r$ can then be applied again. As an example, take $\Sigma=\Sigma',\lef{A\to B}$ and $r=\ruletol$ is applied to $\lef{A\to B}$ and $\iblock{\Phi_1,\iblock{\Psi_1\sand\Psi_2}\sand\Phi_2}$.
\begin{center}
\begin{adjustbox}{max width=.8\textwidth}
\begin{tabular}{c}
\begin{prooftree}
  \hypo{}
  \ellipsis{}{{\Sigma',\lef{A \to B},\iblock{\dow\Phi_1,\rig A, \iblock{\dow\Psi_1 \sand \dow\Psi_2} \sand \dow\Phi_2}}}
  \hypo{}
  \ellipsis{}{{\Sigma',\lef{A \to B},\iblock{\Phi_1, \iblock{\Psi_1 \sand \Psi_2}\sand \Phi_2, \lef B}}}
  \infer2[$r$]{{\Sigma',\lef{A \to B},\iblock{\Phi_1, \iblock{\Psi_1 \sand \Psi_2}\sand \Phi_2}}}
  \infer1[\sf R2]{{\Sigma',\lef{A \to B},\iblock{\Phi_1\sand \Phi_2, \iblock{\Psi_1 \sand \Psi_2}}}}
\end{prooftree}
\quad
$\leadsto$\\
\begin{prooftree}
\hypo{}
  \ellipsis{}{{\Sigma',\lef{A \to B},\iblock{\dow\Phi_1,\rig A, \iblock{\dow\Psi_1 \sand \dow\Psi_2} \sand \dow\Phi_2}}}
  \infer1[IH]{{\Sigma',\lef{A \to B},\iblock{\dow\Phi_1, \rig A \sand \dow\Phi_2,\iblock{\dow\Psi_1\sand \dow\Psi_2}}}}
 \hypo{}
  \ellipsis{}{{\Sigma',\lef{A \to B},\iblock{\Phi_1, \iblock{\Psi_1 \sand \Psi_2}\sand \Phi_2, \lef B}}}
  \infer1[IH]{{\Sigma',\lef{A \to B},\iblock{\Phi_1\sand \Phi_2,\lef B, \iblock{\Psi_1 \sand \Psi_2}}}}
  \infer2[$r$]{{\Sigma',\lef{A \to B},\iblock{\Phi_1\sand \Phi_2, \iblock{\Psi_1 \sand \Psi_2}}}}
\end{prooftree}
\end{tabular}
\end{adjustbox}
\end{center}
If $r=\cut$ then the transformation is as follows:
\begin{center}
\begin{tabular}{c}
    \begin{prooftree}          
    \hypo{}
    \ellipsis{}{\dow\Sigma,\rig A,\iblock{\dow\Phi_1,\iblock{\dow\Psi_1\sand\dow\Psi_2}\sand\dow\Phi_2}}
    \hypo{}
    \ellipsis{}{\Sigma,\lef A,\iblock{\Phi_1,\iblock{\Psi_1\sand\Psi_2}\sand\Phi_2}}
    \infer2[$r$]{\Sigma,\iblock{\Phi_1,\iblock{\Psi_1\sand\Psi_2}\sand\Phi_2}}
    \infer1[$\sf R2$]{\Sigma,\iblock{\Phi_1\sand\Phi_2,\iblock{\Psi_1\sand\Psi_2}}}
    \end{prooftree}
   \quad $\leadsto$\\
   \begin{prooftree}          
    \hypo{}
    \ellipsis{}{\dow\Sigma,\rig A,\iblock{\dow\Phi_1,\iblock{\dow\Psi_1\sand\dow\Psi_2}\sand\dow\Phi_2}}
    \infer1[IH]{\dow\Sigma,\rig A,\iblock{\dow\Phi_1\sand\dow\Phi_2,\iblock{\dow\Psi_1\sand\dow\Psi_2}}}
    \hypo{}
    \ellipsis{}{\Sigma,\lef A,\iblock{\Phi_1,\iblock{\Psi_1\sand\Psi_2}\sand\Phi_2}}
    \infer1[IH]{{\Sigma,\lef A,\iblock{\Phi_1\sand\Phi_2,\iblock{\Psi_1\sand\Psi_2}}}}
    \infer2[$r$]{\Sigma,\iblock{\Phi_1\sand\Phi_2,\iblock{\Psi_1\sand\Psi_2}}}
    \end{prooftree}
    \end{tabular}
\end{center}
\textbf{Case 2.} $r$ is applied in $\Phi_1$.
\begin{itemize}[leftmargin=*]
\item If $r$ is a rule for either $\land$ or $\lor$, no difficulties arise.
    \item Let $\Phi_1 = \Phi_1', \lef{A \to B}$ and $r=\ruletol$ is applied to $\lef{A\to B}$ and $\iblock{\Psi_1\sand\Psi_2}$. We have 
\begin{center}
\begin{adjustbox}{max width=.8\textwidth}
\begin{tabular}{c}
\begin{prooftree}
  \hypo{}
  \ellipsis{}{{\dow\Sigma,\iblock{\dow{\Phi_1'}, \lef{A \to B}, \iblock{\Psi_1^*, \rig A \sand \dow\Psi_2} \sand \dow\Phi_2}}}
  \hypo{}
  \ellipsis{}{{\Sigma,\iblock{\Phi_1', \lef{A \to B}, \iblock{\Psi_1 \sand \Psi_2, \lef B}\sand \Phi_2}}}
  \infer2[$r$]{{\Sigma,\iblock{\Phi_1', \lef{A \to B}, \iblock{\Psi_1 \sand \Psi_2}\sand \Phi_2}}}
  \infer1[\sf R2]{{\Sigma,\iblock{\Phi_1', \lef{A \to B}\sand \Phi_2, \iblock{\Psi_1 \sand \Psi_2}}}}
\end{prooftree}
\quad
$\leadsto$\\
\begin{prooftree}
\hypo{}
  \ellipsis{}{{\dow\Sigma,\iblock{\dow{\Phi_1'}, \lef{A \to B}, \iblock{\Psi_1^*, \rig A \sand \dow\Psi_2} \sand \dow\Phi_2}}}
  \infer1[IH]{{\dow\Sigma,\iblock{\dow{\Phi_1'}, \lef{A \to B} \sand \dow\Phi_2,\iblock{\Psi_1^*, \rig A \sand \dow\Psi_2}}}}
  \infer1[$\rulewl$]{{\dow\Sigma,\iblock{\dow{\Phi_1'}, \lef{A \to B} \sand \dow\Phi_2,\lef{A \to B},\iblock{\Psi_1^*, \rig A \sand \dow\Psi_2}}}}
 \hypo{}
  \ellipsis{}{{\Sigma,\iblock{\Phi_1', \lef{A \to B}, \iblock{\Psi_1 \sand \Psi_2, \lef B}\sand \Phi_2}}}
  \infer1[IH]{{\Sigma,\iblock{\Phi_1', \lef{A \to B}\sand \Phi_2, \iblock{\Psi_1 \sand \Psi_2, \lef B}}}}
  \infer1[$\rulewl$]{{\Sigma,\iblock{\Phi_1', \lef{A \to B}\sand \Phi_2, \lef{A \to B},\iblock{\Psi_1 \sand \Psi_2, \lef B}}}}
  \infer2[$r$]{{\Sigma,\iblock{\Phi_1', \lef{A \to B}\sand \Phi_2, \lef{A \to B}, \iblock{\Psi_1 \sand \Psi_2}}}}
   \infer1[hp-inv. $\ruleliftib$]{{\Sigma,\iblock{\Phi_1', \lef{A \to B}\sand \Phi_2, \iblock{\Psi_1 \sand \Psi_2}}}}
\end{prooftree}
\end{tabular}
\end{adjustbox}
\end{center}

\item Let $\Phi_1 = \Phi_1', \lef{A \to B},\iblock{\Theta_1\sand\Theta_2}$ and $r=\ruletol$ is applied to $\lef{A\to B}$ and $\iblock{\Theta_1\sand\Theta_2}$.

\begin{center}
\begin{adjustbox}{max width=.8\textwidth}
\begin{tabular}{c}
\begin{prooftree}
  \hypo{}
  \ellipsis{}{{\dow\Sigma,\iblock{\dow{\Phi_1'}, \lef{A \to B}, \iblock{\Theta_1^*, \rig A \sand \dow\Theta_2},\iblock{\dow\Psi_1 \sand \dow\Psi_2} \sand \dow\Phi_2}}}
  \hypo{}
  \ellipsis{}{{\Sigma,\iblock{\Phi_1', \lef{A \to B}, \iblock{\Theta_1 \sand \Theta_2, \lef B},\iblock{\Psi_1 \sand \Psi_2}\sand \Phi_2}}}
  \infer2[$r$]{{\Sigma,\iblock{\Phi_1', \lef{A \to B}, \iblock{\Theta_1\sand\Theta_2}, \iblock{\Psi_1 \sand \Psi_2}\sand \Phi_2}}}
  \infer1[\sf R2]{{\Sigma,\iblock{\Phi_1', \lef{A \to B},\iblock{\Theta_1\sand\Theta_2}\sand \Phi_2, \iblock{\Psi_1 \sand \Psi_2}}}}
\end{prooftree}
\quad
$\leadsto$\\
\begin{prooftree}
\hypo{}
  \ellipsis{}{{\dow\Sigma,\iblock{\dow{\Phi_1'}, \lef{A \to B}, \iblock{\Theta_1^*, \rig A \sand \dow\Theta_2},\iblock{\dow\Psi_1 \sand \dow\Psi_2} \sand \dow\Phi_2}}}
  \infer1[IH]{{\Sigma,\iblock{\Phi_1', \lef{A \to B}, \iblock{\Theta_1^*, \rig A \sand \dow\Theta_2} \sand \Phi_2,\iblock{\Psi_1 \sand \Psi_2}}}}
 \hypo{}
  \ellipsis{}{{\Sigma,\iblock{\Phi_1', \lef{A \to B}, \iblock{\Theta_1 \sand \Theta_2, \lef B},\iblock{\Psi_1 \sand \Psi_2}\sand \Phi_2}}}
  \infer1[IH]{{\Sigma,\iblock{\Phi_1', \lef{A \to B}, \iblock{\Theta_1 \sand \Theta_2, \lef B}\sand \Phi_2,\iblock{\Psi_1 \sand \Psi_2}}}}
  \infer2[$r$]{{\Sigma,\iblock{\Phi_1', \lef{A \to B},\iblock{\Theta_1\sand\Theta_2}\sand \Phi_2, \iblock{\Psi_1 \sand \Psi_2}}}}
\end{prooftree}
\end{tabular}
\end{adjustbox}
\end{center}

Other nested cases happening within $\Phi_1$ are similar. 

\item Let $\Phi_1 = \Phi_1', \rig{A \to B}$ and $r=\ruletor$. We consider the following derivation.

\begin{center}
\begin{adjustbox}{max width=.75\textwidth}
\begin{tabular}{ccc}
\begin{prooftree}
  \hypo{}
  \ellipsis{}{{\Sigma,\iblock{\Phi_1', \iblock{\lef{A} \sand \rig{B}}, \iblock{\Psi_1\sand\Psi_2} \sand \Phi_2}}}
  \infer1[$r$]{{\Sigma,\iblock{\Phi_1', \rig{A \to B}, \iblock{\Psi_1 \sand \Psi_2}\sand \Phi_2}}}
  \infer1[\sf R2]{{\Sigma,\iblock{\Phi_1', \rig{A \to B}\sand \Phi_2, \iblock{\Psi_1 \sand \Psi_2}}}}
\end{prooftree}
&
$\leadsto$&
\begin{prooftree}
\hypo{}
  \ellipsis{}{{\Sigma,\iblock{\Phi_1', \iblock{\lef{A} \sand \rig{B}}, \iblock{\Psi_1\sand\Psi_2} \sand \Phi_2}}}
  \infer1[IH]{{\Sigma,\iblock{\Phi_1', \iblock{\lef{A} \sand \rig{B}} \sand \Phi_2,\iblock{\Psi_1\sand\Psi_2}}}}
  \infer1[$r$]{{\Sigma,\iblock{\Phi_1', \rig{A \to B}\sand \Phi_2,\iblock{\Psi_1 \sand \Psi_2}}}}
\end{prooftree}

\end{tabular}
\end{adjustbox}
\end{center}

\item Let $\Phi_1 = \lef{A},\Phi_1'$, with $\lef{A}$ being involved in $r=\ruleliftib$. We consider the following derivation.

\begin{center}
\begin{adjustbox}{max width=.75\textwidth}
\begin{tabular}{ccc}
\begin{prooftree}
  \hypo{}
  \ellipsis{}{{\Sigma,\iblock{\lef{A},\Phi_1',  \iblock{\Psi_1\sand\Psi_2} \sand \lef{A},\Phi_2}}}
  \infer1[$r$]{{\Sigma,\iblock{\lef{A},\Phi_1', \iblock{\Psi_1 \sand \Psi_2}\sand \Phi_2}}}
  \infer1[\sf R2]{{\Sigma,\iblock{\lef{A},\Phi_1'\sand \Phi_2,\iblock{\Psi_1 \sand \Psi_2}}}}
\end{prooftree}
&
$\leadsto$&
\begin{prooftree}
\hypo{}
  \ellipsis{}{{\Sigma,\iblock{\lef{A},\Phi_1',  \iblock{\Psi_1\sand\Psi_2} \sand \lef{A},\Phi_2}}}
\infer1[IH]{{\Sigma,\iblock{\lef{A},\Phi_1' \sand \lef{A},\Phi_2,\iblock{\Psi_1\sand\Psi_2}}}}
  \infer1[$r$]{{\Sigma,\iblock{\lef{A},\Phi_1'\sand \Phi_2,\iblock{\Psi_1 \sand \Psi_2}}}}
\end{prooftree}

\end{tabular}
\end{adjustbox}
\end{center}
\item Let $r=\cut$. We consider the following transformation.
\begin{center}
\begin{tabular}{c}
    \begin{prooftree}          
    \hypo{}
    \ellipsis{}{\dow\Sigma,\iblock{\dow\Phi_1,\rig A,\iblock{\dow\Psi_1\sand\dow\Psi_2}\sand\dow\Phi_2}}
    \hypo{}
    \ellipsis{}{\Sigma,\iblock{\Phi_1,\lef A,\iblock{\Psi_1\sand\Psi_2}\sand\Phi_2}}
    \infer2[$r$]{\Sigma,\iblock{\Phi_1,\iblock{\Psi_1\sand\Psi_2}\sand\Phi_2}}
    \infer1[$\sf R2$]{\Sigma,\iblock{\Phi_1\sand\Phi_2,\iblock{\Psi_1\sand\Psi_2}}}
    \end{prooftree}
\quad   $\leadsto$\\
   \begin{prooftree}          
    \hypo{}
    \ellipsis{}{\dow\Sigma,\iblock{\dow\Phi_1,\rig A,\iblock{\dow\Psi_1\sand\dow\Psi_2}\sand\dow\Phi_2}}
    \infer1[IH]{\dow\Sigma,\iblock{\dow\Phi_1,\rig A\sand\dow\Phi_2,\iblock{\dow\Psi_1\sand\dow\Psi_2}}}
    \hypo{}
    \ellipsis{}{\Sigma,\iblock{\Phi_1,\lef A,\iblock{\Psi_1\sand\Psi_2}\sand\Phi_2}}
    \infer1[IH]{{\Sigma,\iblock{\Phi_1,\lef A\sand\Phi_2,\iblock{\Psi_1\sand\Psi_2}}}}
    \infer2[$r$]{\Sigma,\iblock{\Phi_1\sand\Phi_2,\iblock{\Psi_1\sand\Psi_2}}}
    \end{prooftree}
    \end{tabular}
\end{center}
\end{itemize}
\textbf{Case 3.} $r$ is applied in $\iblock{\Psi_1 \sand \Psi_2}$. We distinguish two subcases, depending on whether $r$ is applied in $\Psi_1$ or in $\Psi_2$. These cases present no particular difficulties. By examining the premises of $r$, we apply the induction hypothesis to each of them in order to relocate the relevant block $\iblock{\Psi_1 \sand \Psi_2}$ into the second component. The desired conclusion is then obtained by reapplying $r$.

\noindent\textbf{Case 4.} $r$ is applied in $\Phi_2$. Similarly to \textbf{Case 2}. 
\\
Now, we turn to $\sf R1$. The proof proceeds in a similar and even simpler way. We only show the case when the premise $\iblock{\Phi_1\sand\Phi_2}$ is derived from $\ruleliftib$. In this case, assume the rule is applied to some $\lef A\in\Phi_1$ and we further denote $\Phi_1=\lef A, \Phi_1'$. Then we have the following transformation of the derivation:
\begin{center}
    \begin{adjustbox}{max width=.9\textwidth}
    \begin{tabular}{ccc}
        \begin{prooftree}
            \hypo{}
            \ellipsis{}{\iblock{\lef A, \Phi_1'\sand\lef A,\Phi_2}}
            \infer1[$\ruleliftib$]{\iblock{\lef A, \Phi_1'\sand\Phi_2}}
            \infer1[$\sf R1$]{\csseq{\lef A, \Phi_1',\Phi_2}}
        \end{prooftree}
        & $\leadsto$ &
        \begin{prooftree}
            \hypo{}
            \ellipsis{}{\iblock{\lef A, \Phi_1'\sand\lef A,\Phi_2}}
            \infer1[IH]{\csseq{\lef A, \lef A, \Phi_1',\Phi_2}}
            \infer1[$\rulecl$]{\csseq{\lef A, \Phi_1',\Phi_2}}
        \end{prooftree}
    \end{tabular}
    \end{adjustbox}
\end{center}

This completes the proof. 
\end{proof}

\begin{restatable}[Cut-admissibility]{theorem}{cutbplus}\label{thm:cut-admissibility}
    $\cut$ is admissible in $\cbplus$.
\end{restatable}

\begin{proof}
    Let us consider an application of $\cut$ of the following form in a derivation $\+D$:
    \begin{center}
        \vspace{-.5cm}    
        \begin{tabular}{c}
            $
            \vlderivation{
            \vliin{}{\cut}{\csseq{\Sigma}}{
            \vltr {\+D_1} {
                    \dowcsseq{\dow{\Sigma},\rig A}
                }{
                \vlhy {\quad}}
                {
                \vlhy {}}
                {
                \vlhy {\quad}}
            }
            {
                \vltr {\+D_2} {
                    \csseq{\lef A,\Sigma}
                }{
                \vlhy {\quad}}
                {
                \vlhy {}}
                {
                \vlhy {\quad}} 
            }
            }
            $
        \end{tabular}
    \end{center}
    We define as usual the rank of each application of $\cut$ as an ordered pair $(|A|,h_1+h_2)$ where $|A|$ denotes the size of the cut formula $A$ and $h_1,h_2$ denote the height of the two premises of $\cut$ in $\+D_1$ and $\+D_2$ respectively.  
    We prove by induction on the rank of $\cut$. 
    
        For the basic step, suppose $A=p$. 
        If the left premise $\csseq{\dow\Sigma,\rig A}$ is an initial sequent, then $\lef A\in\remove{\dow\Sigma}$. Applying contraction to the right premise, we obtain the conclusion from the premise directly. 
        If the right premise $\csseq{\Sigma,\lef A}$ is an initial sequent, then either $\rig A\in\remove{\Sigma}$ or some other nested component containing the output formula is initial. For the former, it implies $\dow\Sigma,\rig p=\Sigma$, so we just take the left premise as it is identical to the conclusion; for the latter, it is easy to see by removing $\lef A$ from the right premise, the sequent which remains $\csseq{\Sigma}$ is still initial, which is exactly the conclusion of $\cut$.  

        For the inductive step, 
        let us further assume the last rule applied in $\+D_1$ and $\+D_2$ are $r_1$ and $r_2$ respectively. 
        We proceed by discussing different roles of $\rig A$ and $\lef A$ in $r_1$ and $r_2$. Note that no rules in the calculus introduce weak formulas in the conclusion, so we only need to consider whether the cut formulas are principal or side. Thus we have the following cases: \textbf{Case 1.} $\rig A$ is principal in $r_1$ and $\lef A$ is principal in $r_2$; \textbf{Case 2.} $\rig A$ is side in $r_1$; \textbf{Case 3.} $\lef A$ is side in $r_2$ and $\rig A$ is principal in $r_1$.  
        We demonstrate each case in turn. \\

        \noindent \textbf{Case 1.} $\rig A$ is principal in $r_1$ and $\lef A$ is principal in $r_2$. 
        We have the following sub-cases according to the shape of $A$: 
        \begin{itemize}[leftmargin=*]
            \item $A=B\wedge C$. Then we have the following transformation of the derivation 
            \vspace{-.3cm}
            \begin{center}
            \begin{adjustbox}{max width=.98\textwidth}
                \begin{tabular}{ccc}
                    \begin{prooftree}
                        \hypo{}
                        \ellipsis{}{\dowcsseq{\dow{\Sigma},\rig {B}}}
                        \hypo{}
                        \ellipsis{}{\dowcsseq{\dow{\Sigma},\rig {C}}}
                        \infer2[$r_1=\rig \wedge$]{\dowcsseq{\dow{\Sigma},\rig{B\wedge C}}}
                        \hypo{}
                        \ellipsis{}{\csseq{\Sigma,\lef{B},\lef C}}
                        \infer1[$r_2=\lef \wedge$]{\csseq{\Sigma,\lef{B\wedge C}}}
                        \infer2{\csseq{\Sigma}}
                    \end{prooftree}
                    &
                    $\leadsto$
                    &
                    \begin{prooftree}
                        \hypo{}
                        \ellipsis{}{\dowcsseq{\dow{\Sigma},\rig {C}}}
                        \hypo{}
                        \ellipsis{}{\dowcsseq{\dow{\Sigma},\rig {B}}}
                        \infer1[$\rulewl$]{\dowcsseq{\dow{\Sigma},\rig {B},\lef C}}
                        \hypo{}
                        \ellipsis{}{\csseq{\Sigma,\lef{B},\lef C}}
                        \infer2[IH]{\csseq{\Sigma,\lef C}}
                        \infer2[IH]{\csseq{\Sigma}}
                    \end{prooftree}
                \end{tabular}
            \end{adjustbox}
            \end{center}
            \item $A=B\vee C$. 
            Assume w.l.o.g the left premise $\dowcsseq{\dow{\Sigma},\rig {B\vee C}}$ is derived from $\dowcsseq{\dow{\Sigma},\rig {B}}$.  
            Then we have the following transformation of the derivation
             \vspace{-.3cm}
            \begin{center}
            \begin{adjustbox}{max width=.9\textwidth}
                \begin{tabular}{ccc}
                    \begin{prooftree}
                        \hypo{}
                        \ellipsis{}{\dowcsseq{\dow{\Sigma},\rig {B}}}
                        \infer1[$r_1=\rig \vee$]{\dowcsseq{\dow{\Sigma},\rig{B\vee C}}}
                        \hypo{}
                        \ellipsis{}{\csseq{\Sigma,\lef{B}}}
                        \hypo{}
                        \ellipsis{}{\csseq{\Sigma,\lef C}}
                        \infer2[$r_2=\lef \vee$]{\csseq{\Sigma,\lef{B\vee C}}}
                        \infer2{\csseq{\Sigma}}
                    \end{prooftree}
                    &
                    $\leadsto$
                    &
                    \begin{prooftree}
                        \hypo{}
                        \ellipsis{}{\dowcsseq{\dow{\Sigma},\rig {B}}}
                        \hypo{}
                        \ellipsis{}{\csseq{\Sigma,\lef{B}}}
                        \infer2[IH]{\csseq{\Sigma}}
                    \end{prooftree}
                \end{tabular}
            \end{adjustbox}
            \end{center}
            \item $A=B\to C$. Assume $\Sigma=\Sigma',\iblock{\Phi_1\sand\Phi_2}$. Then 
            we have the following transformation of the derivation
            \begin{center}
            \vspace{-.5cm}
            \begin{adjustbox}{max width=.98\textwidth}
                \begin{tabular}{c}
                    \begin{prooftree}
                        \hypo{}
                        \ellipsis{}{\dowcsseq{\dow{\Sigma'},\iblock{\dow\Phi_1\sand\dow\Phi_2},\iblock{\lef B\sand \rig C}}}
                        \infer1[$r_1=\rig \to$]{\dowcsseq{\dow{\Sigma'},\iblock{\dow\Phi_1\sand\dow\Phi_2},\rig{B\to C}}}
                        \hypo{}
                        \ellipsis{}{\dowcsseq{\dow{\Sigma'},\iblock{\dow\Phi_1,\rig B\sand\dow\Phi_2}, \lef{B\to C}}}
                        \hypo{}
                        \ellipsis{}{\csseq{\Sigma,\iblock{\Phi_1\sand\Phi_2,\lef C}, \lef{B\to C}}}
                        \infer2[$r_2=\lef \to$]{\csseq{\Sigma',\iblock{\Phi_1\sand\Phi_2}, \lef{B\to C}}}
                        \infer2{\csseq{\Sigma',\iblock{\Phi_1\sand\Phi_2}}}
                    \end{prooftree}
                    \\
                    $\leadsto$ \\[.2cm]
                    \begin{prooftree}
                        \hypo{\dowcsseq{\dow{\Sigma'},\iblock{\dow\Phi_1\sand\dow\Phi_2},\rig{B\to C}}}
                        \hypo{}
                        \ellipsis{}{\dowcsseq{\dow{\Sigma'},\iblock{\dow\Phi_1,\rig B\sand\dow\Phi_2}, \lef{B\to C}}}
                        \infer2[IH]{\dowcsseq{\dow{\Sigma'},\iblock{\dow\Phi_1,\rig B\sand\dow\Phi_2}}}
                        \hypo{}
                        \ellipsis{}{\dowcsseq{\dow{\Sigma'},\iblock{\dow\Phi_1\sand\dow\Phi_2},\iblock{\lef B\sand \rig C}}}
                        \infer1[$\rulemerge$]{\dowcsseq{\dow{\Sigma'},\iblock{\dow\Phi_1,\lef B\sand\dow\Phi_2,\rig C}}}
                        \infer2[IH]{\dowcsseq{\dow{\Sigma'},\iblock{\dow\Phi_1\sand\dow\Phi_2,\rig C}}}
                        \hypo{\dowcsseq{\dow{\Sigma'},\iblock{\dow\Phi_1\sand\dow\Phi_2},\rig{B\to C}}}
                        \hypo{}
                        \ellipsis{}{\csseq{\Sigma,\iblock{\Phi_1\sand\Phi_2,\lef C}, \lef{B\to C}}}
                        \infer2[IH]{\csseq{\Sigma,\iblock{\Phi_1\sand\Phi_2,\lef C}}}
                        \infer2[IH]{\csseq{\Sigma,\iblock{\Phi_1\sand\Phi_2}}}
                    \end{prooftree}
                \end{tabular}
            \end{adjustbox}
            \end{center}
        \end{itemize}

        \noindent \textbf{Case 2.} $\rig A$ is side in $r_1$. 
        Let us discuss the possibilities of $r_1$. Note that in this case $r_1$ cannot be a $\circ$-rule. 
        \begin{itemize}[leftmargin=*]
            \item $r_1\in\{\lef \vee,\lef\wedge, \ruleliftib\}$. Then by hp-invertibility of the rule, we can simulate the same rule application and apply $\cut$ for the premise of $r_1$ and inverted conclusion of $r_2$. 
            For example, assume $\dow\Sigma=\dow{\Sigma'},\lef{B\wedge C}$ and $r_1$ is applied to $\lef {B\wedge C}$. Then we have the following transformation of the derivation 
            \begin{center}
            \vspace{-.5cm}
            \begin{adjustbox}{max width=.9\textwidth}
                \begin{tabular}{ccc}
                    \begin{prooftree}
                        \hypo{}
                        \ellipsis{}{\csseq{\dow{\Sigma'},\lef{B},\lef C,\rig {A}}}
                        \infer1[$r_1=\lef \wedge$]{\csseq{\dow{\Sigma'},\lef{B\wedge C},\rig {A}}}
                        \hypo{}
                        \ellipsis{}{\csseq{\Sigma',\lef{B\wedge C},\lef A}}

                      \infer2{\csseq{\Sigma',\lef{B\wedge C}}}
                    \end{prooftree}
                    &
                    $\leadsto$
                    &
                    \begin{prooftree}
                        \hypo{}
                        \ellipsis{}{\csseq{\dow{\Sigma'},\lef{B},\lef C,\rig {A}}}
                        \hypo{}
                        \ellipsis{}{\csseq{\Sigma',\lef{B\wedge C},\lef A}}
                        \infer1[hp-inver]{\csseq{\Sigma',\lef{B},\lef C,\lef A}}
                        \infer2{\csseq{\Sigma',\lef{B},\lef C}}
                        \infer1{\csseq{\Sigma',\lef{B\wedge C}}}
                    \end{prooftree}
                \end{tabular}
            \end{adjustbox}
            \end{center}
            Other cases when $r_1$ is applied in nested component contained in $\csseq{ \ }$ are similar. 
            \item $r_1=\lef \to$ is applied to some $\lef {B\to C}$. 
            We can transform the derivation by switching the order of $\cut$ and $r_1$. For example, assume w.l.o.g. 
            $\dowcsseq{\dow\Sigma,\rig A}=\dow{\Sigma_0}, \lef {B\to C}, \iblock{\dow\Phi_1,\sand\dow\Phi_2,\iblock{\dow\Psi\sand\dow\Sigma,\rig A}}$ (other cases are treated similarly),
            \begin{center}
            \vspace{-.5cm}
                \begin{adjustbox}{max width=.98\textwidth}
                \begin{tabular}{c}
                    \begin{prooftree}
                        \hypo{}
                        \ellipsis{}{\dow{\Sigma_0}, \lef {B\to C}, \iblock{\dow\Phi_1,\rig B,\sand\dow\Phi_2, \iblock{\dow\Psi\sand\dow\Sigma}}}
                        \hypo{}
                        \ellipsis{}{\dow{\Sigma_0}, \lef {B\to C}, \iblock{\dow\Phi_1\sand\dow\Phi_2, \lef C,\iblock{\dow\Psi\sand\dow\Sigma,\rig A}}}
                        \infer2{\dow{\Sigma_0}, \lef {B\to C}, \iblock{\dow\Phi_1\sand\dow\Phi_2,\iblock{\dow\Psi\sand\dow\Sigma,\rig A}}}
                        \hypo{}
                        \ellipsis{}{{\Sigma_0}, \lef {B\to C}, \iblock{\Phi_1\sand\Phi_2,\iblock{\Psi\sand\Sigma,\lef A}}}
                        \infer2{{\Sigma_0}, \lef {B\to C}, \iblock{\Phi_1\sand\Phi_2,\iblock{\Psi\sand\Sigma}}}
                    \end{prooftree}
                    \\~\\$\leadsto$\\
                    \begin{prooftree}
                        \hypo{}
                        \ellipsis{}{\dow{\Sigma_0}, \lef {B\to C}, \iblock{\dow\Phi_1,\rig B,\sand\dow\Phi_2, \iblock{\dow\Psi\sand\dow\Sigma}}}
                        \hypo{}
                        \ellipsis{}{\dow{\Sigma_0}, \lef {B\to C}, \iblock{\dow\Phi_1\sand\dow\Phi_2, \lef C,\iblock{\dow\Psi\sand\dow\Sigma,\rig A}}}
                        \hypo{}
                        \ellipsis{}{{\Sigma_0}, \lef {B\to C}, \iblock{\Phi_1\sand\Phi_2,\iblock{\Psi\sand\Sigma,\lef A}}}
                        \infer1[$\rulewl$]{{\Sigma_0}, \lef {B\to C}, \iblock{\Phi_1\sand\Phi_2,\iblock{\Psi\sand\Sigma,\lef A},\lef C}}
                        \infer2[IH]{{\Sigma_0}, \lef {B\to C}, \iblock{\Phi_1\sand\Phi_2,\iblock{\Psi\sand\Sigma},\lef C}}
                        \infer2[$r_2=\lef \to$]{{\Sigma_0}, \lef {B\to C}, \iblock{\Phi_1\sand\Phi_2,\iblock{\Psi\sand\Sigma}}}
                    \end{prooftree}  
                \end{tabular}
                \end{adjustbox}
            \end{center}
        \end{itemize}

        \noindent \textbf{Case 3.} $\lef A$ is side in $r_2$ and $\rig A$ is principal in $r_1$. 
        
        In this case, let us discuss different shapes of $A$. If $A=B\vee C$ or $B\wedge C$, then it follows directly by invertibility of $\lef A$ in the right premise of $\cut$. Thus we only need to consider the case when $A=B\to C$. We further discuss the possibilities of $r_2$. 
        \begin{itemize}[leftmargin=*]
            \item $r_2\in\{\lef \vee,\lef\wedge\}$ or $r_2=\ruleliftib$ but not applied to $\lef A$, dual to the first case in \textbf{Case 2}.  
            \item $r_2\in\{\rig \vee,\rig \wedge, \rig\to\}$. In these cases, we just permute the rule order of $r_2$ and $\cut$. 
            \item $r_2=\ruleliftib$ and applied to $\lef A$. In this case we can further assume $\csseq{\Sigma,\lef A}=\Sigma_0,\iblock{\lef A,\Phi_1\sand \Phi_2}$ where $\csseq{ \cdot } =\Sigma_0, \iblock{ \cdot \sand \Phi_2}$ and $\Sigma=\Phi_1$. Then we have the derivation $\+D_2$ is unfolded as 
            \begin{center}
                \vspace{-.5cm}    
                \begin{tabular}{c}
                    $
                    \vlderivation{
                    \vlin{}{r_2=\ruleliftib}{\Sigma_0,\iblock{\lef A,\Phi_1\sand \Phi_2}}
                    {
                        \vltr {\+D_2'} {
                            \Sigma_0,\iblock{\lef A,\Phi_1\sand \lef A,\Phi_2}
                        }{
                        \vlhy {\quad}}
                        {
                        \vlhy {}}
                        {
                        \vlhy {\quad}} 
                    }
                    }
                    $
                \end{tabular}
            \end{center}
            Let us further consider the rule above $r_2$, namely the last rule applied in $\+D_2'$, and call it by $r_3$. 
            If $r_3$ is a $\circ$-rule or another $\ruleliftib$ applied to a formula occurrence rather than $\lef A$, we only need to permute the order of $r_2$ and $r_3$. For example, assume $r_3=\rig \to$ and is applied to $\rig{D\to E}\in\Phi_2$. Let us write $\Phi_2=\dow\Phi_2,\rig{D\to E}$, then we transform the derivation as 
            \begin{center}
            \vspace{-.5cm}
            \begin{adjustbox}{max width=.98\textwidth}
                \begin{tabular}{ccc}
                    \begin{prooftree}
                        \hypo{}
                        \ellipsis{}{\dow\Sigma_0,\iblock{\dow\Phi_1,\rig A\sand\dow\Phi_2}}
                        \hypo{}
                        \ellipsis{}{\Sigma_0,\iblock{\lef A,\Phi_1\sand\lef A,\dow\Phi_2,\iblock{\lef D\sand\rig E}}}
                        \infer1[$r_3=\ruletor$]{\Sigma_0,\iblock{\lef A,\Phi_1\sand\lef A,\dow\Phi_2,\rig{D\to E}}}
                        \infer1[$r_2=\ruleliftib$]{\Sigma_0,\iblock{\lef A,\Phi_1\sand\dow\Phi_2,\rig{D\to E}}}                    \infer2{\Sigma_0,\iblock{\Phi_1\sand\dow\Phi_2,\rig{D\to E}}}
                    \end{prooftree}
                    &
                    $\leadsto$
                    &
                    \begin{prooftree}
                        \hypo{}
                        \ellipsis{}{\dow\Sigma_0,\iblock{\dow\Phi_1,\rig A\sand\dow\Phi_2}}
                        \infer1[$\rulewl$]{\dow\Sigma_0,\iblock{\dow\Phi_1,\rig A\sand\dow\Phi_2,\iblock{\lef D\sand-}}}
                        \hypo{}
                        \ellipsis{}{\Sigma_0,\iblock{\lef A,\Phi_1\sand\lef A,\dow\Phi_2,\iblock{\lef D\sand\rig E}}}
                        \infer1[$r_2=\ruleliftib$]{\Sigma_0,\iblock{\lef A,\Phi_1\sand\dow\Phi_2,\iblock{\lef D\sand\rig E}}}                    
                      \infer2[IH]{\Sigma_0,\iblock{\Phi_1\sand\dow\Phi_2,\iblock{\lef D\sand\rig E}}}
                      \infer1[$r_3=\ruletor$]{\Sigma_0,\iblock{\Phi_1\sand\dow\Phi_2,\rig{D\to E}}}
                    \end{prooftree}
                \end{tabular}
            \end{adjustbox}
            \end{center}
            If $r_3=\ruleliftib$ and is applied to $\lef A$, then we have the following transformation relying on the hp-admissibility of contraction (cf. Proposition \ref{prop:hp-admi-structural}):
            \begin{center}
            \vspace{-.5cm}
            \begin{adjustbox}{max width=.98\textwidth}
                \begin{tabular}{ccc}
                    \begin{prooftree}
                        \hypo{}
                        \ellipsis{}{\dow\Sigma_0,\iblock{\dow\Phi_1,\rig A\sand\dow\Phi_2}}
                        \hypo{}
                        \ellipsis{}{\Sigma_0,\iblock{\lef A,\Phi_1\sand\lef A,\lef A,\Phi_2}}
                        \infer1[$r_3=\ruleliftib$]{\Sigma_0,\iblock{\lef A,\Phi_1\sand\lef A,\Phi_2}}
                        \infer1[$r_2=\ruleliftib$]{\Sigma_0,\iblock{\lef A,\Phi_1\sand\Phi_2}}                    
                      \infer2{\Sigma_0,\iblock{\Phi_1\sand\Phi_2}}
                    \end{prooftree}
                    &
                    $\leadsto$
                    &
                    \begin{prooftree}
                        \hypo{}
                        \ellipsis{}{\dow\Sigma_0,\iblock{\dow\Phi_1,\rig A\sand\dow\Phi_2}}
                        \hypo{}
                        \ellipsis{}{\Sigma_0,\iblock{\lef A,\Phi_1\sand\lef A,\lef A,\Phi_2}}
                        \infer1[$\rulecl$]{\Sigma_0,\iblock{\lef A,\Phi_1\sand\lef A,\Phi_2}}
                        \infer1[$r_2=\ruleliftib$]{\Sigma_0,\iblock{\lef A,\Phi_1\sand\Phi_2}}                    
                      \infer2[IH]{\Sigma_0,\iblock{\Phi_1\sand\Phi_2}}
                    \end{prooftree}
                \end{tabular}
            \end{adjustbox}
            \end{center}
            Lastly, the only tricky case that remains is when $r_3=\lef \to$ and $\lef A$ in the same block of $\Phi_2$ is principal. 
            Let $\Phi_2=\Phi_2', \iblock{\Psi_1\sand\Psi_2}$, then we transform the derivation as 
            \end{itemize}
        \begin{center}
        \vspace{-.5cm}
        \begin{adjustbox}{max width=\textwidth}
        \begin{prooftree}
        \hypo{}
        \ellipsis{}{\dow\Sigma_0,\iblock{\dow\Phi_1,\iblock{\lef B\sand\rig C}\sand\dow{\Phi_2'},\iblock{\dow\Psi_1\sand\dow\Psi_2}}}
        \infer1[$r_1$]{\dow\Sigma_0,\iblock{\dow\Phi_1,\rig{B\to C}\sand\dow{\Phi_2'},\iblock{\dow\Psi_1\sand\dow\Psi_2}}}
        \hypo{}
        \ellipsis{}{\dow\Sigma_0,\iblock{\lef{B\to C},\dow\Phi_1\sand \lef{B\to C},\dow{\Phi_2'},\iblock{\dow\Psi_1,\rig B\sand\dow\Psi_2}}}
        \hypo{}
        \ellipsis{}{\Sigma_0,\iblock{\lef{B\to C},\Phi_1\sand \lef{B\to C},{\Phi_2'},\iblock{\Psi_1\sand\Psi_2,\lef C}}}
        \infer2[$r_3=\lef \to$]{\Sigma_0,\iblock{\lef{B\to C},\Phi_1\sand \lef{B\to C},\Phi_2',\iblock{\Psi_1\sand\Psi_2}}}
        \infer1[$r_2=\ruleliftib$]{\Sigma_0,\iblock{\lef{B\to C},\Phi_1\sand{\Phi_2'},\iblock{\Psi_1\sand\Psi_2}}}
        \infer2{\Sigma_0,\iblock{\Phi_1\sand\dow{\Phi_2'},\iblock{\Psi_1\sand\Psi_2}}}
        \end{prooftree}
        \end{adjustbox}
        
        ~\\
        $\leadsto$

        \begin{adjustbox}{max width=1.05\textwidth}
        \begin{prooftree}
        \hypo{\dow{\Sigma_0} ,\iblock{\dow\Phi_1,\rig{B\to C}\sand\dow{\Phi_2'}, \iblock{\Psi_1\sand\Psi_2}}}
        \hypo{}
        \ellipsis{}{\dow\Sigma_0,\iblock{\lef{B\to C},\dow\Phi_1\sand \lef{B\to C},\dow{\Phi_2'},\iblock{\dow\Psi_1, \rig B \sand \dow\Psi_2}}}
        \infer1[$\ruleliftib$]{\dow\Sigma_0,\iblock{\lef{B\to C},\dow\Phi_1\sand \dow{\Phi_2'},\iblock{\dow\Psi_1, \rig B \sand \dow\Psi_2}}}
        \infer2[IH]{\dow\Sigma_0,\iblock{\dow\Phi_1\sand \dow{\Phi_2'},\iblock{\dow\Psi_1, \rig B \sand \dow\Psi_2}}}
     
        \hypo{\dow{\Sigma_0},\iblock{\dow\Phi_1,\iblock{\lef B\sand \rig C}\sand\dow{\Phi_2'}, \iblock{\dow\Psi_1\sand\dow\Psi_2}}}
        \infer1[$\sf R2$]{\dow{\Sigma_0} ,\iblock{\dow\Phi_1\sand\dow{\Phi_2'}, \iblock{\lef B\sand \rig C}, \iblock{\dow\Psi_1\sand\dow\Psi_2}}}
        \infer1[$\rulemerge$]{\dow{\Sigma_0} ,\iblock{\dow\Phi_1\sand\dow{\Phi_2'},\iblock{\dow\Psi_1,\lef B\sand\dow\Psi_2,\rig C}}}
        \infer2[IH]{\dow{\Sigma_0} ,\iblock{\dow\Phi_1\sand\dow{\Phi_2'},\iblock{\dow\Psi_1 \sand \dow\Psi_2,\rig C}}}
     
       \hypo{}
       \ellipsis{$\+D'$}{\Sigma_0,\iblock{\Phi_1\sand \Phi_2',\iblock{\Psi_1,  \sand \Psi_2, \lef C}}}
        \infer2[IH]{{\Sigma_0} ,\iblock{\Phi_1\sand\Phi_2', \iblock{\Psi_1\sand \Psi_2}}}
        \end{prooftree}
        \end{adjustbox}
        \end{center} 

        And the conclusion of $\+D'$ is obtained by the following derivation:

        \begin{center}
            \begin{prooftree}
             \hypo{\dow{\Sigma_0} ,\iblock{\dow\Phi_1,\rig{B\to C}\sand\dow{\Phi_2'}, \iblock{\dow\Psi_1\sand\dow\Psi_2}}}
        \hypo{}
        \ellipsis{}{\Sigma_0,\iblock{\lef{B\to C},\Phi_1\sand \lef{B\to C},\Phi_2',\iblock{\Psi_1,  \sand \Psi_2, \lef C}}}
        \infer1[$\ruleliftib$]{\Sigma_0,\iblock{\lef{B\to C},\Phi_1\sand\Phi_2',\iblock{\Psi_1,  \sand \Psi_2, \lef C}}}
     
        \infer2[IH]{\Sigma_0,\iblock{\Phi_1\sand \Phi_2',\iblock{\Psi_1,  \sand \Psi_2, \lef C}}}
            \end{prooftree}
        \end{center}
        This completes the proof. 
\end{proof}

As a direct consequence of cut-free completeness, we can syntactically prove the verification lemma (cf. Lemma \ref{lem:verification}). 
Let us consider the following rules:
$$
\begin{aligned}
\begin{prooftree}
  \hypo{\iblock{\lef A,\rig B\sand -}}
  \infer[double]1[$\sf V1$]{\rig{A\to B}}
\end{prooftree}
&\quad &
\begin{prooftree}
  \hypo{\iblock{-\sand \lef A,\rig B}}
  \infer[double]1[$\sf V2$]{\rig{A\to B}}
\end{prooftree}
\end{aligned}
$$
where double line in an inference rule means the premise and conclusion are interchangeable. 

Notice that the two rules $\sf V1, V2$ can be regarded the syntactic version of the verification lemma. This is due to the following fact, which is easily verified using the frame condition (F1).  

\begin{proposition}\label{prop:seq-equiv}
    Let $\+M=(\+O,K,R,V)$ be a model. The following statements are equivalent: 
    \begin{enumerate}
    \item[(i).] $\iblock{\lef A,\rig B\sand -}$ is valid in $\+M$;
    \item[(ii).] $\iblock{-\sand \lef A,\rig B}$ is valid in $\+M$;
    \item[(iii).] for every $a\in K$, $a\Vdash A$ implies $a\Vdash B$.
    \end{enumerate}
\end{proposition}

\begin{proof}
    The equivalence of the three are shown by proving $(iii)$ is equivalent to both $(i)$ and $(ii)$. We only show the former case as the latter is similar. 
    \begin{itemize}
        \item $(i)\Rightarrow(iii)$: Take arbitrary $a\in K$ and assume $a\Vdash A$. By (F1), there exists $x\in\+O$ s.t. $Rxaa$. Since $x\Vdash \iblock{\lef A,\rig B\sand-}$, we have $a\Vdash \lef A,\rig B$ and hence $a\Vdash B$. 
        \item $(iii)\Rightarrow(i)$: Assume $\iblock{\lef A,\rig B\sand -}$ is not valid in $\+M$, then there exists $x\in\+O$ s.t. $x\nVdash \iblock{\lef A,\rig B\sand -}$, meaning that there are $y,z\in K$ s.t. $Rxyz$ and $y\Vdash A$ and $y\nVdash B$, a contradiction. 
    \end{itemize}
    This completes the proof. 
\end{proof}

From this proposition, by Lemma \ref{lem:verification}, we have the following immediately: 

\begin{corollary}
    The two rules $\sf V1,V2$ are valid. 
\end{corollary}

The next proposition shows that the two rules above are admissible in $\cbplus$. 

\begin{proposition}
    The two rules $\sf V1,V2$ are admissible in $\cbplus$. 
\end{proposition}

\begin{proof}
    We only show the admissibility of $\sf V1$ as the proof for $\sf V2$ is similar. First, assume $\rig{A\to B}$ is derivable in $\cbplus$. Since $\cbplus$ is cut-free and $\rig{A\to B}$ itself cannot be initial, we have $\iblock{\lef A\sand \rig B}$ derivable as well. Then, by $\sf R1$, we derive $\iblock{\lef A,\rig B\sand-}$. 
    Conversely, assume $\iblock{\lef A,\rig B\sand-}$ is derivable, we have the following short derivation of $\rig {A\to B}$:
    \begin{center}
    \begin{prooftree}
        \hypo{\iblock{\lef A,\rig B\sand -}}
        \hypo{\iblock{\lef A,\lef B\sand \lef B, \rig B}}
        \infer1[$\ruleliftib$]{\iblock{\lef A,\lef B\sand \rig B}}
        \infer2[$\cut$]{\iblock{\lef A\sand \rig B}}
        \infer1[$\rig \to$]{\rig{A\to B}}
    \end{prooftree}
    \end{center}
    Since $\cut$ is admissible in $\cbplus$, we conclude $\rig{A\to B}$ has a $\cut$-free derivation. 
\end{proof}

\section{Tracking relevance via nested sequents}\label{sec:relevance-property}

In this section, we establish several meta-logical properties regarding relevance by means of the cut-free calculus obtained in the previous section. We provide detailed proofs of variable-sharing property and depth relevance property and describe how these proofs can be easily adapted to their polarized versions. 

The \emph{variable-sharing property} is regarded as the fundamental feature for relevant logic~\cite{Standefer2025vsp}, and it can be strengthened to the 
\emph{strong variable sharing property} and the \emph{depth relevance property}~\cite{Brady:1984-depthrl} which put further requirements on the polarity and location of the shared variable. 
However, all these properties have only complicated semantic proofs to date~(cf. e.g, \cite{Brady:1984-depthrl}) and typically rely on specially designed valuations or matrices. 
Although there are several existing calculi for the logic, such as Hilbert calculus, display calculus and several labeled sequent calculi, none of them is capable to capture these essential properties syntactically. 
The reason is somewhat intuitive. On one hand, it is always hard to track variables in a derivation in a Hilbert system due to its non-analytic appearance; on the other hand, display and labeled calculi are too general to provide a precise answer to what is really necessary for a proof. 

In contrast, our proof is internal to the cut-free nested calculus. Thanks to the tree structure and the single output syntax of $\cbplus$, it is natural to characterize the relationship between different nodes and to track the flow of variable occurrences along a backward derivation. 
Variable-sharing property follows almost straightforward from a basic feature of the tree structure and depth relevance follows from a simple invariant combining nesting depth with implication depth. And the proof is a modular extension of the former. 
The key notion used in the proofs is called \emph{reachability}, which extracts sub-branches in a nested tree. By means of this notion, we provide the first syntactic proofs for all these fundamental relevance properties, which provides strong evidence that $\cbplus$ is a suitable calculus for the logic, as it well captures the essence of relevance. 

\subsection{Variable-sharing property}

\smallskip

\begin{definition}[Variable-sharing property]
    A logic $\mathsf{L}$ is said to have the \emph{variable-sharing property} (VSP for short) if whenever $A\to B$ is a theorem in $\mathsf{L}$, $\var{A}\cap\var{B}\neq\varnothing$. 
\end{definition}

We show that a generalized form of variable-sharing property is preserved by backward rule applications of $\cbplus$. Before that, we first prove some basic facts on provable equivalence. 

\begin{proposition}\label{prop:lift-eq}
    Let $\Phi$ be a multi-set of $\bullet$-formulas and $\Psi$ a nested sequent. Then $\iblock{\Phi\sand\Psi}$ is provable if and only if $\iblock{-\sand\Phi,\Psi}$ is provable. 
\end{proposition}

\begin{proof}
    The $(\Leftarrow)$ direction follows from applications weakening and $\ruleliftib$; for the converse, it is easily proved by induction on the height of a derivation of $\iblock{\Phi\sand\Psi}$. Assume the last rule applied in a derivation of $\iblock{\Phi\sand\Psi}$ is $r$. 
    We only show the following two non-trivial sub-cases:
    \begin{itemize}
        \item $r=\ruleliftib$. Assume $\Phi=\Phi',\lef A$ and $\iblock{\Phi',\lef A\sand\Psi}$ is derived from $\iblock{\Phi',\lef A\sand\lef A,\Psi}$. By IH, we have $\iblock{-\sand\Phi',\lef A,\lef A,\Psi}$ is derivable. By contraction, we obtain $\iblock{-\sand\Phi',\lef A,\Psi}$ as required. 
        \item $r=\lef\to$. Note that $\Phi$ is block-free, so $\lef\to$ only involves formulas and blocks in $\Psi$. We only need to apply IH first and then apply $\lef\to$ to the same formulas. 
    \end{itemize}
    This completes the proof. 
\end{proof}

We call a nested sequent in the form $\iblock{-\sand \Phi}$ \emph{base-free} or \emph{0-free} sequent. It is easy to see that $\ruleliftib$ is not applicable to such a sequent, and other rules can only move formulas within $\Phi$, therefore, the following holds trivially: 

\begin{fact}
    For any derivation in $\cbplus$, if it is rooted by a 0-free sequent then the derivation only contains 0-free sequents. 
\end{fact}

The following proposition allows to transform a implication formula into an equivalent 0-free sequent. 

\begin{proposition}\label{prop:imp-eq}
    The following are equivalent in $\cbplus$:
    \begin{enumerate}
        \item[(i).] $A\to B$ is provable;
        \item[(ii).] $\iblock{\lef A\sand\rig B}$ is provable;
        \item[(iii).] $\iblock{-\sand\lef A,\rig B}$ is provable.
    \end{enumerate}
\end{proposition}

\begin{proof}
    $(i)\Rightarrow (ii)$ is due to the cut-free feature of $\cbplus$; the converse follows by applying $\rig\to$. 
    The equivalence of $(ii)$ and $(iii)$ is a corollary of Proposition \ref{prop:lift-eq}. 
\end{proof}

From now on, in the rest of the section, we will mainly consider 0-free nested sequents and their proofs. 

Next, we define the notion of reachability, which describes relations between nodes occurring in the tree structure of a nested sequent and is the key notion for proving the relevance properties. First, we need some relations to  describe the nesting structure between blocks. 
We adopt the same notation from \cite{CSL-2024,gao2026bi-ik}. 

\begin{definition}[$\in^+_0, \in^+$-relation]
    Let $\Sigma$ be a nested sequent. For any (full or left-sided) nested sequents $\Phi$ and $\Psi$, We say 
    $\Psi_i$ ($i\in\{1,2\}$) directly belongs to $\Phi$ (in $\Sigma$), written as $\Psi_i\in^+_0 \Phi$, if  
    $$
    \begin{aligned}
      \Sigma=\csseq{\Phi} \quad \text{and} \quad \Phi=\Phi',\iblock{\Psi_1\sand\Psi_2} \quad \text{for~some~}\csseq{ \ }
    \end{aligned}
    $$
    The relation $\in^+$ is defined as the transitive and reflexive closure of $\in^+_0$. 
\end{definition}

According to the definition, for a given sequent $\Sigma$. it holds that $\Phi\in^+\Sigma$ for any nested component $\Phi$ in it. For example, let $\Sigma=\lef {p\vee q},\iblock{\lef p, \iblock{\lef q\sand\rig p}\sand \lef t}$ and denote 
$$
\begin{aligned}
    \Phi_1=\lef p, \iblock{\lef q\sand\rig p} & \quad &
    \Phi_2=\lef t & \quad &
    \Phi_3=\lef p & \quad &
    \Phi_4=\rig p
\end{aligned}
$$
we have $\Phi_i\in^+_0\Phi_1\in^+_0\Sigma$ for $i\in\{3,4\}$ and $\Phi_2\in^+\Sigma$ but not $\Phi_1\in^+_0 \Phi_2$. \\

Recall the $\sim$-operator in Definition~\ref{def:nested-operators}, we define reachability as follows. 

\begin{definition}[Reachable nodes]\label{def:reachability}
    Let $\Sigma$ be a nested sequent and $\Phi,\Psi\in^+\Sigma$. We say $\remove{\Phi}$ is \emph{direct reachable} to $\remove{\Psi}$ (or $[\remove{\Phi},\remove{\Psi}]$ forms a direct reachable pair) 
    if $\Phi\in^+_0\Psi$ or $\Psi\in^+_0\Phi$; and $\remove{\Phi}$ is \emph{reachable} to $\remove{\Psi}$ 
    if $\Phi\in^+\Psi$ or $\Psi\in^+\Phi$. 
\end{definition}

According to the definition, reachability is reflexive and generated by nested blocks; however, the two arguments in one single binary block are not reachable. Intuitively, if $\remove{\Phi}$ is \emph{reachable} to $\remove{\Psi}$, it implies there is a sub-branch in the tree of $\Sigma$ connecting $\remove{\Phi}$ and $\remove{\Psi}$. 

It is worth noticing the distinction between a node of a nested tree and a component of a nested sequent in the two definitions above. The relations $\in^+$ and $\in^+_0$ are defined between nested components, that are the whole content in a component of a relevant block, hence might contain nested blocks. However, nodes in the tree of a nested sequent are flat, obtained by forgetting the further nested structure with the $\remove{ \ }$-operator. 
In the previous example, we have $\remove{\Phi_1}=\lef p$ and $\remove{\Phi_i}=\Phi_i$ for $i\in\{2,3,4\}$. The nodes $\remove{\Phi_3}$ and $\remove{\Phi_1}$ are reachable to each other but $\Phi_1$ and $\remove{\Phi_2}$ as two components in the same block are not reachable. 
Moreover, every node is reachable to the root node, which in the example is $\remove{\Sigma}=\lef{p\vee q}$. 

For convenience, we shall call the node that contains the unique output formula the \emph{output node} and others the \emph{input nodes}. 

Now we can show a generalized form of VSP based on 0-free sequents. 

\begin{lemma}\label{lem:seq-vsp}
  Let $\Gamma=\iblock{-\sand \Sigma}$ be a 0-free nested sequent with the output formula $\theta$. 
  If it is provable, 
  then there exist $\Phi,\Psi\in^+\Sigma$ such that 
  \begin{itemize}
    \item $\rig \theta\in \remove{\Phi}$;
    \item $\remove{\Phi}$ is reachable to $\remove{\Psi}$;
    \item there exists $\lef \rho\in \remove{\Psi}$ such that $\var{\theta} \cap \var{\rho}\neq\varnothing$. 
  \end{itemize}
\end{lemma}

\begin{proof}
    By induction on the height $h$ of the derivation of $\Gamma$. 
    First, if $h=0$, meaning that $\Gamma$ is an axiom, so $\Gamma$ is of the form $\csseq{\Lambda,\lef p,\rig p}$ for some context $\csseq{ \ }$. 
    Since reachability is reflexive, we simply take $\lef p$ as the required $\lef\rho$, $\rig p$ as the required $\rig \theta$, and the node $\remove{\Lambda},\lef p,\rig p$ as both $\remove{\Phi}$ and $\remove{\Psi}$, 
    and the condition on variable holds trivially. 

    Next, if $h>0$, let us consider the last rule $r$ applied in the derivation and verify each rule of $\cbplus$ in turn. 
    First $\ruleliftib$is vacuous in 0-free sequents; 
    the cases regarding $\lef \vee, \lef \wedge,\rig\vee,\rig\wedge$ are straightforward, 
    as the location of the output formula and the situation of reachability to the output node is not affected by backward rule application. 
    We only show the following two cases.  
    \begin{itemize}
        \item $r=\lef \to$. Assume $\Gamma=\csseq{\Sigma,\lef{C\to D},\iblock{\Phi_1\sand\Phi_2}}$ and is derived from the two premises 
        $\Gamma_1=\dowcsseq{\dow\Sigma,\lef{C\to D},\iblock{\dow\Phi_1,\rig C\sand\dow\Phi_2}}$ 
        and  
        $\Gamma_2=\csseq{\Sigma,\lef{C\to D},\iblock{\Phi_1\sand\Phi_2,\lef D}}$. 
        Note that for $\Gamma_2$, the reachability of the output node (call it $\Theta$) and the output formula $\rig \theta$ are the same as the conclusion $\Gamma$. This implies if the required $\lef \rho$ is not the displayed $\lef D$ produced by $\lef \to$, 
        then we only need to apply IH and take the same $\rho$ for the conclusion. 
        Therefore, it suffices to check the remaining case when the required $\lef \rho$ for $\Gamma_2$ is $\lef D$. According to the conditions in the lemma, this means $\Theta$ is reachable to $\remove{\Phi_2},\lef D$. Since the unique ancestor of $\remove{\Phi_2},\lef D$ is $\remove{\Sigma},\lef{C\to D}$, by the tree structure, $\Theta$ is also reachable to $\remove{\Sigma},\lef{C\to D}$. Since the output node in the conclusion is also $\Theta$, we can take $\lef{C\to D}$ as $\rho$ for the conclusion. By IH, $\var{\theta}\cap\var{D}\neq\varnothing$, so $\var{\theta}\cap\var{C\to D}\neq\varnothing$ as well. 
        \item $r=\rig \to$. Assume $\Gamma=\csseq{\Sigma,\rig{C\to D}}$ and is derived from $\Gamma'=\csseq{\Sigma,\iblock{\lef C\sand\rig D}}$. 
        By IH, $\rig D$ is $\rig \theta$ in the statement of the lemma;
        by Definition \ref{def:reachability}, the component containing $\lef C$ is not reachable to the one containing $\rig D$, so the required $\lef \rho$ occurs either in $\Sigma$ or somewhere in the context $G\{\ \}$ and $\var{\rho}\cap \var{D}\neq \varnothing$. 
        It follows directly $\var{\rho}\cap \var{C\to D}\neq \varnothing$ as well. 
        For the reachability condition, notice that the same reachability is preserved for the conclusion, as both (1) the output node of the conclusion $\remove{\Sigma},\rig{C\to D}$ is reachable to itself and (2) any other node in the context $G\{\ \}$ reachable to the output node $\rig D$ in the premise is also reachable to $\Sigma$ hold trivially. 
    \end{itemize}
    This completes the proof. 
\end{proof}

\begin{theorem}[VSP for $\bplus$]\label{thm:vsp}
    If $A\to B$ is a theorem in $\bplus$, then $\var{A}\cap\var{B}\neq\varnothing$. 
\end{theorem}

\begin{proof}
    Since $A\to B$ is a theorem in $\bplus$ and $\cbplus$ is complete, it follows $\rig {A\to B}$ is provable in $\cbplus$. 
    By Proposition \ref{prop:imp-eq}, $\iblock{-\sand\lef A,\rig B}$ is provable as well. 
    Note that the only reachable node for the output node is itself, and $\lef A$ is the only $\bullet$-formula therein, 
    by Lemma \ref{lem:seq-vsp}, we have  
    $\var{A}\cap\var{B}\neq\varnothing$. 
\end{proof}

\begin{rem}
    Since $\bplus$ is constant-free, if it has Craig interpolation, then VSP follows as well, but not vice versa. 
\end{rem}

\subsection{Depth relevance property}

To distinguish a specific formula occurrence and a formula, we use $\hat{F}$ to denote an occurrence of $F$. 
As a formula might have several formula occurrences, we further denote them by $\hat{F_1},\hat{F_2}, \ldots$ and so on. Here the subscript is only used to specify a certain occurrence (serving like a name of the occurrence) but has no meaning with respect to an order. We shall also omit subscripts when the occurrence is clear from the context. 
Moreover, we call the two displayed occurrences of $\hat{B}@\hat{C}$ by \emph{direct sub-occurrences} of $\widehat{B@C}$ for $@\in\{\to,\vee,\wedge\}$. 

We define the notion of depth of a formula occurrence in a given formula as follows. 

\begin{definition}[Depth of formula occurrence of a given formula]
    Let $F$ be a formula and $\hat{A}$ occurs in $F$, we define the depth of $\hat{A}$ in $F$, denoted $\dep{F}{\hat{A}}$, as follows:
    
    \begin{itemize}
        \item $\dep{F}{\hat{F}}:=0$
        \item 
        $\dep{F}{\hat{B}}:=\dep{F}{\hat{C}}=n$, 
        if they are direct sub-occurrences of $\widehat{{B@ C}}$ and $\dep{F}{\widehat{{B@ C}}}=n$ where $@\in\{\vee,\wedge\}$
        \item $\dep{F}{\hat{B}}=\dep{F}{\hat{C}}=n+1$ if they are direct sub-occurrences of $\widehat{{B\to C}}$ and $\dep{F}{\widehat{{B\to C}}}=n$
    \end{itemize}

    For two formula occurrences $\hat{A},\hat{G}$ in $F$ such that $\hat{A}$ occurs in $\hat{G}$, we write $\dep{G}{\hat{A}}$ for the depth of $\hat{A}$ in $\hat{G}$ regarding $G$ as an independent formula. 
\end{definition}

Intuitively, the depth of $\hat{A}$ in $F$ means the length of the branch from $\hat{A}$ to $F$ in the decomposing tree of $F$, where $\vee,\wedge$-decompositions are local, child nodes are only generated by $\to$.  

By definition, for the two occurrences $\hat{A},\hat{G}$ in $F$, the depth of $\hat{A}$ in $\hat{G}$ is only determined by $\hat{A}$ and $\hat{G}$ themselves, regardless of the precise location of $\hat{G}$, i.e. of which formula $F$ it occurs as a sub-occurrence. 

\begin{example}
    Let $F=p\to (p \to ((s\wedge t)\vee q))$ and $G=p \to ((s\wedge t)\vee q)$. By definition $\dep{F}{\hat{t}}=2, \dep{G}{\hat{t}}=1$ and $\dep{F}{\hat{G}}=1$.
\end{example}

\begin{proposition}\label{prop:depth-sub}
    Let $F$ be a formula and $\hat{A},\hat{G}$ be two formula occurrences in $F$. If $\hat{A}$ occurs in $\hat{G}$, then $\dep{F}{\hat{A}}=\dep{F}{\hat{G}}+\dep{G}{\hat{A}}$. 
\end{proposition}

\begin{proof}
    By induction on the depth of $\hat{G}$ in $F$. 
\end{proof}

We can naturally extend this notion to a sequent, distinguishing either the formula is explicit in some node of the sequent, or occurs as a subformula. To achieve this, we first define the notion of depth of nested components, which measures the depth of nesting of a nested block in a given sequent. Then we generically define the depth of formula occurrence in a sequent regardless of it is polarized with $\bullet$ or $\circ$. 

\begin{definition}[Depth in a given sequent]\label{def:depth-nested-nodes}
    Let $\Sigma$ be a sequent. For any $\Phi\in^+\Sigma$, we define $\dep{\Sigma}{\Phi}$ as follows:
    \begin{itemize}
        \item $\dep{\Sigma}{\Sigma}=0$
        \item $\dep{\Sigma}{\Phi}=n+1$ if $\Phi\in^+_0\Psi$ and $\dep{\Sigma}{\Psi}=n$
    \end{itemize}

    For any $A\in\remove{\Phi}$ for some $\Phi\in^+\Sigma$, we define $\dep{\Sigma}{A}:=\dep{\Sigma}{\Phi}$. 
    Furthermore, for any formula occurrences $\hat{B}$ in $A$, we define its depth in $\Sigma$ with respect to $A$ as $\indep{\Sigma}{A}{\hat{B}}:=\dep{\Sigma}{A}+\dep{A}{\hat{B}}$.
\end{definition}

Regarding $\hat{A}$ as the maximal formula occurrence in $A$, we observe that $\indep{\Sigma}{A}{\hat{A}}=\dep{\Sigma}{A}+\dep{A}{\hat{A}}=\dep{\Sigma}{A}$. 
It is also worth noticing that, in a sequent, when a formula explicitly occurs in a multiset, the notions of formula and formula occurrences coincide, but the latter is meaningful for those which are not independently displayed as formulas. 

The following is the depth relevance property for 0-free sequents, which is a stronger version of Lemma~\ref{lem:seq-vsp} by taking depth into consideration. The proof is structured in the same way and we only need to verify the equality of depth. In the case of $\lef\to$ and $\rig \to$, we observe that the change in depth at a given node is exactly the opposite of the change in depth in the formula; therefore, the depth of the shared variable as the witness remains unchanged.

\begin{lemma}\label{lem:seq-drp}
    Let $\iblock{-\sand \Sigma}$ be a 0-free sequent with the output formula $\theta$. If it is provable, then there exist $\Phi,\Psi\in^+\Sigma$ such that 
    \begin{enumerate}
    \item[(a).] $\rig \theta\in \remove{\Phi}$;
    \item[(b).] $\remove{\Psi}$ is reachable to $\remove{\Phi}$;
    \item[(c).] there exists $\lef \chi\in \remove{\Psi}$ and a variable $p$ such that $p\in \var{\theta} \cap \var{\chi}$ and there are two occurrences $\hat{p}_1,\hat{p}_2$ of $p$ such that $\indep{\Sigma}{\theta}{\hat{p}_1}=\indep{\Sigma}{\chi}{\hat{p}_2}$. 
    \end{enumerate}
\end{lemma}

\begin{proof}
    By induction on the height $h$ of the derivation of $\Gamma$. 
    First, if $h=0$, meaning that $\Gamma$ is an axiom, so $\Gamma$ is of the form $\csseq{\Lambda,\lef p,\rig p}$ for some context $\csseq{ \ }$. 
    We simply take $p$ as the required $\theta$ and $\Lambda,\lef p,\rig p$ as both $\Phi$ and $\Psi$ as required. 
    Then all the three conditions hold trivially. 

    Next, if $h>0$, let us consider the last rule $r$ applied in the derivation and verify each rule of $\cbplus$ in turn. 
    We only show the following three cases.  
    \begin{itemize}
        \item $r=\rig\vee$. Assume $\Gamma=\csseq{\Lambda, \rig{A\vee B}}$ is obtained from $\Gamma'=\csseq{\Lambda, \rig{A}}$. It is easy to see that all the depth information of variables in $\Gamma'$ is precisely preserved in $\Gamma$. 
        By IH, there exist $\Psi\in^+\Gamma'$ such that $\remove{\Psi}$ is reachable to $\remove{\Lambda},\rig{A}$; and $\lef\chi\in \remove{\Psi}$ with $p$ in it satisfies (c) in the lemma. 
        Since from $\Gamma'$ to $\Gamma$, only the output formula is changed, so we can take the same $\lef \chi$ in $\Gamma$ and verify condition (b) and (c) for the updated $\rig \theta=\rig{A\vee B}$. The reachability is the same, hence (b) holds. Since $p\in\var{A}\cap\var{\chi}$, it follows $p\in\var{A\vee B}\cap\var{\chi}$ as well. 
        According to (c) for $\Gamma'$, there are two occurrences $\hat{p}_1$ and $\hat{p}_2$ of $p$ such that $\indep{\Gamma'}{A}{\hat{p}_1}=\indep{\Gamma'}{\chi}{\hat{p}_2}$. 
        By definition, $\indep{\Gamma'}{A}{\hat{p}_1}=\dep{\Gamma'}{\Lambda,\rig A}+\dep{A}{\hat{p}_1}$ and 
        $\indep{\Gamma'}{\chi}{\hat{p}_2}=\dep{\Gamma'}{\Psi}+\dep{\chi}{\hat{p}_2}$. 
        Also, $\indep{\Gamma}{A\vee B}{\hat{p}_1}=\dep{\Gamma}{\Lambda,\rig {A\vee B}}+\dep{A\vee B}{\hat{p}_1}$ 
        and $\indep{\Gamma}{\chi}{\hat{p}_2}=\dep{\Gamma}{\Psi}+\dep{\chi}{\hat{p}_2}$. 
        By Proposition~\ref{prop:depth-sub}, we have $\dep{A\vee B}{\hat{p}_1}=\dep{A\vee B}{\hat{A}}+\dep{A}{\hat{p_1}}=\dep{A}{\hat{p_1}}$. 
        Since $\dep{\Gamma}{\Lambda,\rig {A\vee B}}=\dep{\Gamma'}{\Lambda,\rig {A}}$ and $\dep{\Gamma}{\Psi}=\dep{\Gamma'}{\Psi}$, 
        we conclude $\indep{\Gamma}{A\vee B}{\hat{p}_1}=\indep{\Gamma}{\chi}{\hat{p}_2}$. 

        \item $r=\lef \to$. Assume $\Gamma=\csseq{\Sigma,\lef{C\to D},\iblock{\Phi_1\sand\Phi_2}}$ and is derived from the two premises 
        $\Gamma_1=\dowcsseq{\dow\Sigma,\lef{C\to D},\iblock{\dow\Phi_1,\rig C\sand\dow\Phi_2}}$ 
        and  
        $\Gamma_2=\csseq{\Sigma,\lef{C\to D},\iblock{\Phi_1\sand\Phi_2,\lef D}}$. 
        Note that for $\Gamma_2$, the reachability of the output node $\Phi$ and the output formula $\rig \theta$ are the same as the conclusion $\Gamma$. 
        We distinguish two sub-cases. 

        (i) Assume the required $\lef \chi$ for $\Gamma_2$ is not the displayed $\lef D$ produced by $\lef \to$.  Then we only need to apply IH and take the same $\chi$ for the conclusion at the same depth. The proof proceeds similarly to the previous case of $\rig \vee$. 
        
        (ii) Otherwise, $\lef \chi$ for $\Gamma_2$ is $\lef D$. 
        For the same reason in the corresponding case of Lemma~\ref{lem:seq-vsp}, 
        we can take $\lef{C\to D}$ as $\chi$ for the conclusion $\Gamma$. 
        By IH, there is $\hat{p}_1$ in $\lef D$ in $\Gamma_2$ satisfying the condition. We take the corresponding occurrence $\hat{p}_1'$ of $\hat{p}$ in $\hat{D}$ in $\lef{C\to D}$. 
        It suffices to show $\indep{\Gamma}{C\to D}{\hat{p}_1'}=\indep{\Gamma_2}{D}{\hat{p}_1}$. 
        By definition, $\indep{\Gamma}{C\to D}{\hat{p}_1}=\dep{\Gamma}{\Sigma,\lef{C\to D},\iblock{\Phi_1\sand\Phi_2}}+\dep{C\to D}{\hat{p}_1'}$ and $\indep{\Gamma_2}{D}{\hat{p}_1}=\dep{\Gamma_2}{\Phi_2,\lef D}+\dep{D}{\hat{p}_1}$. We further observe $\dep{\Gamma_2}{\Phi_2,\lef D}=\dep{\Gamma}{\Sigma,\lef{C\to D},\iblock{\Phi_1\sand\Phi_2}}+1$ and $\dep{C\to D}{\hat{p}_1'}=\dep{C\to D}{D}+\dep{D}{\hat{p}_1}=\dep{D}{\hat{p}_1}+1$, thus $\indep{\Gamma}{C\to D}{\hat{p}_1'}=\indep{\Gamma_2}{D}{\hat{p_1}}$.  
        
        \item $r=\rig \to$. Assume $\Gamma=\csseq{\Sigma,\rig{C\to D}}$ and is derived from $\Gamma'=\csseq{\Sigma,\iblock{\lef C\sand\rig D}}$. 
        Dual to (ii) in the case of $\lef \to$, by IH, there is $\hat{p}_2$ in $\rig D$ in $\Gamma'$ satisfying the condition. We take the corresponding occurrence $\hat{p}_2'$ of $\hat{p}_2$ in $\hat{D}$ in $\rig{C\to D}$.  
        It suffices to show $\indep{\Gamma}{C\to D}{\hat{p}_2'}=\indep{\Gamma'}{D}{\hat{p}_2}$. By definition, $\indep{\Gamma}{C\to D}{\hat{p}_2'}=\dep{\Gamma}{\Sigma,\rig{C\to D}}+\dep{C\to D}{\hat{p}_2'}$ and $\indep{\Gamma'}{D}{\hat{p}_2}=\dep{\Gamma'}{\Sigma,\iblock{\lef C\sand\rig D}}+\dep{D}{\hat{p}_2}$. It is easy to see $\dep{\Gamma}{\Sigma,\rig{C\to D}}=\dep{\Gamma'}{\Sigma,\rig{D}}-1$ and $\dep{C\to D}{\hat{p}_2'}=\dep{D}{\hat{p}_2}+1$, 
        so the equality holds as required. 
    \end{itemize}
    Then we are done. 
\end{proof}

Now in contrast to the previous semantic proof~\cite{Brady:1984-depthrl}, we can show the DRP for $\bplus$ syntactically. 

\begin{theorem}[DRP for $\bplus$]
    If $A\to B$ is a theorem in $\bplus$, then there exists a variable $p$ such that $p\in \var{A}\cap\var{B}$ and there are two occurrences $p_1,p_2$ of $p$ in $A$ and $B$ respectively with $\dep{A}{\hat{p}_1}=\dep{B}{\hat{p}_2}$. 
\end{theorem}

\begin{proof}
    Due to the same reason of the first lines of the proof of Theorem \ref{thm:vsp}, we can apply Lemma \ref{lem:seq-drp}. 
    Let $\Sigma_0=\iblock{-\sand\lef A,\rig B}$. 
    Then there exists a variable $p$ such that $p\in \var{A} \cap \var{B}$ and 
    and there are two occurrences $\hat{p}_1,\hat{p}_2$ of $p$ 
    $\indep{\Sigma_0}{A}{\hat{p}_1}=\indep{\Sigma_0}{B}{\hat{p}_2}$. 
    Let us denote the nested component $\lef A,\rig B$ by $\Phi$. By definition, $\indep{\Sigma_0}{A}{\hat{p}_1}=\dep{\Sigma_0}{\Phi}+\dep{A}{\hat{p}_1}$ and $\indep{\Sigma_0}{B}{\hat{p}_2}=\dep{\Sigma_0}{\Phi}+\dep{B}{\hat{p}_2}$, thus $\dep{A}{\hat{p}_1}=\dep{B}{\hat{p}_2}$. 
\end{proof}

\subsection{Remarks on polarized relevance property and the general methodology}

Following the proof of Lemma \ref{lem:seq-vsp}, we can further derive the so-called \emph{strong variable sharing property} (SVSP)
and \emph{strong depth relevance property} (SDRP)
almost for free. 
DVSP further requires the obtained shared variable of VSP are of opposite polarities and SDRP is defined based on DRP accordingly. 

To show the SVSP for $\bplus$, we only need to assign polarities $\bullet,\circ$ to each formula occurrence within a polarized formula $\rig F$ or $\lef F$, 
and require opposite polarities for the shared variable. According to the syntex of sequents, formulas are already polarized, it suffices to define the polarities of inner occurrences within a polarized formula. 

\newcommand\pol[2]{\mathsf{pol}_{#1}(#2)}

\begin{definition}[Polarity of formula occurrence of a given formula]
    Let $F^\dagger$ be a formula where $\dagger\in\{\circ,\bullet\}$ and $\hat{A}$ occurs in $F$. We define the polarity of $\hat{A}$ in $F$, denoted $\pol{F}{\hat{A}}$, as follows:
    
    \begin{itemize}
        \item $\pol{F}{\hat{F}}=\dagger$
        \item 
        $\pol{F}{\hat{B}}=\pol{F}{\hat{C}}=\pol{F}{\widehat{B@ C}}$, 
        if they are direct sub-occurrences of $\widehat{{B@ C}}$ where $@\in\{\vee,\wedge\}$
        \item $\pol{F}{\hat{B}}=\bullet$ and $\pol{F}{\hat{C}}=\circ$
        if they are direct sub-occurrences of $\widehat{{B\to C}}$ and $\pol{F}{\widehat{{B\to C}}}=\circ$
        \item $\pol{F}{\hat{B}}=\circ$ and $\pol{F}{\hat{C}}=\bullet$
        if they are direct sub-occurrences of $\widehat{{B\to C}}$ and $\pol{F}{\widehat{{B\to C}}}=\bullet$
    \end{itemize}
\end{definition}

For a given sequent $\Sigma$, the polarity of formula occurrence $\hat{A}$ is defined according to whether it is an explicit formula. If $A$ directly occurs in some component of $\Sigma$, namely $A\in\remove{\Phi}$ for some $\Phi\in^+\Sigma$, then $\hat{A}$ already has the polarity assigned by the sequent rules; otherwise, we take the definition above where $\hat{A}$ occurs in $F$ and $F$ directly occurs in some component of $\Sigma$. 

Following a similar proof of VSP or DRP, we can show $\bplus$ has the SVSP. The proof can be easily extended to cover SDRP as well. 

\begin{theorem}[SVSP and SDRP for $\bplus$]
    For $\mathrm{SVSP}$, if $A\to B$ is a theorem in $\bplus$, then there exists a variable $p$ such that $p\in \var{A}\cap\var{B}$ and there are two occurrences $p_1,p_2$ of $p$ in $A$ and $B$ respectively of opposite polarities, namely $\pol{A}{\hat{p}_1}\neq \pol{B}{\hat{p}_2}$. 
    For $\mathrm{SDRP}$, it further requires $\dep{A}{\hat{p}_1}=\dep{B}{\hat{p}_2}$. 
\end{theorem}

We end this section by a brief discussion on the difference between the existing semantic proofs and the syntactic proof presented here. 

\begin{rem}
    Existing semantic proofs of these relevance properties are all based on contraposition, which means, e.g., to show a relevant logic has the VSP, they have to show that if the formula does not have shared variables, it is not valid. This requires a tricky construction of a countermodel and could not provide explicit evidence of shared variables for provable formulas. 
    Moreover, once the condition is changed, the countermodel must be reconstructed according to the new condition, so the proof for DRP is not an extension of that for VSP, but an entirely new proof. 
    
    However, in our syntactic proof, the nested structure allows us to track the flow of variable occurrences directly by means of the rules, as a consequence, the proof is based on a derivation of a valid formula and is constructive. And the resulting proofs are modular with respect to different properties in the following pattern: 
    $$
    \begin{aligned}
    \text{Reachability}& \quad +  & \text{polarity}&\quad +&\text{depth}&\quad  \Rightarrow\quad &\text{proof of SDRP}\\
    \text{Reachability}&\quad +& & &\text{depth} &\quad\Rightarrow\quad& \text{proof of DRP}\\
    \text{Reachability}&\quad +&\text{polarity}& & &\quad \Rightarrow\quad&\text{proof of SVSP}\\
    \text{Reachability} & & & & &\quad \Rightarrow \quad& \text{proof of VSP}\\
    \end{aligned}
    $$
    This means, e.g., with a proof $\mathsf{P1}$ of a stronger property say DRP, we can directly obtain another proof $\mathsf{P2}$ of VSP by removing depth as the extra parameter in the invariance measure occurring in $\mathsf{P1}$. 
\end{rem}
\section{Interpolation property for $\bplus$: a syntactic proof without arbitrary partition
}\label{sec:cip}

Interpolation is a rare property for relevant logics; it is shown in the literature that for many relevant logics stronger than $\logicb$, the logics as well as their positive fragment, lack this property~\cite{urquhart1993failure,fussner2025interpolation}. 
However, when it turns to the base logic $\logicb$ and its positive fragment $\bplus$, the question whether they have the \cip~remains open~\cite{Knudstorp2026,standefer2026relevant}. 
In this section, we prove the Craig interpolation property (\cip) for the logic $\bplus$ by means of the cut-free calculus $\cbplus$. 
This result shows that $\bplus$ is not only a significant fragment of the base logic $\logicb$, but also an important independent system in the family of relevant logic of its own value. 

Recall that in the previous section, in order to track the flow of variable occurrences, we have already restricted our attention from arbitrary nested sequents to only 0-free nested sequents. Now, to establish interpolation, we need an additional machinery called \emph{focused sequents} to trace the flow of the unique output formula taken from the root along a derivation. As we will see later in this section, the key lemmas towards interpolation will only take the class of focused 0-free sequents into consideration. The correctness of this restriction will be ensured by a completeness result of normalization. Moreover, when constructing the interpolant inductively along a derivation, we do not need to preserve the whole nested sequent but only take into consideration the \emph{innermost} block. This helps us to avoid the usual complex semantic construction used in multi-component calculi in order to construct interpolants (cf. e.g., \cite{kuznets2018multicomponent}). Split in this sense is uniquely determined by each focused sequent and is adequate for the inductive argument. 

\subsection{Focused sequents and existence of focused proofs}

In this subsection we introduce focused sequents and show the class of focused sequents is adequate for provability. Before that, we show some basic facts regarding general forms of proofs. 

\begin{proposition}\label{prop:0-free-gen}
    If a 0-free sequent $\iblock{-\sand \Phi}$ is provable at height $n$ and $\Phi$ is focused, then $\iblock{-\sand\csseq{\Phi}}$ is also provable at height $n$ for any context $\csseq{ \ }$. 
\end{proposition}

\begin{proof} Let $G\{\cdot\}$ be an arbitrary context. We define the context $H\{\cdot\}$ as $\iblock{-\sand G\{\cdot\}} $. Assume that $\iblock{-\sand \Phi}$ is provable. 
By ${\sf R1}$ with the context $H\{\cdot\}$, we have
$$ \begin{prooftree}
\hypo{\iblock{-\sand \Phi}} 
\infer1[{\sf R1}]{H\{\Phi\}} 
\end{prooftree} $$
Hence $\dsseq{\Phi}=\iblock{-\sand G\{ \Phi\}}$ is provable. 
By Proposition~\ref{prop:hp-admi-r1-r2}, $\mathsf{R1}$ is hp-admissible, so the proof of $\dsseq{\Phi}$ has the same height as $\iblock{-\sand\Phi}$. 
\end{proof}

As a special case, we have $\dsseq{\Phi,\rig F}$ is provable with the same height as the proof of $\iblock{-\sand\Phi,\rig F}$. 

We also need the following, which can be easily verified by induction on the height of $\+D$.  

\begin{proposition}\label{prop:non-empty-component}
    Let $\Sigma_0=\iblock{-\sand\lef A,\rig B}$ and $\+D$ be a derivation of it. Then for any sequent $\Sigma$ in $\+D$ and any $\Omega=\Phi,\iblock{\Psi_1\sand\Psi_2}\in^+\Sigma$, if $\remove{\Phi}$ is contains input formula, so does $\Psi_1$. 
\end{proposition}

The next proposition shows if the first component of a relevant block is full and the corresponding second component is block-free, then we can keep only the full component and forget all the rest without losing provability. 

\begin{proposition}\label{prop:component-remove}
    Let $\Sigma=\csseq{\Phi,\iblock{\Psi_1\sand\Psi_2}}$ be a 0-free sequent where $\Psi_1$ is full and $\Psi_2$ is block-free. 
    If $\Sigma$ is provable of height $\leq n$, then 
    $\Sigma_0=\iblock{-\sand \Psi_1}$ 
    is provable of height $\leq n$ as well. 
\end{proposition}

\begin{proof}
We proceed by induction on the height $h$ of the derivation of the 0-free sequent $\Sigma$.\\
\textit{Base Case}: $h=0$. 
$\Sigma=\csseq{\Phi, \iblock{\Psi_1 \sand \Psi_2}}$ is an initial sequent. 
Since $\Psi_1$ is full, it contains the unique output formula $\rig{p}$, and the matching input formula $\lef{p}$ is located in $\Psi_1$ as well, 
This means $\Psi_1 = \dsseq{\Psi_1', \lef{p}, \rig{p}}$ for some context $H$, which is an initial sequent. Hence $\Sigma_0 = \iblock{-\sand \Psi_1}$ is also an initial sequent.

\noindent \textit{Inductive Step}: $h>0$.
Let $r$ be the last rule applied in the derivation of $\Sigma$.
  
  \begin{itemize}
    \item If $r = \ruletor$, then the formula $\rig{A \to B}$ 
    occurs within 
    $\Psi_1$, as $\Psi_1$ contains the unique output formula (whereas $G\{\cdot\}, \Phi$ and $\Psi_2$ are purely input). The application of $\ruletor$ introduces a new block $\iblock{\lef A \sand \rig B}$ inside $\Psi_1$, leaving $\Psi_2$ completely unchanged and block-free. By the induction hypothesis applied to the premise, we obtain that $\iblock{-\sand \Psi_1,\iblock{\lef A \sand \rig B}}$ is provable. Re-applying $\ruletor$ directly inside $\Psi_1$ yields the target sequent $\Sigma_0 = \iblock{-\sand \Psi_1, \rig{A\to B}}$.

    \item If $r = \ruletol$ and $r$ is applied to $\lef{A\to B}$, we have the following cases: 
    \begin{itemize}[leftmargin=*]
      \item $\lef{A\to B}$ occurs in $\Phi$. We distinguish two sub-cases:
      \begin{itemize}
   \item[i.] Assume $\Phi=\Phi',\lef{A\to B}$. 
      Consider the left and right premises, namely, $\Sigma_{\textsf{L}} = \csseq{\dow{\Phi'},\lef{A\to B}, \iblock{\rig{A},\dow\Psi_1 \sand \dow\Psi_2}}$ and $\Sigma_{\textsf{R}} = \csseq{\Phi',\lef{A\to B}, \iblock{\Psi_1 \sand \Psi_2, \lef B}}$. 
      By the induction hypothesis applied to $\Sigma_{\textsf{R}}$, we directly obtain that $\Sigma'_{\textsf{R}} = \iblock{-\sand \Psi_1}$, which is our required sequent.
      \item[ii.] $\Phi=\Phi',\lef{A\to B},\iblock{\Omega_1\sand\Omega_2}$. Suppose $\ruletol$ is applied to $\lef{A\to B}$ together with the block $\iblock{\Omega_1\sand\Omega_2}$ rather than $\iblock{\Psi_1\sand\Psi_2}$. The left premise is $\Sigma_{\textsf{L}} = \dow{G}\{\dow{\Phi'}, \iblock{\dow{\Omega}_1, \rig A \sand \dow{\Omega}_2}, \iblock{\dow{\Psi}_1 \sand \dow{\Psi}_2}\}$ and the right premise is $\Sigma_{\textsf{R}} = \csseq{\Phi',\lef{A\to B}, \iblock{\Omega_1 \sand \Omega_2, \lef B}, \iblock{\Psi_1 \sand \Psi_2}}$. In the latter premise, the component $\Psi_1$ is full and $\Psi_2$ is block-free. By the induction hypothesis applied to $\Sigma_{\textsf{R}}$, we obtain that $\Sigma'_{\textsf{R}} = \iblock{-\sand \Psi_1}$ is provable, being our desired conclusion.
          \end{itemize}
      \item $\lef{A\to B}$ appears somewhere in $G$. Similar as above. 
    \item $\lef{A\to B}$ occurs in $\Psi_1$.
        Assume w.l.o.g. $\Psi_1=\Psi_1',\lef{A\to B},\iblock{\Omega_1\sand \Omega_2}$ and $r$ involves $\iblock{\Omega_1\sand \Omega_2}$. 
        First, in the left premise $\Sigma_{\textsf{L}} = \dow{G}\{\dow{\Phi}, \iblock{\dow{\Psi_1'}, \iblock{\dow{\Omega}_1, \rig A \sand \dow{\Omega}_2} \sand \dow{\Psi}_2}\}$, the component $\dow{\Psi_1'}, \iblock{\dow{\Omega}_1, \rig A \sand \dow{\Omega}_2}$ remains full and $\dow{\Psi}_2$ remains block-free. Applying the induction hypothesis yields $\Sigma_{\textsf{L}0} = \iblock{-\sand \dow{\Psi_1'}, \iblock{\dow{\Omega}_1, \rig A \sand \dow{\Omega}_2}}$. 
        Second, in the right premise $\Sigma_{\textsf{R}} = \csseq{\Phi, \iblock{\Psi_1', \iblock{\Omega_1 \sand \Omega_2, \lef B} \sand \Psi_2}}$, the component $\Psi_1', \iblock{\Omega_1 \sand \Omega_2, \lef B}$ is full and $\Psi_2$ is unchanged and block-free. Applying the induction hypothesis yields $\Sigma_{\textsf{R}0} = \iblock{-\sand \Psi_1', \iblock{\Omega_1 \sand \Omega_2, \lef B}}$. 
        Finally, re-applying $\ruletol$ to $\Sigma_{\textsf{L}0}$ and $\Sigma_{\textsf{R}0}$ yields $\Sigma_0 = \iblock{-\sand \Psi_1', \lef{A\to B}, \iblock{\Omega_1\sand \Omega_2}} = \iblock{-\sand \Psi_1}$.

    \item $\lef{A\to B}$ occurs in $\Psi_2$. Since $\Psi_2$ is block-free, we have $\Psi_2=\Psi_2',\lef{A\to B}$. This case is impossible since the application of $\ruletol$ requires a block $\iblock{\Omega_1 \sand \Omega_2}$ associated with $\lef{A \to B}$ to propagate the subformulas of $A\to B$ into the block's components. 
    \end{itemize}

  \end{itemize}

  All other rules behave locally and their cases follow straightforwardly. 
\end{proof}

\begin{rem}
    Note that this proposition does not have a dual version on the second component of a relevant block. This is due to the reason that the two components in the same relevant block are not symmetric, which further implies the asymmetric feature of the second and third arguments in the ternary Routley-Meyer relations. 
\end{rem}

\begin{definition}[Focused sequents]
    A sequent $\Sigma$ is called \emph{focused} if for every nested block $\iblock{\Phi_1\sand\Phi_2}$ in it, either $\Phi_1$ or $\Phi_2$ is full. 
    We define $\focus{\Sigma}$ as the \emph{focused part} of $\Sigma$, which refers to the resulting sequent by removing every input block in any nested component in $\Sigma$. 
    We say a proof is \emph{focused} if it contains focused sequents only. 
\end{definition}

\begin{example}
    Let $\Sigma_1=\lef p,\iblock{\rig q\sand\lef p,\iblock{\lef q\sand\lef q}}$ and $\Sigma_2=\lef p,\iblock{\lef q\sand\lef p,\iblock{\lef q\sand\rig q}}$. By definition, $\Sigma_1$ is not a focused sequent but $\Sigma_2$ is, and we further have the focused part of $\Sigma_1$ is $\focus{\Sigma_1}=\lef p,\iblock{\rig q\sand\lef p}$. 
\end{example}

By definition, focused sequents are sequents without purely input blocks and the output formula always occurs in the deepest block. 
Recall Definition \ref{def:depth-nested-nodes}, we denote the \emph{depth} of a sequent as the maximal depth of a nested component occurring in it.
We can enumerate components of a focused sequent $\Gamma$ with depth $n$ as 
$$
\Gamma_{11}, \Gamma_{12}, \ldots, \Gamma_{n1}, \Gamma_{n2},
$$
where $\Gamma_{h1},\Gamma_{h2}$ are the two components of depth $h$, so that $\Gamma_{11}=\varnothing$ and 
$\Gamma=\iblock{-\sand\Gamma_{12}}$, and, for every $2\leq h<n$, 
$\exists! i\in\{1,2\}$ s.t. $\iblock{\Gamma_{(h+1)1}\sand\Gamma_{(h+1)2}}\in \Gamma_{hi}$.

\begin{proposition}\label{prop:focused-part}
    If a 0-free sequent is provable with a proof of height $\leq n$, so is its focused part. 
\end{proposition}

\begin{proof}
    We prove this by induction on the height $h$ of a derivation $\+D$ of $\Sigma$. If $h=0$, $\Sigma$ is an initial sequent, 
    which can be written as $\csseq{\Lambda,\lef p,\rig p}$ for some context and $\Lambda$. By definition, $\focus{\csseq{\Lambda,\lef p,\rig p}}=\dsseq{\Lambda', \lef p,\rig p}$ for some other context $\dsseq{ \ }$ and $\Lambda'$, but is still initial. Otherwise, $h>0$, let us consider the last rule $r$ applied in $\+D$. 
    \begin{itemize}
        \item $r$ is a $\circ$-rule. Notice that $\circ$-rules will neither introduce purely input blocks, nor transform existing output block into input block, 
        so we only need to apply IH to the premise(s) and apply the same rule again. 
        \item $r\in\{\lef \vee,\lef\wedge\}$. We only show the case of $\lef \vee$ and the other is similar. Assume the application is of the form 
        \begin{center}
        \begin{prooftree}
            \hypo{\csseq{\Sigma,\lef A}}
            \hypo{\csseq{\Sigma,\lef B}}
            \infer2[$\ruleorl$]{\csseq{\Sigma,\lef{A\vee B}}}
        \end{prooftree}
        \end{center}
        Let $\Phi=\focus{\csseq{\Sigma,\lef{A\vee B}}}, \Phi_1=\focus{\csseq{\Sigma,\lef{A}}}$ and $\Phi_2=\focus{\csseq{\Sigma,\lef{B}}}$. 
        If the displayed $\lef{A\vee B}$ on the conclusion is removed by $\focus{ \ }$, so are both $\lef A$ and $\lef B$ in the respective premise. So $\Phi=\Phi_1=\Phi_2$. 
        By IH, $\Phi_1$ and $\Phi_2$ are provable, so is $\Phi$. 
        Otherwise, the displayed $\lef{A\vee B}$ on the conclusion is preserved in $\Phi$, so are both $\lef A$ and $\lef B$ in the respective premise. 
        Further assume $\Phi=\dsseq{\Sigma',\lef{A\vee B}}$, then $\Phi_1=\dsseq{\Sigma',\lef{A}}$ and $\Phi_2=\dsseq{\Sigma',\lef{B}}$. By IH and $\lef \vee$, we derive $\Phi$ again. 
        \item $r=\lef \to$. 
        Assume the application is of the form:
        \begin{center}
        \begin{prooftree}
        \hypo{\dowcsseq{\dow\Psi,\lef{A\to B},\iblock{\dow\Psi_1,\rig A\sand\dow\Psi_2}}}
        \hypo{\csseq{\Psi,\lef{A\to B},\iblock{\Psi_1\sand\Psi_2,\lef B}}}
        \infer2[$\ruletol$]{\csseq{\Psi,\lef{A\to B},\iblock{\Psi_1\sand\Psi_2}}}
        \end{prooftree}
        \end{center}
        Let us denote $\Phi=\focus{\csseq{\Phi',\lef{A\to B},\iblock{\Psi_1\sand\Psi_2}}}, \Phi_1=\focus{\csseq{\Phi',\lef{A\to B},\iblock{\dow\Psi_1,\rig A\sand\Psi_2}}}$ and $\Phi_2=\focus{\csseq{\Phi',\lef{A\to B},\iblock{\Psi_1\sand\Psi_2,\lef B}}}$. If the displayed $\lef B$ is removed by $\focus{\cdot}$ on $\Phi_2$, then we can easily see $\Phi=\Phi_2$ and we only need to apply IH to the right premise. 
        Otherwise, the displayed $\lef{B}$ on the right premise is preserved in $\Phi_2$, which means $\iblock{\Psi_1\sand\Psi_2,\lef B}$ is a full block. Hence all the displayed $\lef{A\to B},\rig A,\lef B$ are preserved in the respective $\Phi,\Phi_1,\Phi_2$. Similar to the case of $\lef\vee$, by IH and $\lef \vee$, with probably some weakening (since $\Phi_1$  might contain less blocks than $\Phi_2$), then we derive $\Phi$ again. 
    \end{itemize}
    This completes the proof. 
\end{proof}

We call a backward rule application \emph{focused-preserving} if whenever its conclusion is focused, so is any of its premises. Conversely, we call an application of $r$ with focused conclusion \emph{focused-breaking} if it is not focused-preserving. 

\begin{fact}\label{fact:focus-preserving}
Let $\+D$ be a derivation of $\iblock{-\sand\lef A,\rig B}$ and $\Gamma$ be the conclusion of a rule application $r$ which is a focused $0$-free sequent with 
output formula of depth $n$. Then the following hold:
\begin{itemize}
    \item If $r\neq\ruletol$, the application is focused-preserving. 
    \item If $r=\ruletol$ with principal formula $\lef{A\to B}\in\remove{\Gamma_{mi}}$ and it acts on the block $\iblock{\Gamma_{(m+1)1}\sand\Gamma_{(m+1)2}}$, then it is focused-preserving if and only if $n=m+1$. 
\end{itemize}
\end{fact}

\begin{proof}
Every rule other than $\ruletol$ is obviously focused-preserving as they will never introduce purely input blocks in any premise. 
As for $\ruletol$, by definition, the only block occurring in $\Gamma_{mi}$ is $\iblock{\Gamma_{(m+1)1}\sand\Gamma_{(m+1)2}}$, its right premise only adds $\lef B$ to
$\Gamma_{(m+1)2}$ and is focused, and its left premise deletes the output formula and adds $\rig A$ to $\Gamma_{(m+1)1}$, which makes the block of depth $m+1$ full and leaves no deeper block when $n=m+1$, whereas for $n\geq m+2$ it turns the blocks of depth $m+2,\ldots,n$ into purely input ones.
\end{proof}
 
Consequently, a focused-breaking application of $\ruletol$ is one such that the principal implication formula occurs at depth $m$ and the output formula occurs at depth $n$ where $n\geq m+2$.

Now we are going to prove the completeness of focused proofs. 

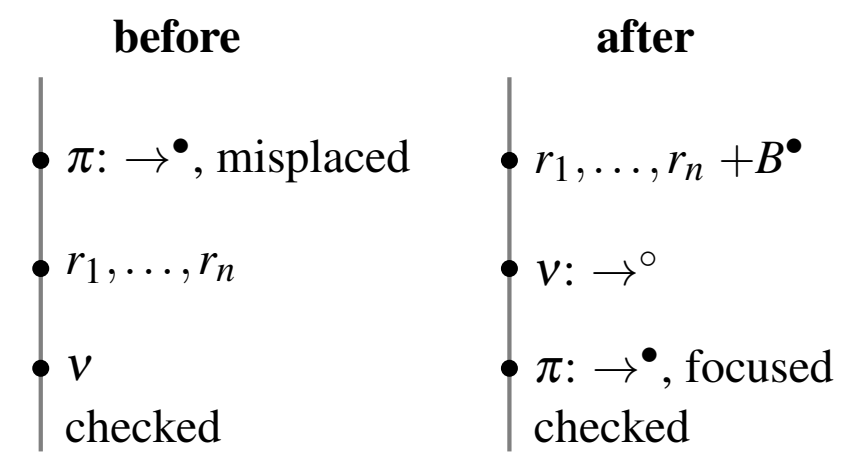
\begin{figure}[t]
    \centering
\begin{tikzpicture}[x=1cm,y=0.55cm,
  lab/.style={font=\footnotesize,anchor=west,inner sep=1pt}]
\node[font=\footnotesize\bfseries] at (0.9,5.0) {before};
\draw[thick,gray] (0,0)--(0,4.5);
\foreach \y in {1,2.2,3.5} \fill (0,\y) circle (1.6pt);
\node[lab] at (0.12,0.3) {checked};
\node[lab] at (0.12,1) {$\nu$};
\node[lab] at (0.12,2.2) {$r_1,\dots,r_n$};
\node[lab] at (0.12,3.5) {$\pi$: $\lef\to$, misplaced};
\begin{scope}[xshift=3.1cm]
\node[font=\footnotesize\bfseries] at (0.9,5.0) {after};
\draw[thick,gray] (0,0)--(0,4.5);
\foreach \y in {1,2.2,3.5} \fill (0,\y) circle (1.6pt);
\node[lab] at (0.12,0.3) {checked};
\node[lab] at (0.12,1) {$\pi$: $\lef\to$, focused};
\node[lab] at (0.12,2.2) {$\nu$: $\rig\to$};
\node[lab] at (0.12,3.5) {$r_1,\dots,r_n$ $+\lef{B}$};
\end{scope}
\end{tikzpicture}
\caption{Terminating transformation contained in the proof of Lemma~\ref{lem:focus-proof}}
\label{fig:termination-focused}
\end{figure}

\begin{lemma}[Existence of focused proof]\label{lem:focus-proof}
    If $\iblock{-\sand\lef A,\rig B}$ is provable in $\cbplus$, then it has a focused proof. 
\end{lemma}

\begin{proof}
Based on the notion of \emph{focused-preserving} and \emph{focused-breaking} rule applications in Fact~\ref{fact:focus-preserving}, we prove the lemma by transforming every focused-breaking rule application into a focused-preserving one. We check all the rules iteratively and the goal is to show the transformation procedure terminates.

Let us call a focused-breaking application in a given derivation \emph{lowest} if no focused-breaking application occurs below it, and let $\+U(\+D)$ be the set of the subderivations of $\+D$ rooted at the conclusions of its lowest focused-breaking applications. Define  
$$
\mu(\+D)=\{\{h(\+E):\+E\in\+U(\+D)\}\}
$$
as a finite multiset of natural numbers, ordered by the multiset extension of $<$ on $\mathbb{N}$, which is a well-founded order. 
The step below applies whenever
$\+U(\+D)\neq\varnothing$, it produces a derivation of the same sequent with the same property, it deletes one element $N_0$ from $\mu(\+D)$, adds only elements $<N_0$ and leaves the remaining ones
unchanged. 
Hence $\mu$ strictly decreases, the procedure terminates and it terminates with $\mu=\varnothing$, that is, with a derivation without focused-breaking applications.

Assume $\pi$ be the lowest focused-breaking application at the root of some $\+E\in\+U(\+D)$ and let $h(\+E)=N_0$. By Fact~\ref{fact:focus-preserving}, it is of the following form: 
$$
\begin{prooftree}
\hypo{\Theta_1=\dowcsseq{\dow{\Sigma},\lef{A\to B},\iblock{\dow\Phi_1,\rig A\sand\dow\Phi_2}}}
\hypo{\Theta_2=\csseq{\Sigma,\lef{A\to B},\iblock{\Phi_1\sand\Phi_2,\lef B}}}
\infer2[$\lef\to$]{\Theta_0=\csseq{\Sigma,\lef{A\to B},\iblock{\Phi_1\sand\Phi_2}}}
\end{prooftree}
$$
where $\Sigma,\lef{A\to B}$ is the component of depth $m$ and the output formula has depth $n\geq m+2$. 

We trace back to where the block of $\iblock{\Phi_1\sand\Phi_2}$ was introduced and denote it by $\nu$. 
Let $\Theta^{\ell}$ be its conclusion. We claim that the principal formula of $\nu$ is the output
formula $\rig{E\to F}$ of $\Theta^{\ell}$, of depth $m+1$, and that
$$
\Theta^{\ell}=\csseq{\Sigma',\lef{A\to B},\iblock{\Phi_1'\sand\Phi_2',\rig{E\to F}}}
\quad\text{or}\quad
\Theta^{\ell}=\csseq{\Sigma',\lef{A\to B},\iblock{\Phi_1',\rig{E\to F}\sand\Phi_2'}}.
$$ 


We display the case in which the output formula $\rig{E\to F}$ of $\Theta^{\ell}$ occurs in the second component of the block of depth $m+1$; the other case when $\rig{E\to F}$ occurs in the first component of that block is similar. 

Consider the applications lying between $\Theta^{\ell}$ and $\Theta_0$; they are all below $\pi$, hence focused-preserving. We further argue the depth of the principal formula of each rule as follows. 
Since each application of $\ruletol$ among them is focused-preserving, by Fact~\ref{fact:focus-preserving}. its
principal formula has depth $\geq m+1$ and, by the shape of the rule, it modifies only components of depth $\geq m+2$. 
The principal formula of the applications of $\ruletor,\ruleandr,\ruleorr$ among them is the output formula, of depth $\geq m+2$, with the single exception of $\nu$, whose principal formula $\rig{E\to F}$ has depth $m+1$. 
Further, we assume among $r_1,\ldots,r_n$, both $\lef \vee$ and $\lef\wedge$ are only allowed to be applied at depth $\geq m+2$. This is ensured by the height-preserving inversion (cf. Proposition~\ref{prop:invertibility}). 
This prevents possible $\lef \vee$ and $\lef\wedge$ rules applied within $\csseq{ \ }, \Sigma,\Phi_1$ and $\remove{\Phi_2}$, otherwise, we can simply switch the order of such rule and $\lef\to$ and the height goes down without further steps. 
Thus $\+D$ has the shape displayed below, where $r_1,\ldots,r_n$ are the applications lying between
$\Theta^{\ell}$ and $\Theta_0$, which by the claim modify components of depth $\geq m+2$ only, and
$\+D_0$ is the part of $\+D$ below $\Theta^{\ell}$. 
\vspace{-.4cm}

$$
\begin{prooftree}
\hypo{}
\ellipsis{$\+D_1$}{\Theta_1=\csseq{\Sigma,\lef{A\to B},\iblock{\Phi_1,\rig A\sand\dow\Phi_2}}}
\hypo{}
\ellipsis{$\+D_2$}{\Theta_2=\csseq{\Sigma,\lef{A\to B},\iblock{\Phi_1\sand\Phi_2,\lef B}}}
\infer2[$\pi:\ruletol$]{\Theta_0=\csseq{\Sigma,\lef{A\to B},\iblock{\Phi_1\sand\Phi_2}}}
\ellipsis{$\+D_3$, containing $r_1,\ldots,r_n$}{}
\infer1[]{\csseq{\Sigma,\lef{A\to B},\iblock{\Phi_1\sand\Phi_2',\iblock{\lef E\sand\rig F}}}}
\infer1[$\nu:\ruletor$]{\Theta^{\ell}=\csseq{\Sigma,\lef{A\to B},\iblock{\Phi_1\sand\Phi_2',\rig{E\to F}}}}
\ellipsis{$\+D_0$}{}
\infer[no rule]1{}
\end{prooftree}
$$

Notice the application of $\ruletol$ to $\lef{A\to B}$ at $\Theta^{\ell}$ is focused-preserving, and we replace the sub-derivation of $\+D$ rooted at $\Theta^{\ell}$, which we call $\+D^{\ell}$, by the following one, where $\nu$ is applied \emph{after} $\pi$ (the idea is illustrated in Figure~\ref{fig:termination-focused}):
\vspace{-.4cm}

$$
\begin{prooftree}
\hypo{}
\ellipsis{$\+D_1'$}{\Theta^{\ell}_1=\csseq{\Sigma,\lef{A\to B},\iblock{\Phi_1,\rig A\sand\Phi_2'}}}
\hypo{}
\ellipsis{$\+D_2$}{\Theta_2=\csseq{\Sigma,\lef{A\to B},\iblock{\Phi_1\sand\Phi_2,\lef B}}}
\ellipsis{$\+D_3'$, containing $r_1,\ldots,r_n$}{}
\infer1[]{\csseq{\Sigma,\lef{A\to B},\iblock{\Phi_1\sand\Phi_2',\lef B,\iblock{\lef E\sand\rig F}}}}
\infer1[$\nu:\ruletor$]{\Theta^{\ell}_2=\csseq{\Sigma,\lef{A\to B},\iblock{\Phi_1\sand\Phi_2',\lef B,\rig{E\to F}}}}
\infer2[$\pi:\ruletol$, normal]{\Theta^{\ell}=\csseq{\Sigma,\lef{A\to B},\iblock{\Phi_1\sand\Phi_2',\rig{E\to F}}}}
\ellipsis{$\+D_0$}{}
\infer[no rule]1{}
\end{prooftree}
$$

\noindent \textbf{The left premise.} For the left premise $\Theta_1$, first by Proposition~\ref{prop:focused-part}, the focused part of $\Theta_1$ is provable at $N_0-1$. Then by Proposition~\ref{prop:component-remove}, we have $\iblock{-\sand \Phi_1,\rig A}$ is also provable at height $N_0-1$. By Proposition~\ref{prop:0-free-gen}, we have a proof for $\Theta_1^\ell$ at the same height, denoted by $\+D_1'$. 

\noindent \textbf{The right premise.} $\Theta^{\ell}_2$ is obtained from $\Theta^{\ell}$ by adding $\lef B$ to the second component of its block of depth
$m+1$ and permuting the order of $\rig\to$ and other deeper rules. 
The derivation $\+D^{\ell}$ of $\Theta^{\ell}_2$ thus is a derivation of the same height in which each sequent is the corresponding sequent of $\+D^{\ell}$ with that extra $\lef B$ in that component; in particular the sequent occurring in it at the position of $\Theta_0$ is
$\csseq{\Sigma,\lef{A\to B},\iblock{\Phi_1\sand\Phi_2,\lef B}}$, which is $\Theta_2$. 

Now we verify that the measure goes down after the transformation. 
Let $\+D'_2$ be that derivation with the part above $\Theta_2$ replaced by $\+D_2$. 
Every application of $\+D'_2$ below $\+D_2$ is an application of $\+D^{\ell}$ with the same depth of
the principal formula and of the output component, hence is focused-preserving, and so is the new
application of $\ruletol$. Therefore, the focused-breaking applications of the new
derivation are those of $\+D'_1$, those of $\+D_2$ and those of $\+D^{\ell}$ lying neither above
$\Theta_0$ nor below it, that is, those in the side branches of the applications between
$\Theta^{\ell}$ and $\Theta_0$. The elements of $\+U$ rooted in the first two have height $\leq
N_0-1$, those rooted in the side branches are unchanged, for the same reason, the elements of
$\+U(\+D)$ outside $\+D^{\ell}$ are untouched, and the element $N_0$ contributed by $\+E$ is gone.
\end{proof}

\begin{definition}[Normal $\to$-application]
    A $\rig \to$-application of the following form is called \emph{normal}, 

    \begin{center}
    \begin{prooftree}
        \hypo{\csseq{\Sigma,\lef C, \iblock{\lef A\sand \rig B}}}
        \infer1[$\ruletor$]{\csseq{\Sigma,\lef C, \rig{A\to B}}}
    \end{prooftree}
    \end{center}
    
    A $\lef \to$-application of one of the following forms is called \emph{normal},

    \begin{center}
        \begin{prooftree}
        \hypo{\csseq{\Sigma,\lef{C\to D},\iblock{\Phi_1,\rig C\sand\Phi_2}}}
        \hypo{\csseq{\Sigma,\lef{C\to D},\iblock{\Phi_1\sand\Phi_2,\rig F,\lef D}}}
        \infer2[$\ruletol$]{\csseq{\Sigma,\lef{C\to D},\iblock{\Phi_1\sand\Phi_2,\rig F}}}
        \end{prooftree}
    \end{center}
    \begin{center}
        \begin{prooftree}
        \hypo{\csseq{\Sigma,\lef{C\to D},\iblock{\Phi_1,\rig C\sand\Phi_2}}}
        \hypo{\csseq{\Sigma,\lef{C\to D},\iblock{\Phi_1,\rig F\sand\Phi_2,\lef D}}}
        \infer2[$\ruletol$]{\csseq{\Sigma,\lef{C\to D},\iblock{\Phi_1,\rig F\sand\Phi_2}}}
        \end{prooftree}
    \end{center}
\end{definition}

According to the definition above, normal $\rig \to$ is applied when the output node contains input formulas; and normal $\lef\to$ says the rule can only be applied to the direct reachable node of the output node. 

\begin{lemma}\label{lem:normal-imp}
    Every $\to$-application contained in a focused proof of $\iblock{-\sand\lef A,\rig B}$ is normal. 
\end{lemma}

\begin{proof}
    Let us consider the first backward non-normal $\to$-application which has a focused sequent as the conclusion, 
    and show that such an application must produce either non-focused premise or further non-focused sequents in the branch. The case of $\lef \to$ is a corollary of Fact~\ref{fact:focus-preserving} and hence we only need to consider $\rig \to$. 
        Assume there is a non-normal $\rig \to$ in such a focused proof, so it is of the following form ($\rig{A\to B}$ occurs on the first component of a relevant block is similar): 
    \begin{center}
    \vspace{-.5cm}    
    \begin{tabular}{c}
        $
        \vlderivation{
        \vlin{}{\ruletor}{\csseq{\Sigma,\iblock{\Phi\sand\rig{A\to B}}}}
        {
        \vltr {\+D'} {
        \csseq{\Sigma,\iblock{\Phi\sand\iblock{\lef A\sand\rig B}}}
        }{
        \vlhy {\quad}}
        {
        \vlhy {}}
        {
        \vlhy {\quad}} 
        }
        }
        $
        \end{tabular}
    \end{center}
    Let us take an arbitrary branch $\+B$ in $\+D'$. It is easy to see that if there is no $\lef\to$ applied in $\+B$, then it cannot end up with an initial sequent, as no other rules can move the output formula nor add input formulas to the output node. 
    Notice that, starting from the existing $\iblock{\lef A\sand\rig B}$, new inner blocks can be produced but no $\lef\to$ can be applied therein, since $\ruleliftib$ is not applicable and no input formula can be moved to the same component of a block. 
    Also, the direct reachable node of $\rig B$ contains no input formula; thus if there is any $\lef \to$ applied in $\+B$, it can only be applied somewhere between $\csseq{ \ },\Sigma,\Phi$. However, this will introduce pure input blocks on one premise by removing the output formula, making the premise not focused, a contradiction.  
\end{proof}

\subsection{Craig interpolation via active unit}

The remainder of the subsection is devoted to a syntactic proof of the Craig interpolation property of $\bplus$ by means of the nested calculus $\cbplus$. With the proposition above, to find an interpolant for a theorem $\rig{A\to B}$, we can restrict our attention only to focused proofs of $\iblock{-\sand\lef A,\rig B}$. 

\begin{definition}[Craig interpolation property]
    A logic $\mathsf{L}$ is said to have the \emph{Craig interpolation property} (CIP for short) if whenever $A\to B$ is a theorem in $\mathsf{L}$, there exists a formula $C$ such that (i) both $A\to C$ and $C\to B$ are theorems in $\mathsf{L}$; and (ii) $\var{C}\subseteq \var{A}\cap\var{B}$. Such a formula $C$ is called an \emph{interpolant} of $A\to B$. 
\end{definition}

The following notions are needed in order to introduce the key lemma of interpolation property with respect to focused sequents. 

\begin{definition}\label{def:MO-MRO}
    Let $\Sigma$ be a focused sequent. 
    The \emph{active unit} $\act{\Sigma}$ is the minimal nested component along the output spine containing the minimal output block. 
    We further define the \emph{left and right active context} $\actl{\Sigma}{ \ }$ and $\actr{\Sigma}{\ }$ of the active unit in the 0-free form as well as the \emph{active variables} 
    accordingly: 
        \begin{center}
        \begin{adjustbox}{max width=\textwidth}
        $
        \begin{array}{c|c|c|c|c|c}
        & \act{\Sigma} & \actl{\Sigma}{ \ } & \actr{\Sigma}{ \ } & \vactl{\Sigma} & \vactr{\Sigma}\\
        \hline
        (a) & \Psi,\iblock{\Phi_1,\rig F\sand\Phi_2} & \iblock{-\sand\Phi_1,\{ \ \}} & \iblock{-\sand\rig F, \{ \ \}} & \var{\Phi_1} & \var{F}
        \\
        \hline
        (b1) & \Psi,\iblock{\Phi_1\sand\Phi_2,\rig F} &  
        \iblock{-\sand\Psi,\iblock{\Phi_1\sand\Phi_2,\{ \ \}}} & 
        \iblock{-\sand\rig F, \{ \ \}} & \var{\Phi_2,\Psi} & \var{F}
        \\
        \hline
        (b2) & \Psi,\iblock{\Phi_1\sand\rig F} & \iblock{-\sand\Psi, \{ \ \}} & \iblock{-\sand\{ \ \}, \iblock{\Phi_1\sand\rig F}} & \var{\Psi} & \var{F,\Phi_1}
        \end{array}
        $
        \end{adjustbox}
        \end{center}
        where in (b1) $\Phi_2$ is non-empty. 
\end{definition}

According to the definition of focused sequents, it is easy to see that all the $\Psi,\Phi_1,\Phi_2$ displayed in the table above are block-free, namely multi-sets of input formulas.  


Now, with all the ingredients prepared already, we formalize the interpolation lemma for sequents as below. 

\begin{lemma}\label{lem:interpolation-calculus}
    Let $\+D$ be a focused proof of $\iblock{-\sand\lef A,\rig B}$ and let $\Sigma$ occur in $\+D$. 
    Then there exist a formula $\theta$ 
    such that all the following hold:
    \begin{itemize}
        \item[(C1)] 
        both $\actr{\Sigma}{\lef \theta}$ and $\actl{\Sigma}{\rig \theta}$ are provable;
        \item[(C2)] 
        $\var{\theta}\subseteq\vactl{\Sigma}\cap\vactr{\Sigma}$;
        \item[(C3)] if $\act{\Sigma}$ is of the form (b1), then for any $\alpha$ such that $\iblock{-\sand\Phi_1,\rig{\alpha}}$ is provable, there exists $\beta$ such that $\iblock{-\sand\Phi_1,\rig{\beta}}$, $\iblock{-\sand\rig \alpha, \lef \beta}$ and $\iblock{-\sand\Psi,\iblock{\lef{\beta}\sand \Phi_2,\rig{\theta}}}$ are provable with $\var{\beta}\subseteq\var{\alpha}\cup\bigl(\var{\Phi_1}\cap\var{\Psi}\bigr)$. 
    \end{itemize}
\end{lemma}

\begin{proof}
    We prove by induction on the height of a focused proof of $\Sigma$, the idea is to verify that each rule of $\cbplus$ preserves the properties stated in the lemma. 
    Assume the last rule applied in the derivation is $r$, we only show the cases of initial sequents and inductive steps of $\lef\to$ and $\rig\to$ here, as the other cases are straightforward. 
    \begin{itemize}[leftmargin=*]
        \item Assume $\Sigma$ is an instance of $\idone$, which implies the active unit can only be of the form (a) or (b1). 
        If it is in (a), $\Sigma=\csseq{\Psi,\iblock{\Lambda,\lef p, \rig p\sand\Theta}}$ for some context. By definition,  
        $\actl{\Sigma}{ \ }=\iblock{-\sand \Lambda, \lef p \{ \ \} }$ and 
        $\actr{\Sigma}{ \ }=\iblock{-\sand \rig p, \{\ \}}$. 
        We simply take $A$ as $\theta$ 
        then both $\actl{\Sigma}{\rig \theta}$ and $\actr{\Sigma}{\lef \theta}$ are initial sequents, 
        The three conditions (C1)-(C3) follow directly. 

        If it is in (b1), $\Sigma=\csseq{\Psi,\iblock{\Theta\sand \Lambda,\lef p, \rig p}}$ for some context. By definition,  
        $\actl{\Sigma}{ \ }=\iblock{-\sand \Psi,\iblock{\Theta\sand \Lambda,\lef p, \{ \ \}}}$ and 
        $\actr{\Sigma}{ \ }=\iblock{-\sand \rig p, \{ \ \}}$. We simply take $A$ as $\theta$ for (C1)-(C2) and $\beta=\alpha$, the proof goes the same as the case of (a). For (C3), take $\beta=\alpha$, then the required sequent 
        $\iblock{-\sand\Psi,\iblock{\lef \alpha\sand\Lambda,\lef p, \rig p}}$ is an initial sequent. 
        
        \item $r=\lef\to$. By Lemma \ref{lem:normal-imp}, the application is either of the two following forms:  
        \begin{center}
        \begin{prooftree}
        \hypo{\Sigma_1=\csseq{\Lambda,\lef{C\to D},\iblock{\Phi_1,\rig C\sand\Phi_2}}}
        \hypo{\Sigma_2=\csseq{\Lambda,\lef{C\to D},\iblock{\Phi_1\sand\Phi_2,\rig F,\lef D}}}
        \infer2[$\ruletol$]{\Sigma_0=\csseq{\Lambda,\lef{C\to D},\iblock{\Phi_1\sand\Phi_2,\rig F}}}
        \end{prooftree}
        \end{center}
        \begin{center}
        \begin{prooftree}
        \hypo{\Sigma_1=\csseq{\Lambda,\lef{C\to D},\iblock{\Phi_1,\rig C\sand\Phi_2}}}
        \hypo{\Sigma_2=\csseq{\Lambda,\lef{C\to D},\iblock{\Phi_1,\rig F\sand\Phi_2,\lef D}}}
        \infer2[$\ruletol$]{\Sigma_0=\csseq{\Lambda,\lef{C\to D},\iblock{\Phi_1,\rig F\sand\Phi_2}}}
        \end{prooftree}
        \end{center}
        By Proposition \ref{prop:non-empty-component}, in both cases $\Phi_1$ is not empty. 
        
        In the first case, we further distinguish two sub-cases when $\Phi_2$ is empty or not as it makes a difference to the MO of the conclusion. 
        \begin{enumerate}
        \item[(i).] $\Phi_2$ is empty. By definition, 
        for the left premise, $\actl{\Sigma_1}{ \ }=\iblock{-\sand\Phi_1,\{ \ \}}$ and $\actr{\Sigma_1}{ \ }=\iblock{-\sand \{ \ \},\rig C}$; 
        for the right premise, $\actl{\Sigma_2}{ \ }=\iblock{-\sand\Lambda, \lef{C\to D},\iblock{\Phi_1\sand\lef D, \{ \ \}}}$ and $\actr{\Sigma_2}{ \ }=\iblock{-\sand \{ \ \},\rig F}$.         
        Meanwhile, for the conclusion, $\actl{\Sigma_0}{ \ }=\iblock{-\sand \Lambda,\lef{C\to D},\{ \ \}}$ and $\actr{\Sigma_0}{ \ }=\iblock{-\sand\{ \ \}, \iblock{\Phi_1\sand \rig F}}$. 

        By IH to the left premise, there is a formula $\theta_1$ such that both $\actl{\Sigma_1}{\rig{\theta_1}}=\iblock{-\sand\Phi_1,\rig{\theta_1}}$ and $\actr{\Sigma_1}{\lef{\theta_1}}=\iblock{-\sand\lef{\theta_1},\rig C}$ are provable and $\var{\theta_1}\subseteq \var{C}\cap \var{\Phi_1}$. 
        By IH to the right premise, there is a formula $\theta_2$ such that both $\actl{\Sigma_2}{\rig{\theta_2}}=\iblock{-\sand\Lambda,\lef{C\to D},\iblock{\Phi_1\sand\lef D,\theta_2}}$ and $\actr{\Sigma_2}{\lef{\theta_2}}=\iblock{-\sand\lef{\theta_2},\rig C}$ are provable and $\var{\theta_2}\subseteq \var{F}\cap \var{\Lambda,C\to D,D}=\var{F}\cap\var{\Lambda,C,D}$. 
        Meanwhile, since $\act{\Sigma_2}$ is of the form (b1), 
        take $\alpha=\theta_1$, 
        we further have a formula $\beta$ such that $\iblock{-\sand\Phi_1,\rig \beta}, \iblock{-\sand\lef \beta,\rig{\theta_1}}$ and $\iblock{-\sand \Lambda,\lef{C\to D},\iblock{\lef \beta\sand\Phi_1,\rig {\theta_1}}}$ are provable and $\var{\beta}\subseteq \var{\theta_1}\cup(\var{\Phi_1}\cap\var{\Lambda,\lef{C\to D}})$. Thus $\var{\beta}\subseteq\var{C,D,\Lambda}\cap\var{\Phi_1}$. 
        
        Take $\theta=\beta\to\theta_2$, we need to show both 
        (C1-a) $\actl{\Sigma_0}{\rig \theta}=\iblock{-\sand \Lambda,\lef{C\to D},\rig{\beta\to\theta_2}}$ and 
        (C1-b) $\actr{\Sigma_0}{\lef \theta}=\iblock{-\sand\lef{\beta\to\theta_2},\iblock{\Phi_1\sand\rig F}}$ are provable and (C2) $\var{\theta}\subseteq\var{\Lambda,\lef{C\to D}}\cap\var{\Phi_1}= \var{C,D,\Lambda}\cap\var{\Phi_1}$. 
        Notice that (C3) holds immediately as $\act{\Sigma_0}$ is not of the form (b1). 
        (C1-a) is obtained by the following derivation: 
        \begin{center}
            \begin{prooftree}
            \hypo{}
            \infer1[IH, right, C3]{\iblock{-\sand\lef \beta,\rig{\theta_1}}}
            \hypo{}
            \infer1[IH, left, C1]{\iblock{-\sand \lef{\theta_1},\rig {C}}}
            \infer2[$\cut$]{\iblock{-\sand\lef\beta,\rig C}}
            \infer1[Prop. \ref{prop:0-free-gen}]{\iblock{-\sand \Lambda,\lef{C\to D},\iblock{\lef\beta,\rig C\sand-}}}
            \hypo{}
            \infer1[IH, right, C3]{\iblock{-\sand \Lambda,\lef{C\to D},\iblock{\lef\beta\sand\rig{\theta_2},\lef D}}}
            \infer2{\iblock{-\sand \Lambda,\lef{C\to D},\iblock{\lef\beta\sand\rig{\theta_2}}}}
            \infer1{\iblock{-\sand \Lambda,\lef{C\to D},\rig{\beta\to\theta_2}}}
            \end{prooftree}
        \end{center}

        Similarly, (C1-b) is obtained by 
        \begin{center}
            \begin{prooftree}
            \hypo{}
            \infer1[IH, right, C3]{\iblock{-\sand\Phi_1,\rig\beta}}
            \infer1[Prop. \ref{prop:0-free-gen}]{\iblock{-\sand\lef{\beta\to\theta_2},\iblock{\Phi_1,\rig\beta\sand-}}}
            \hypo{}
            \infer1[IH, right, C1]{\iblock{-\sand\lef{\theta_2},\rig F}}
            \infer1[Prop. \ref{prop:0-free-gen}]{\iblock{-\sand\lef{\beta\to\theta_2},\iblock{\Phi_1\sand\rig F,\lef{\theta_2}}}}
            \infer2{\iblock{-\sand\lef{\beta\to\theta_2},\iblock{\Phi_1\sand\rig F}}}
            \end{prooftree}
        \end{center}
        
        Lastly, (C2) follows from both $\var{\beta}\subseteq\var{C,D,\Lambda}\cap\var{\Phi_1}\subseteq\var{C,D,\Lambda}\cap\var{\Phi_1,F}$ and $\var{\theta_2}\subseteq\var{F}\cap\var{C,D,\Lambda}\subseteq\var{C,D,\Lambda}\cap\var{\Phi_1,F}$. 

        \item[(ii).]  $\Phi_2$ is non-empty. 
        By definition, 
        for the left premise, $\actl{\Sigma_1}{ \ }=\iblock{-\sand\Phi_1,\{ \ \}}$ and $\actr{\Sigma_1}{ \ }=\iblock{-\sand \{ \ \},\rig C}$; 
        for the right premise, $\actl{\Sigma_2}{ \ }=\iblock{-\sand\Lambda, \lef{C\to D},\iblock{\Phi_1\sand\Phi_2, \lef D, \{ \ \}}}$ and $\actr{\Sigma_2}{ \ }=\iblock{-\sand \{ \ \},\rig F}$.         
        Meanwhile, for the conclusion, we have $\actl{\Sigma_0}{ \ }=\iblock{-\sand\Lambda, \lef{C\to D},\iblock{\Phi_1\sand\Phi_2, \{ \ \}}}$ and $\actr{\Sigma_0}{ \ }=\actr{\Sigma_2}{ \ }$. 
        By IH to $\Sigma_1$, there is a formula $\theta_1$ such that both 
        $\actl{\Sigma_1}{\rig{\theta_1}}=\iblock{-\sand\Phi_1,\rig{\theta_1}}$ and $\actr{\Sigma_1}{\lef{\theta_1}}=\iblock{-\sand \lef{\theta_1},\rig C}$ are provable and $\var{\theta_1}\subseteq \var{C}\cap \var{\Phi_1}$. 
        By IH to $\Sigma_2$, there is a formula $\theta_2$ such that both 
        $\actl{\Sigma_2}{\rig{\theta_2}}=\iblock{-\sand\Lambda, \lef{C\to D},\iblock{\Phi_1\sand\Phi_2, \lef D, \rig {\theta_2}}}$ and $\actr{\Sigma_2}{\lef{\theta_2}}=\iblock{-\sand \lef{\theta_2},\rig F}$ are provable and $\var{\theta_2}\subseteq \var{F}\cap \var{\Phi_2, \Lambda,C\to D}$. 

        Since the left premise $\Sigma_1$ is provable, by Proposition \ref{prop:component-remove}, $\iblock{-\sand\Phi_1,\rig C}$ is provable. 
        We take $\theta=\theta_2$ and we only need to prove $\actl{\Sigma_0}{\theta}=\iblock{-\sand\Lambda, \lef{C\to D},\iblock{\Phi_1\sand\Phi_2, \rig{\theta_2}}}$, which is derived from 
        \begin{center}
            \begin{prooftree}
                \hypo{}
                \infer1{\iblock{-\sand\Phi_1,\rig C}}
                \infer1[Prop. \ref{prop:0-free-gen}]{\iblock{-\sand\Lambda, \lef{C\to D},\iblock{\Phi_1,\rig C\sand\Phi_2}}}
                \hypo{}
                \infer1[IH, right, C1]{\iblock{-\sand\Lambda, \lef{C\to D},\iblock{\Phi_1\sand\Phi_2, \rig{\theta_2},\lef D}}}
                \infer2[$\lef \to$]{\iblock{-\sand\Lambda, \lef{C\to D},\iblock{\Phi_1\sand\Phi_2, \rig{\theta_2}}}}
            \end{prooftree}
        \end{center}
    
        Lastly, for (C3), notice both $\act{\Sigma_2}$ and $\act{\Sigma_0}$ are of form (b1). It suffices to show for $\beta'$ given by (C3) of $\Sigma_2$, there exists a corresponding $\beta$ satisfies (C3) of $\Sigma_0$. We take $\beta=\beta'\wedge\theta_1$ and need to show (1) $\iblock{-\sand \Phi_1,\rig{\beta'\wedge\theta_1}}$, (2) $\iblock{-\sand\rig \alpha,\lef{\beta'\wedge\theta_1}}$ and (3) $\iblock{-\sand \Lambda,C\to D, \iblock{\lef{\beta'\wedge\theta_1}\sand \Phi_2,\rig{\theta_2}}}$ are provable and (4) $\var{\beta}\subseteq\var{\alpha}\cup(\var{\Phi_1}\cap\var{\Lambda,C\to D})$. 
        By IH, both $\iblock{-\sand \Phi_1,\rig{\beta'}}$ and $\iblock{-\sand \Phi_1,\rig{\theta_1}}$ are provable, which yields (1). By IH, $\iblock{-\sand\rig \alpha,\lef {\beta'}}$ and $\iblock{-\sand \Lambda,C\to D, \iblock{\lef{\beta'}\sand \Phi_2,\rig{\theta_2}}}$ are provable, then (2) and (3) follow by weakening and $\lef\wedge$. For the condition of variables (4), by IH, we have $\var{\beta'}\subseteq\var{\alpha}\cup(\var{\Phi_1}\cap\var{\Lambda,C\to D})$ and $\var{\theta_1}\subseteq\var{\Phi_1}\cap\var{C}\subseteq\var{\Phi_1}\cap\var{\Lambda,C\to D}$, so (4) follows directly. 
        \end{enumerate}

        In the second case, by definition, $\act{\Sigma_0}$ is not of the form of  (b1), the condition of (C3) holds trivially. Thus we only need to check (C1) and (C2). 
        Since $\Phi_1$ is non-empty, by definition, $\actl{\Sigma_0}{ \ }=\actl{\Sigma_2}{ \ }=\iblock{-\sand\Phi_1,\{ \ \}}$ and $\actr{\Sigma_0}{ \ }=\actr{\Sigma_2}{ \ }=\iblock{-\sand\{ \ \},\rig F}$. 
        By IH to $\Sigma_2$, there is $\theta_2$ such that $\iblock{-\sand \Phi_1, \rig {\theta_2}}$ and $\iblock{-\sand \lef{\theta_2},\rig F}$ are provable and $\var{\theta_2}\subseteq\var{\Phi_1}\cap\var{F}$. 
        We take $\theta_2$ as the required $\theta$ for the conclusion and the conditions follow directly. 
        
        \item $r=\rig \to$. By Lemma \ref{lem:normal-imp}, the application is of the following form:  
        \begin{center}
        \begin{prooftree}
            \hypo{\Sigma_1=\csseq{\Psi,\iblock{\lef E\sand \rig F}}}
            \infer1[$\ruletor$]{\Sigma_0=\csseq{\Psi,\rig{E\to F}}}
        \end{prooftree}
        where $\Psi$ is non-empty. 
        \end{center}
        In this case, by definition, for the premise, $\actl{\Sigma_1}{ \ }=\iblock{-\sand\Psi,\{ \ \}}$ and $\actr{\Sigma_1}{ \ }=\iblock{-\sand\{ \ \},\iblock{\lef E\sand\rig F} }$. By IH, there is $\theta_1$ such that both $\actl{\Sigma_1}{\rig{\theta_1}}=\iblock{-\sand\Psi,\rig{\theta_1}}$ and $\actr{\Sigma_1}{\lef{\theta_1}}=\iblock{-\sand \lef{\theta_1},\iblock{\lef E\sand\rig F}}$ are provable and $\var{\theta_1}\subseteq\var{\Psi}\cap\var{E,F}$. 
        Meanwhile, for the conclusion, $\Sigma_0$ is of the form (i) 
        $\dsseq{\Lambda,\iblock{\Phi\sand\Psi,\rig{E\to F}}}$ such that $\dsseq{\Lambda,\iblock{\Phi\sand \{ \ \}}}=\csseq{ \ }$; or 
        (ii) $\dsseq{\Lambda,\iblock{\Psi,\rig{E\to F}\sand\Phi}}$ such that  $\dsseq{\Lambda,\iblock{\{ \ \}\sand\Phi}}=\csseq{ \ }$. 

        For (i), $\act{\Sigma_0}=\Lambda,\iblock{\Phi\sand\Psi,\rig{E\to F}}$. 
        Take $\theta=\theta_1$, we first show (C1) $\actl{\Sigma_0}{\rig \theta}=\iblock{-\sand \Lambda,\iblock{\Phi\sand\Psi,\rig{\theta_1}}}$ and $\actr{\Sigma_0}{\lef \theta}=\iblock{-\sand \lef{\theta_1}, \rig{E\to F}}$ are provable. $\iblock{-\sand \Lambda,\iblock{\Phi\sand\Psi,\rig{\theta_1}}}$ follows directly from $\iblock{-\sand\Psi,\rig{\theta_1}}$ by Proposition \ref{prop:0-free-gen}. $\iblock{-\sand \lef{\theta_1}, \rig{E\to F}}$ is derived from $\iblock{-\sand \lef{\theta_1},\iblock{\lef E\sand\rig F}}$ by $\ruletor$. 
        Next, (C2) is satisfied due to the fact $\var{\theta}=\var{\theta_1}\subseteq\var{\Psi}\cap\var{E,F}\subseteq \var{\Lambda,\Psi}\cap\var{E\to F}$. 
        Lastly for (C3), we take $\beta=\alpha$ which is given by the assumption of (C3). The variable condition and the provability of $\iblock{-\sand \Phi_1,\rig \beta}$ and $\iblock{-\sand \rig \alpha,\lef\beta}$ hold trivially. $\iblock{-\sand\Lambda,\iblock{\Phi\sand\Psi,\rig{\theta}}}$ follows directly from $\iblock{-\sand\Psi,\rig{\theta_1}}$ by Proposition \ref{prop:0-free-gen}.  

        For (ii), the conditions of (C1) and (C2) follow the similar argument as in (i); moreover, since in (ii) $\act{\Sigma_0}$ is not of the form of  (b1), the condition of (C3) holds trivially. 
        
    \end{itemize}
    This completes the proof. 
\end{proof}

\begin{theorem}[CIP for $\bplus$]
    $\bplus$ enjoys the \cip. 
\end{theorem}

\begin{proof}
    Assume $A\to B$ is a theorem of $\bplus$. We have $\iblock{-\sand\lef A,\rig B}$ is provable. By Lemma~\ref{lem:focus-proof}, it has a focused proof. 
    Let us take the root $\Sigma=\iblock{-\sand\lef A,\rig B}$, by definition, 
    $\act{\Sigma}=\iblock{-\sand\lef {A},\rig{B}}$ with 
    $\actr{\Sigma}{ \ }=
    \iblock{-\sand\{ \ \},\rig B}$ and $\actl{\Sigma}{ \ }=\iblock{-\sand\iblock{-\sand\lef A, \{ \ \}}}$. 
    By Lemma~\ref{lem:interpolation-calculus}, there exists a formula $\theta$ such that $\iblock{-\sand\iblock{-\sand\lef A, \rig\theta}}$ and $\iblock{-\sand\lef\theta, \rig B}$ are provable and $\var{\theta}\subseteq\var{A}\cap\var{B}$. 
    By Proposition \ref{prop:0-free-gen}, $\iblock{-\sand\lef A, \rig\theta}$ is also provable. 
    By Proposition \ref{prop:imp-eq}, we have both $\rig {A\to\theta}$ and $\rig{\theta\to B}$ are provable in $\cbplus$, and hence are theorems in $\bplus$. 
\end{proof}
\section{Concluding remarks}\label{sec:con}
We have proposed the first nested calculus for $\bplus$ by relying on binary  relevant blocks, and used it to establish variants of the variable-sharing property and Craig interpolation property for $\bplus$. Several topics regarding both the logic and the proof-theoretic framework are left for further investigation. 

\noindent \textbf{Decidability and complexity.} \ 
We do not treat decidability and complexity by means of our calculus in this paper, since the logic $\bplus$ is known to be decidable and an $\+O(n\log n)$-space upper bound has been provided in~\cite{Vigano2000} using another proof system. We conjecture our calculus also supports a terminating proof search for $\bplus$, but it might not entail the same bound in an obvious way. 

\noindent \textbf{Extensions of $\bplus$.} 
We intend to study similar nested calculi for various extensions of $\bplus$.  
We do not address the full logic $\mathsf{B}$ in this work, since accommodating relevant negation within nested sequents is non-trivial and would likely require additional structural operators. We expect the negation-free framework developed here to be both sufficiently general and able to capture a wide range of logics, including negation-free fragments of other relevant logics and, more broadly, 
distributive substructural logics with relational semantics, or other logics with $n$-ary style relational semantics. 
Lastly, we are also interested in modal logics with a relevant base. Similar interacting frame conditions as intuitionistic modal logic are considered for some logics in this family, but their proof theory is far from developed.

\section*{AI Disclosure}

Claude and Gemini were used in this work to 
check grammar and improve language, fix \LaTeX~errors, and draw Figure~\ref{fig:frame-condition} and \ref{graph:sequent}. 
Any LLM was not used to produce any scientific content. The authors reviewed the final manuscript and are therefore responsible for all the content of this paper. 

\section*{Acknowledgement}

\noindent We thank Nicola Olivetti and Marianna Girlando for explaining the underlying idea of using nested sequents in conditional logics, which helped us to make the comparison in Remark~\ref{rem:conditional}. 
Fabio De Martin Polo was supported by the project OP JAK: TRUST Knowledge in the Age of Distrust, CZ.02.01.01/00/23\_025/0008711. Han Gao was supported by Czech Science Foundation (GACR) grant 25-18306M, INTERACT.


\appendix

\section{Appendix}

\completenesscbplus*

\begin{proof}
We proceed to derive the axioms and rules of $\cbplus$. Since the derivations are routine, we restrict attention to a few illustrative cases.

$\vdash_{\cbplus} \rig{A \land (B \lor C)  \to (A \land B)  \lor (A \land C)}$


\begin{center}
            \begin{adjustbox}{max width=.8\textwidth}
                \begin{tabular}{c}
\begin{prooftree}
    \hypo{}
    \infer1[$\idone$]{\iblock{\lef{A} ,\lef{B}\sand \lef A, \rig{A}}}
    \infer1[$\ruleliftib$]{\iblock{\lef{A} ,\lef{B}\sand \rig{A}}}
 \hypo{}
    \infer1[$\idone$]{\iblock{\lef{A} ,\lef{B}\sand \lef B, \rig{B}}}
    \infer1[$\ruleliftib$]{\iblock{\lef{A} ,\lef{B}\sand \rig{B}}}
    \infer2[$\ruleandr$]{\iblock{\lef{A} ,\lef{B}\sand \rig{A \land B}}}
    \infer1[$\ruleorr$]{\iblock{\lef{A} ,\lef{B}\sand \rig{(A \land B) \lor   (A \land C)}}}
\hypo{}
    \infer1[$\idone$]{\iblock{\lef{A} ,\lef{C}\sand \lef A,  \rig{A }}}
    \infer1[$\ruleliftib$]{\iblock{\lef{A} ,\lef{C}\sand  \rig{A }}}
\hypo{}
    \infer1[$\idone$]{\iblock{\lef{A} ,\lef{C}\sand \lef C,  \rig{ C}}}
    \infer1[$\ruleliftib$]{\iblock{\lef{A} ,\lef{C}\sand   \rig{ C}}}
    \infer2[$\ruleandr$]{\iblock{\lef{A} ,\lef{C}\sand \rig{A \land C}}}
    \infer1[$\ruleorr$]{\iblock{\lef{A} ,\lef{C}\sand \rig{(A \land B)\lor(A \land C)}}}
        \infer2[$\ruleorl$]{\iblock{\lef{A}, \lef{B \lor C}\sand \rig{(A \land B)  \lor (A \land C)}}}
    \infer1[$\ruleandl$]{\iblock{\lef{A \land (B \lor C)}\sand \rig{(A \land B)  \lor (A \land C)}}}
    \infer1[$\ruletor$]{\rig{A \land (B \lor C)  \to (A \land B)  \lor (A \land C)}}
\end{prooftree}
                \end{tabular}
            \end{adjustbox}
            \end{center}

For {\sf mp}, we show that if $\vdash_{\cbplus} \rig {A\to B}$ and  $\vdash_{\cbplus} \rig{A} $, then $\vdash_{\cbplus} \rig{B} $. 


\begin{center}
            \begin{adjustbox}{max width=.75\textwidth}
                \begin{tabular}{c}
\begin{prooftree}

    \hypo{\rig A}
    \hypo{\rig{A\to B}}
    \hypo{}
    \infer1[$\idone$]{\lef{A\to B},\iblock{\lef A, \rig A\sand -}}
     \hypo{}
    \infer1[$\idone$]{\lef{A\to B},\iblock{\lef A\sand \lef B, \rig B}}
        \infer2[$\ruletol$]{\lef{A\to B},\iblock{\lef A\sand \rig B}}
    \infer2[$\cut$]{\iblock{\lef A\sand \rig B}}
    \infer1[$\sf R1$]{\lef A, \rig B}
    \infer2[$\cut$]{\rig{B}}
\end{prooftree}
                \end{tabular}
            \end{adjustbox}
            \end{center}

For $\sf aff$, we prove that if $\vdash_{\cbplus} \rig{A\to B}$ and $\vdash_{\cbplus}\rig{C\to D}$, then $\vdash_{\cbplus}\rig{(B \to C)  \to (A \to D)}$.

 
 \begin{center}
            \begin{adjustbox}{max width=.9\textwidth}
                \begin{tabular}{c}
\begin{prooftree}
\hypo{\rig{A\to B}}
    \hypo{}
    \infer1[$\idone$]{\lef{A\to B},\iblock{\lef{A},\rig{A}\sand -}}
    \hypo{}
    \infer1[$\idone$]{\lef{A\to B},\iblock{\lef{A}\sand \lef{B},\rig{B}}}
        \infer2[$\ruletol$]{\lef{A\to B},\iblock{\lef{A}\sand \rig{B}}}
    \infer2[$\cut$]{\iblock{\lef{A}\sand \rig{B}}}
    \infer1[$\sf R1$]{\iblock{\lef{B \to C},\iblock{\lef{A}, \rig{B} \sand - } \sand - }}
    \hypo{\rig{C\to D}}
 \hypo{}
      \infer1[$\idone$]{\lef{C\to D},\iblock{\lef{C}, \rig{C}\sand -}}
      \hypo{}
      \infer1[$\idone$]{\lef{C\to D},\iblock{\lef{C}\sand \lef{D},\rig{D}}}
      \infer2[$\ruletol$]{\lef{C\to D},\iblock{\lef{C}\sand\rig{D}}}
      \infer2[$\cut$]{\iblock{\lef{C}\sand\rig{D}}}
    \infer1[$\sf R1$]{\iblock{\lef{B \to C},\iblock{\lef{A} \sand \lef{C},\rig{D}}\sand - }}
    \infer2[$\ruletol$]{\iblock{\lef{B \to C},\iblock{\lef{A} \sand\rig{D}}\sand - }}
        \infer1[$\sf R2$]{\iblock{\lef{B \to C} \sand \iblock{\lef{A} \sand\rig{D}}}}
    \infer1[$\ruletor$]{\iblock{\lef{B \to C} \sand \rig{A \to D}}}
    \infer1[$\ruletor$]{\rig{(B \to C)  \to (A \to D)}}
\end{prooftree}
                \end{tabular}
            \end{adjustbox}
            \end{center}
            This completes the proof. 
\end{proof}

\nocite{*}
\bibliographystyle{plain}
\bibliography{refs}

@article{standefer2022connective,
  title={What is a relevant connective?},
  author={Standefer, Shawn},
  journal={Journal of Philosophical Logic},
  volume={51},
  number={4},
  pages={919--950},
  year={2022},
}

@inproceedings{Knudstorp2026,
  author    = {Knudstorp, S{\o}ren Brinck},
  title     = {Possibly Relevant Translations},
  booktitle = {Advances in Modal Logic 2026},
  editor    = {B{\'i}lkov{\'a}, Marta and Gattinger, Malvin and {van der Giessen}, Iris and Girlando, Marianna and Wang, Yanjing},
  volume    = {447},
  pages     = {567--583},
  year      = {2026},
}

@article{MarchioniMetcalfe2012,
  author    = {Marchioni, Enrico and Metcalfe, George},
  title     = {Craig interpolation for semilinear substructural logics},
  journal   = {Mathematical Logic Quarterly},
  volume    = {58},
  number    = {6},
  pages     = {468--481},
  year      = {2012},
}

@article{FussnerTedder2026,
  author  = {Fussner, Wesley and Tedder, Andrew},
  title   = {Revisiting Interpolation in Relevant Logics},
  journal = {Bulletin of the Section of Logic},
  year    = {2026},
  note    = {Published online}
}

@article{seki2003completeness,
  title={A {S}ahlqvist theorem for relevant modal logics},
  author={Seki, Takahiro},
  journal={Studia Logica},
  volume={73},
  number={3},
  pages={383--411},
  year={2003},
}

@article{FabioDMP2023,
  title={Modular labelled calculi for relevant logics},
  author={{De Martin Polo}, Fabio},
  journal={The Australasian Journal of Logic},
  volume={20},
  number={1},
  pages={47--87},
  year={2023}
}

@article{seki2003,
  title={General frames for relevant modal logics},
  author={Seki, Takahiro},
  journal={Notre Dame Journal of Formal Logic},
  volume={44},
  number={2},
  pages={93--109},
  year={2003},
}

@article{fuhrmann1990,
  title={Models for relevant modal logics},
  author={Fuhrmann, Andr{\'e}},
  journal={Studia Logica},
  volume={49},
  number={4},
  pages={501--514},
  year={1990},
}

@article{giambrone.red,
	author = {Giambrone, Steve},
	journal = {Notre Dame journal of formal logic},
	number = {3},
	pages = {442--449},
	title = {Real reduced models for relevant logics without {WI}},
	volume = {33},
	year = {1992}}

@article{slaney.red,
	author = {Slaney, John K},
	journal = {Notre Dame Journal of Formal Logic},
	number = {3},
	pages = {395--407},
	title = {Reduced models for relevant logics without {WI}},
	volume = {28},
	year = {1987}}

@incollection{routleymeyer73,
	author = {Routley, Richard and Meyer, Robert K},
	booktitle = {Truth, Syntax, and Modality (Proceedings Of The Temple University Conference on Alternative Semantics)},
	editor = {Leblanc, Hugues},
	pages = {199--243},
	publisher = {North-Holland Publishing Company},
	title = {The Semantics of Entailment {I}},
	year = {1973}}

@book{rl.rivals,
	author = {Routley, Richard and Plumwood, Val and Meyer, Robert K and Brady, Ross T},
	publisher = {Ridgeview Publishing Company},
	title = {Relevant Logics and their Rivals},
	volume = { 1},
	year = {1982}}

@article{urquhart.proof.semil.,
	author = {Giambrone, Steve and Urquhart, Alasdair},
	journal = {Zeitschrift F{\"u}r Mathematische Logik Und Grundlagen Der Mathematik},
	number = {5},
	pages = {433--439},
	title = {Proof theories for semilattice logics},
	volume = {33},
	year = {1987}}

@incollection{kurokawa.negri,
	author = {Kurokawa, Hidenori and Negri, Sara},
	booktitle = {Contemporary Logic and Computing},
	pages = {259--306},
	publisher = {College publications},
	title = {Relevant logics: from semantics to proof systems},
	year = {2020}}

@book{vigano,
	author = {Vigan{\`o}, Luca},
	publisher = {Springer},
	title = {Labelled Non-classical Logics},
	year = {2000}}

@book{belnap.anderson,
	author = {Anderson, Alan R and Belnap, Nuel D},
	publisher = {Princeton University Press},
	title = {Entailment: The Logic of Relevance and Necessity},
	volume = { I},
	year = {1975}}

@book{bimbobook,
	author = {Bimb{\'o}, Katalin},
	publisher = {CRC Press},
	title = {Proof theory: Sequent Calculi and Related Formalisms},
	year = {2014}}

@InProceedings{CSL-2024,
  author =	{Balbiani, Philippe and Gao, Han and Gencer, {\c{C}}i\u{g}dem and Olivetti, Nicola},
  title =	{{A Natural Intuitionistic Modal Logic: Axiomatization and Bi-Nested Calculus}},
  booktitle =	{32nd EACSL Annual Conference on Computer Science Logic (CSL 2024)},
  pages =	{13:1--13:21},
  year =	{2024},
  volume =	{288}
}

@inproceedings{IJCAR-2024,
  title={Local Intuitionistic Modal Logics and Their Calculi},
  author={Balbiani, Philippe and Gao, Han and Gencer, {\c{C}}i{\u{g}}dem and Olivetti, Nicola},
  booktitle={International Joint Conference on Automated Reasoning},
  pages={78--96},
  year={2024},
  organization={Springer}
}

@article{Fitting:2014,
  author       = {Melvin Fitting},
  title        = {Nested Sequents for Intuitionistic Logics},
  journal      = {Notre Dame J. Formal Log.},
  volume       = {55},
  number       = {1},
  pages        = {41--61},
  year         = {2014},
  doi          = {10.1215/00294527-2377869}
}

@inproceedings{Strassburger:2013-ik-cut,
  title={Cut elimination in nested sequents for intuitionistic modal logics},
  author={Stra{\ss}burger, Lutz},
  booktitle={Foundations of Software Science and Computation Structures: 16th International Conference, FOSSACS 2013},
  pages={209--224},
  year={2013},
  organization={Springer}
}

@article{Brunner:2009,
  author       = {Kai Br{\"{u}}nnler},
  title        = {Deep sequent systems for modal logic},
  journal      = {Arch. Math. Log.},
  volume       = {48},
  number       = {6},
  pages        = {551--577},
  year         = {2009},
  doi          = {10.1007/S00153-009-0137-3}
}

@article{RoutleyMeyer1972,
  author  = {Routley, Richard and Meyer, Robert K},
  title   = {The Semantics of Entailment {III}},
  journal = {Journal of Philosophical Logic},
  volume  = {1},
  year    = {1972},
  pages   = {192--208}
}

@article{Meyer2004,
  title={Ternary relations and relevant semantics},
  author={Meyer, Robert K},
  journal={Annals of Pure and Applied Logic},
  volume={127},
  number={1-3},
  pages={195--217},
  year={2004},
}

@article{PalMeyer2005,
  title={Basic relevant theories for combinators at levels {I} and {II}},
  author={Pal, Koushik and Meyer, Robert K},
  journal={The Australasian Journal of Logic},
  volume={3},
  year={2005}
}

@incollection{Gochet2005,
  title={Algorithms for Relevant Logic},
  author={Paul Gochet and Pascal Gribomont and Didier Rossetto},
  booktitle={Logic, Thought and Action},
  pages={479--496},
  year={2005},
  publisher={Springer}
}

@article{Vigano2000,
  title={An $O$($n$ log $n$)-space decision procedure for the relevance logic {B+}},
  author={Vigan{\`o}, Luca},
  journal={Studia Logica},
  volume={66},
  number={3},
  pages={385--407},
  year={2000},
  publisher={Springer}
}

@book{AndersonBelnapDunn1992,
  author    = {Anderson, Alan R and Dunn, Jon M and Belnap, Nuel D},
  title     = {Entailment: The Logic of Relevance and Necessity},
  volume    = { II},
  year      = {1992},
  publisher = {Princeton University Press}
}

@article{kuznets2019maehara,
  title={Maehara-style modal nested calculi},
  author={Kuznets, Roman and Stra{\ss}burger, Lutz},
  journal={Archive for Mathematical Logic},
  volume={58},
  number={3},
  pages={359--385},
  year={2019},
  publisher={Springer}
}

@article{DunnMeyer1997Combinators,
  title={Combinators and structurally free logic},
  author={Dunn, Jon M and Meyer, Robert K},
  journal={Logic Journal of IGPL},
  volume={5},
  number={4},
  pages={505--537},
  year={1997},
}

@incollection{LellmannPoggiolesi2024,
  author    = {Lellmann, Bj{\"o}rn and Poggiolesi, Francesca},
  title     = {Nested Sequents or Tree-Hypersequents -- A Survey},
  booktitle = {Saul Kripke on Modal Logic},
  publisher = {Springer},
  year      = {2024},
  pages     = {243--301},
  doi       = {10.1007/978-3-031-57635-5_11}
}

@article{Lyon2021Correspondence,
  title={On the correspondence between nested calculi and semantic systems for intuitionistic logics},
  author={Lyon, Tim S},
  journal={Journal of Logic and Computation},
  volume={31},
  number={1},
  pages={213--265},
  year={2021},
}

@incollection{FabioDMP2025,
  author    = {De Martin Polo, Fabio},
  title     = {Fusion, Fission, and {A}ckermann’s Truth Constant in Relevant Logics: A Proof-Theoretic Investigation},
  booktitle = {New Directions in Relevant Logic},
  publisher = {Springer},
  year      = {2025},
  pages     = {267-316},
  doi       = {10.1007/978-3-031-69940-5_12}
}

@article{Fischer:Servi:1984,
  title={Axiomatizations for some intuitionistic modal logics},
  author={Fischer Servi, Gis{\`e}le},
  journal={Rendiconti del Seminario Matematico Universit{\`a} e Politecnico di Torino},
  volume={42},
  year={1984}
}

@article{Simpson1994-thesis,
  title={The proof theory and semantics of intuitionistic modal logic},
  author={Simpson, Alex K},
  year={1994},
  publisher={University of Edinburgh. School of College of Science and Engineering. }
}

@article{Galmiche:Salhi:2010,
author = {Didier Galmiche and Yakoub Salhi},
title = {Label-free natural deduction systems for intuitionistic and classical modal logics},
journal = {Journal of Applied Non-Classical Logics},
volume = {20},
number = {4},
pages = {373-421},
year = {2010},
publisher = {Taylor & Francis},
doi = {10.3166/jancl.20.373-421}
}

@article{AlendaOlivettiPozzato2016,
  author  = {Alenda, R{\'e}gis and Olivetti, Nicola and Pozzato, Gian Luca},
  title   = {Nested Sequent Calculi for Normal Conditional Logics},
  journal = {Journal of Logic and Computation},
  volume  = {26},
  number  = {1},
  pages   = {7--50},
  month   = feb,
  year    = {2016},
  doi     = {10.1093/logcom/ext034}
}

@book{RoblesMendez2018RM,
  title={Routley-Meyer Ternary Relational Semantics for Intuitionistic-type Negations},
  author={Robles, Gemma and M{\'e}ndez, Jos{\'e} M},
  year={2018},
  publisher={Academic Press}
}

@incollection{Standefer2025vsp,
  title={Variable-sharing as relevance},
  author={Standefer, Shawn},
  booktitle={New directions in relevant logic},
  pages={97--117},
  year={2025},
  publisher={Springer}
}

@ARTICLE{kuznets-lellmann2016-grafted,
  author={Kuznets, Roman and Lellmann, Björn},
  journal={Logic Journal of the IGPL}, 
  title={Grafting hypersequents onto nested sequents}, 
  year={2016},
  volume={24},
  number={3},
  pages={375-423},
  doi={10.1093/jigpal/jzw005}}

@article{Brady:1984-depthrl,
 author = {Ross T. Brady},
 journal = {Studia Logica},
 number = {1/2},
 pages = {63--73},
 publisher = {Springer},
 title = {Depth Relevance of Some Paraconsistent Logics},
 urldate = {2026-08-10},
 volume = {43},
 year = {1984}
}

@article{urquhart1993failure,
  title={Failure of interpolation in relevant logics},
  author={Urquhart, Alasdair},
  journal={Journal of Philosophical Logic},
  volume={22},
  number={5},
  pages={449--479},
  year={1993},
  publisher={Springer}
}

@incollection{fussner2025interpolation,
  title={Interpolation in non-classical logics},
  author={Fussner, Wesley},
  booktitle={Theory and Applications of Craig Interpolation},
  publisher={Ubiquity Press},
  year={2025}
}

@article{gao2026bi-ik,
  title={A bi-nested calculus for intuitionistic {K}: proofs and countermodels},
  author={Gao, Han and Girlando, Marianna and Olivetti, Nicola},
  journal={Journal of Applied Non-Classical Logics},
  pages={1--40},
  year={2026},
  publisher={Taylor \& Francis}
}

@article{logan2021strong,
  title={Strong depth relevance},
  author={Logan, Shay Allen},
  journal={The Australasian Journal of Logic},
  volume={18},
  number={6},
  pages={645--656},
  year={2021}
}

@article{gore1998substructural,
  title={Substructural logics on display},
  author={Gor{\'e}, Rajeev},
  journal={Logic Journal of IGPL},
  volume={6},
  number={3},
  pages={451--504},
  year={1998},
  publisher={Oxford University Press}
}

@article{belnap1982display,
  title={Display logic},
  author={Belnap, Nuel D},
  journal={Journal of philosophical logic},
  volume={11},
  number={4},
  pages={375--417},
  year={1982},
  publisher={JSTOR}
}

@book{standefer2026relevant,
  title={Relevant Logics: Implication, Modality, Quantification},
  author={Standefer, Shawn},
  year={2026},
  publisher={Cambridge University Press}
}

@article{kuznets2018multicomponent,
  title={Multicomponent proof-theoretic method for proving interpolation properties},
  author={Kuznets, Roman},
  journal={Annals of Pure and Applied Logic},
  volume={169},
  number={12},
  pages={1369--1418},
  year={2018},
  publisher={Elsevier}
}

\end{document}